\documentclass[12pt,twoside]{article}
\usepackage{verbatim}

\begin{document}

\begin{titlepage}
\begin{center}
{\Huge\bf A Field Guide to Recent} \\[0.5cm]
{\Huge\bf Work on the Foundations of}\\[0.5cm]
{\Huge\bf Statistical Mechanics}\\[2cm]

Forthcoming in Dean Rickles (ed.): The Ashgate Companion to
Contemporary Philosophy of Physics. London: Ashgate. \\[8cm]

{\Large\bf Roman Frigg \\ [0.3cm]
London School of Economics}\\[0.3cm]
r.p.frigg@lse.ac.uk\\[0.3cm]
www.romanfrigg.org\\[1cm]

March 2008 \\ [0.2cm]

\end{center}
\end{titlepage}


\tableofcontents

\newpage

\section{Introduction}

\subsection{Statistical Mechanics---A Trailer}\label{trailer}

Statistical mechanics (SM) is the study of the connection between
micro-physics and macro-physics.\footnote{Throughout this chapter I
use `micro' and `macro' as shorthands for `microscopic' and
`macroscopic' respectively.} Thermodynamics (TD) correctly describes
a large class of phenomena we observe in macroscopic systems. The
aim of statistical mechanics is to account for this behaviour in
terms of the dynamical laws governing the microscopic constituents
of macroscopic systems and probabilistic assumptions.\footnote{There
is a widespread agreement on the broad aim of SM; see for instance
Ehrenfest \& Ehrenfest (1912, 1), Khinchin (1949, 7), Dougherty
(1993, 843), Sklar (1993, 3), Lebowitz (1999, 346), Goldstein (2001,
40), Ridderbos (2002, 66) and Uffink (2007, 923).}

This project can be divided into two sub-projects, equilibrium SM
and non-equilibrium SM.\index{Equilibrium statistical mechanics}This
distinction is best illustrated with an example. Consider a gas
initially confined to the left half of a box (see Figure 1.1):

\vspace{1cm}

\noindent [Insert Figure 1.1 (appended at the end of the document).]

\vspace{1cm}

\noindent {\footnotesize{Figure 1.1 Initial state of a gas, wholly
confined to the left compartment of a box separated by a
barrier.}\label{frigg1}}

\vspace{1cm}

\noindent This gas is in equilibrium as all natural processes of
change have come to an end and the observable state of the system is
constant in time, meaning that all macroscopic parameters such as
local temperature and local pressure assume constant values. Now we
remove the barrier separating the two halves of the box. As a
result, the gas is no longer in equilibrium and it quickly disperses
(see Figures 1.2 and 1.3):

\vspace{1cm}

\noindent [Insert Figures 1.2 and 1.3 (appended at the end of the
document).]

\vspace{1cm}

\noindent {\footnotesize{Figures 1.2 and 1.3: When the barrier is
removed the gas is no longer in equilibrium and disperses.}}

\vspace{1cm}

\noindent This process of dispersion continues until the gas
homogeneously fills the entire box, at which point the system will
have reached a new equilibrium state (see Figure 1.4):

\vspace{1cm}

\noindent [Insert Figure 1.4 (appended at the end of the document).]

\vspace{1cm}

\noindent {\footnotesize{Figure 1.4: Gas occupies new equilibrium
state.}}

\vspace{1cm}

\noindent From an SM point of view, equilibrium needs to be
characterised in microphysical terms. What conditions does the
motion of the molecules have to satisfy to ensure that the
macroscopic parameters remain constant as long as the system is not
subjected to perturbations from the outside (such as the removal of
barriers)? And how can the values of macroscopic parameters like
pressure and temperature be calculated on the basis of such a
microphysical description? Equilibrium SM provides answers to these
and related questions.

Non-equilibrium SM deals with systems out of
equilibrium.\index{Non-equilibrium statistical mechanics} How does a
system approach equilibrium when left to itself in a non-equilibrium
state and why does it do so to begin with? What is it about
molecules and their motions that leads them to spread out and assume
a new equilibrium state when the shutter is removed? And crucially,
what accounts for the fact that the reverse process won't happen?
The gas diffuses and spreads evenly over the entire box; but it
won't, at some later point, spontaneously move back to where it
started. And in this the gas is no exception. We see ice cubes
melting, coffee getting cold when left alone, and milk mix with tea;
but we never observe the opposite happening. Ice cubes don't
suddenly emerge from lukewarm water, cold coffee doesn't
spontaneously heat up, and white tea doesn't un-mix, leaving a
spoonful of milk at the top of a cup otherwise filled with black
tea. Change in the world is \emph{unidirectional}: systems, when
left alone, move towards equilibrium but not away from it. Let us
introduce a term of art and refer to processes of this kind as
`irreversible'.\index{Irreversible processes} The fact that many
processes in the world are irreversible is enshrined in the
so-called Second Law of thermodynamics\index{Second law of
thermodynamics}, which, roughly, states that transitions from
equilibrium to non-equilibrium states cannot occur in isolated
systems. What explains this regularity? It is the aim of
non-equilibrium SM to give a precise characterisation of
irreversibility and to provide a microphysical explanation of why
processes in the world are in fact irreversible.\footnote{Different
meanings are attached to the term `irreversible' in different
contexts, and even within thermodynamics itself (see Denbigh 1989a
and Uffink 2001, Section 3). I am not concerned with these in what
follows and always use the term in the sense introduced here.}

The issue of irreversibility is particularly perplexing because (as
we will see) the laws of micro physics have no asymmetry of this
kind built into them. If a system can evolve from state $A$ into
state $B$, the inverse evolution, from state $B$ to state $A$, is
not ruled out by any law governing the microscopic constituents of
matter. For instance, there is nothing in the laws governing the
motion of molecules that prevents them from gathering again in the
left half of the box after having uniformly filled the box for some
time. But how is it possible that irreversible behaviour emerges in
systems whose components are governed by laws which are not
irreversible? One of the central problems of non-equilibrium SM is
to reconcile the asymmetric behavior of irreversible thermodynamic
processes with the underlying symmetric dynamics.


\subsection{Aspirations and Limitations}

This chapter presents a survey of recent work on the foundations of
SM from a systematic perspective. To borrow a metaphor of Gilbert
Ryle's, it tries to map out the logical geography of the field,
place the different positions and contributions on this map, and
indicate where the lines are blurred and blank spots occur.
Classical positions, approaches, and questions are discussed only if
they have a bearing on current foundational debates; the
presentation of the material does not follow the actual course of
the history of SM, nor does it aim at historical accuracy when
stating arguments and positions.\footnote{Those interested in the
long and intricate history of SM are referred to Brush (1976), Sklar
(1993, Ch. 2), von Plato (1994) and Uffink (2007).}

Such a project faces an immediate difficulty. Foundational debates
in many other fields can take as their point of departure a
generally accepted formalism and a clear understanding of what the
theory is. Not so in SM. Unlike quantum mechanics and relativity
theory, say, SM has not yet found a generally accepted theoretical
framework, let alone a canonical formulation. What we find in SM is
a plethora of different approaches and schools, each with its own
programme and mathematical apparatus, none of which has a legitimate
claim to be more fundamental than its competitors.

For this reason a review of foundational work in SM cannot simply
begin with a concise statement of the theory's formalism and its
basic principles, and then move on to the different interpretational
problems that arise. What, then, is the appropriate way to proceed?
An encyclopaedic list of the different schools and their programme
would do little to enhance our understanding of the workings of SM.
Now it might seem that an answer to this question can be found in
the observation that across the different approaches equilibrium
theory is better understood than non-equilibrium theory, which might
suggest that a review should begin with a presentation and
discussion of equilibrium, and then move on to examining
non-equilibrium.

Although not uncongenial, this approach has two serious drawbacks.
First, it has the disadvantage that the discussion of specific
positions (for instance the ergodic approach) will be spread out
over different sections, and as a result it becomes difficult to
assess these positions as consistent bodies of theory. Second, it
creates the wrong and potentially misleading impression that
equilibrium theory can (or even should be) thought of as an
autonomous discipline. By disconnecting the treatment of equilibrium
from a discussion of non-equilibrium we lose sight of the question
of how and in what way the equilibrium state constitutes the final
point towards which the dynamical evolution of a system converges.

In what follows I take my lead from the fact that all these
different schools (or at any rate those that I discuss) use (slight
variants) of either of two theoretical frameworks, one of which can
be associated with Boltzmann (1877) and the other with Gibbs (1902),
and can thereby classify different approaches as either
`Boltzmannian' or `Gibbsian'. The reliance on a shared formalism
(even if the understanding of the formalism varies radically)
provides the necessary point of reference to compare these accounts
and assess their respective merits and drawbacks. This is so because
the problems that I mentioned in Subsection \ref{trailer} can be
given a precise formulation only within a particular mathematical
framework. Moreover it turns out that these frameworks give rise to
markedly different characterisations both of equilibrium and of
non-equilibrium, and accordingly the problems that beset accounts
formulated within either framework are peculiar to one framework and
often do not have a counterpart in the other. And last but not
least, the scope of an approach essentially depends on the framework
in which it is formulated, and, as we shall see, there are
significant differences between two (I return to this issue in the
conclusion).

Needless to say, omissions are unavoidable in an chapter-size
review. I hope that any adversity caused by these omissions is
somewhat alleviated by the fact that I clearly indicate at what
point they occur and how the omitted positions or issues fit into
the overall picture; I also provide ample references for those who
wish to pursue the avenues I bypass.

The most notable omission concerns the macro theory at stake,
thermodynamics. The precise formulation of the theory, and in
particular the Second Law, raises important questions. These are
beyond the scope of this review; Appendix B provides a brief
statement of the theory and flags the most important problems that
attach to it.

What is the relevant microphysical theory? A natural response would
be to turn to quantum theory, which is generally regarded as the
currently best description of micro entities. The actual debate has
followed a different path. With some rare exceptions, foundational
debates in SM have been, and still are, couched in terms of
classical mechanics (which I briefly review in Appendix A). I adopt
this point of view and confine the discussion to classical
statistical mechanics. This, however, is not meant to suggest that
the decision to discuss foundational issues in a classical rather
than a quantum setting is unproblematic. On the contrary, many
problems that occupy centre stage in the debate over the foundations
of SM are intimately linked to aspects of classical mechanics and it
seems legitimate to ask whether, and, if so, how, these problems
surface in quantum statistical mechanics. (For a review of
foundational issues in quantum SM see Emch (2007)).


\section{The Boltzmann Approach}\label{The Boltzmann Approach}

Over the years Boltzmann \index{Boltzmann, L.} developed a multitude
of different approaches to SM. However, contemporary Boltzmannians
(references will be given below), take the account introduced in
Boltzmann (1877) and streamlined by Ehrenfest \& Ehrenfest (1912) as
their starting point. For this reason I concentrate on this
approach, and refer the reader to Klein (1973), Brush (1976), Sklar
(1993), von Plato (1994), Cercignani (1998) and Uffink (2004, 2007)
for a discussion of Boltzmann's other approaches and their tangled
history, and to de Regt (1996), Blackmore (1999) and Visser (1999)
for discussions of Boltzmann's philosophical and methodological
presuppositions at different times.


\subsection{The Framework}\label{The Framework}

Consider a system consisting of $n$ particles with three degrees of
freedom,\footnote{The generalisation of what follows to systems
consisting of objects with any finite number of degrees of freedom
is straightforward.} which is confined to a container of finite
volume $V$ and has total energy $E$.\footnote{The version of the
Boltzmann framework introduced in this subsection is the one
favoured by Lebowitz (1993a, 1993b, 1999), Goldstein (2001), and
Goldstein \& Lebowitz (2004). As we shall see in the next
subsection, some authors give different definitions of some of the
central concepts, most notably the Boltzmann entropy.} The system's
fine-grained micro-state is given by a point in its $6n$ dimensional
phase space $\Gamma_{\gamma}$.\footnote{The choice of the somewhat
gawky notation `$\Gamma_{\gamma}$' will be justified in the next
subsection.} In what follows we assume that the system's dynamics is
governed by Hamilton's equations of motion,\footnote{For brief
review of classical mechanics see Appendix A. From a technical point
of view the requirement that the system be Hamiltonian is
restrictive because the Boltzmannian machinery, in particular the
combinatorial argument introduced in the next subsection, can be
used also in some cases of non-Hamiltonian systems (for instance the
Baker's gas and the Kac ring). However, as long as one believes that
classical mechanics is the true theory of particle motion (which is
what we do in classical SM), these other systems are not relevant
from a foundational point of view.} and that the system is isolated
from its environment.\footnote{This assumption is not
uncontroversial; in particular, it is rejected by those who advocate
an interventionist approach to SM; for a discussion see Subsection
\ref{Interventionism}.} Hence, the system's fine-grained micro-state
$x$ lies within a finite sub-region $\Gamma_{\gamma, \,a}$ of
$\Gamma_{\gamma}$, the so-called `accessible region' of
$\Gamma_{\gamma}$. This region is determined by the constraints that
the motion of the particles is confined to volume $V$ and that the
system has constant energy $E$---in fact, the latter implies that
$\Gamma_{\gamma, \,a}$ entirely lies within a $6n-1$ dimensional
hypersurface $\Gamma_{E}$, the so-called `energy hypersurface',
which is defined by the condition $H(x)=E$, where $H$ is the
Hamiltonian of the system and $E$ the system's total energy. The
phase space is endowed with a Lebesgue measure $\mu_{_{L}}$, which
induces a measure $\mu_{_{L, \, E}}$ on the energy hypersurface via
Equation (\ref{Restricted Lebesgue Measure}) in the Appendix.
Intuitively, these measures associate a `volume' with subsets of
$\Gamma_{\gamma}$ and  $\Gamma_{E}$; to indicate that this `volume'
is not the the familiar volume in three dimensional physical space
it is often referred to as `hypervolume'.

Hamilton's equations define a measure preserving flow $\phi_{t}$ on
$\Gamma_{\gamma}$, meaning that $\phi_{t}: \Gamma_{\gamma}
\rightarrow \Gamma_{\gamma}$ is a one-to-one mapping and
$\mu_{_{L}}(R) = \mu_{_{L}}(\phi_{t}(R))$ for all times $t$ and all
regions $R \subseteq \Gamma_{\gamma}$, from which it follows that
$\mu_{_{L, \, E}}(R_{E}) = \mu_{_{L, \, E}}(\phi_{t}(R_{E}))$ for
all regions $R_{E} \subseteq \Gamma_{E}$.

Let $M_{i}$, $i=1, ..., m$,  be the system's
macro-states.\index{Macrostates} These are characterised by the
values of macroscopic variables such as local pressure, local
temperature, and volume.\footnote{Whether index $i$ ranges over a
set of finite, countably infinite, or uncountably infinite
cardinality depends both on the system and on how macro-states are
defined. In what follows I assume, for the sake of simplicity, that
there is a finite number $m$ of macro-states.} It is one of the
basic posits of the Boltzmann approach that a system's macro-state
supervenes \index{Supervenience} on its fine-grained micro-state,
meaning that a change in the macro-state must be accompanied by a
change in the fine-grained micro-state (i.e. it is not possible,
say, that the pressure of a system changes while its fine-grained
micro-state remains the same).\index{Macrostates!as supervenient on
microstates} Hence, to every given fine-grained micro-state $x \in
\Gamma_{E}$ there corresponds \emph{exactly one} macro-state. Let us
refer to this macro-state as $M(x)$.\footnote{This is not to claim
that \emph{all} macroscopic properties of a gas supervene on its
mechanical configuration; some (e.g. colour and smell) do not.
Rather, it is an exclusion principle: if a property does not
supervene on the system's mechanical configuration then it does not
fall within the scope of SM.} This determination relation is not
one-to-one\index{Determination}; in fact many different $x \in
\Gamma_{E}$ can correspond to the same macro-state (this will be
illustrated in detail in the next subsection). It is therefore
natural to define

\begin{equation}\label{macro regions}
    \Gamma_{M_{i}}\, := \, \{x\, \in \Gamma_{E} \, | \, M_{i} \, =
\, M(x)\}, i=1, ..., m,
\end{equation}

\noindent the subset of $\Gamma_{E}$ consisting of all fine-grained
micro-states that correspond to macro-state $M_{i}$. The proposition
that a system with energy $E$ is in macro-state $M_{i}$ and the
proposition that the system's fine-grained micro-state lies within
$\Gamma_{M_{i}}$ always have the same truth value; for this reason,
$M_{i}$ and $\Gamma_{M_{i}}$ alike are sometimes referred to as
`macro-states'. However, at some points in what follows it is
important to keep the two separate and so I do not follow this
convention; I reserve the term `macro-state' for the $M_{i}$'s and
refer to the $\Gamma_{M_{i}}$'s as
`macro-regions'.\index{Macrostates!and macro-regions}

The $\Gamma_{M_{i}}$ don't overlap because macro-states supervene on
micro-states: $\Gamma_{M_{i}} \cap \Gamma_{M_{j}} = \oslash $ for
all $i \neq j$ and $i, j=1, ....m$. For a complete set of
macro-states the $\Gamma_{M_{i}}$ also jointly cover the accessible
region of the energy hypersurface: $\Gamma_{M_{1}} \cup ... \cup
\Gamma_{M_{m}} = \Gamma_{\gamma, a}$ (where `$\cup$', `$\cap$' and
`$\oslash$' denote set theoretic union, intersection and the empty
set respectively). In this case the $\Gamma_{M_{i}}$ form a
partition of $\Gamma_{\gamma, \, a}$.\footnote{Formally,
$\{\alpha_{1}, ... , \alpha_{k}\}$, where $\alpha_{i} \subseteq A$
for all $i$, is a partition of $A$ iff $\alpha_{1} \cup ... \cup
\alpha_{k} = A$ and $\alpha_{i} \cap \alpha_{j} = \oslash $ for all
$i \neq j$ and $i, j=1, ..., k$.}

The Boltzmann entropy \index{Boltzmann entropy} of a macro-state
$M_{i}$ is defined as\footnote{Goldstein (2001, 43), Goldstein and
Lebowitz (2004, 57), Lebowitz (1993a, 34; 1993b, 5; 1999, 348).}

\begin{equation}\label{G-L Boltzmann Entropy Macro-state}
    S_{_{B}}(M_{i}) \, \, = \, \,  k_{_{B}} \, \log [\mu_{_{L, \,E}}(\Gamma_{M_{i}})],
\end{equation}

\noindent where $k_{_{B}}$ is the so-called Boltzmann constant. For
later discussions, in particular for what Boltzmannians have to say
about non-equilibrium and reductionism, a small `cosmetic' amendment
is needed. The Boltzmann entropy as introduced in Equation (\ref{G-L
Boltzmann Entropy Macro-state}) is a property of a macro-state.
Since a system is in exactly one macro-state at a time, the
Boltzmann entropy can equally be regarded as a property of a system
itself. Let $M(x(t))$ be the system's macro-state at time $t$ (i.e.
$M(x(t))$ is the $M_{i}$ in which the system's state $x$ happens to
be at time $t$), then the \emph{system's} Boltzmann entropy at time
$t$ is defined as

\begin{equation}\label{G-L Boltzmann Entropy system}
    S_{_{B}}(t) := S_{_{B}}[M(x(t))].
\end{equation}

\noindent By definition, the equilibrium state is the macro-state
for which the Boltzmann entropy is maximal. Let us denote that state
by $M_{eq}$ (and, without loss of generality, choose the labelling
of macro-states such that $M_{eq} = M_{m}$). Justifying this
definition is one of the main challenges for the Boltzmannian
approach, and I return to this issue below in Subsection
\ref{Limitations}.

We now need to explain the approach to equilibrium. As phrased in
the introduction, this would amount to providing a mechanical
explanation of the Second Law of thermodynamics. It is generally
accepted that this would be aiming too high; the best we can hope
for within SM is to get a justification of a `probabilistic version'
of the Second Law, which I call `Boltzmann's Law' (BL)
\index{Boltzmann's Law!the H-theorem} (Callender 1999; Earman 2006,
401-403):

\begin{quote}
Boltzmann's Law: Consider an arbitrary instant of time $t=t_{1}$ and
assume that the Boltzmann entropy of the system at that time,
$S_{_{B}}(t_{1})$, is far below its maximum value. It is then highly
probable that at any later time $t_{2} > t_{1}$ we have
$S_{_{B}}(t_{2}) \geq S_{_{B}}(t_{1})$.\footnote{BL is sometimes
referred to as the `statistical H-Theorem' or the `statistical
interpretation of the $H$-theorem' because in earlier approaches to
SM Boltzmann introduced a quantity $H$, which is basically equal to
$-S_{_{B}}$, and aimed to prove that under suitable assumptions it
decreased monotonically. For discussions of this approach see the
references cited at the beginning of this subsection.}
\end{quote}

\noindent Unlike the Second Law, which is a universal
law\index{Universal law} (in that it does not allow for exceptions),
BL only makes claims about what is very likely to
happen.\index{Boltzmann's Law!contrast with second law of
thermodynamics} Whether it is legitimate to replace the Second Law
by BL will be discussed in Subsection \ref{Reductionism -
Boltzmann}. Even if this question is answered in the affirmative,
what we expect from SM is an argument for the conclusion that BL,
which so far is just a conjecture, holds true in the relevant
systems, and, if it does, an explanation of why this is so. In order
to address \emph{this} question, we need to introduce probabilities
into the theory to elucidate the locution `highly
probable'.\index{Probability!in statistical mechanics}

There are two different ways of introducing probabilities into the
Boltzmannian framework. The first assigns probabilities directly to
the system's macro-states; the second assigns probabilities to the
system's micro-state being in particular subsets of the macro-region
corresponding to the system's current macro-state.\footnote{This is
not to say that the these two kinds of probabilities are
incompatible; in fact they could be used in conjunction. However,
this is not what happens in the literature.} For want of better
terms I refer to these as `macro-probabilities' and
`micro-probabilities' respectively. Although implicit in the
literature, the distinction between macro-probabilities and
micro-probabilities has never been articulated, and it rarely, if
ever, receives any attention. This distinction plays a central
r\^{o}le in the discussion both of BL and the interpretation of SM
probabilities, and it is therefore important to give precise
definitions.

\emph{Macro-Probabilities}:\index{Probability!in statistical
mechanics!of the macro-state} A way of introducing probabilities
into the theory, invented by Boltzmann (1877) and advocated since
then by (among others) those working within the ergodic
programme\index{Ergodic programme} (see Subsection \ref{The
Ergodicity Programme}), is to assign probabilities to the
macro-states $M_{i}$ of the system. This is done by introducing the
postulate that the probability of a macro-state $M_{i}$ is
proportional to the measure of its corresponding macro-region:

\begin{equation}\label{Proportionality Postulate}
    p(M_{i})\, \, :=\,  c\,  \mu_{_{L, \, E}}(\Gamma_{M_{i}}),
\end{equation}

\noindent where $c$ is a normalisation constant. I refer to this as
the `Proportionality Postulate' (PP). From this postulate and
Equation (\ref{G-L Boltzmann Entropy Macro-state}) it follows
immediately that the most likely macro-state is the macro-state with
the highest Boltzmann entropy and the one that occupies the largest
part of the (accessible) phase space.

From this point of view it seems natural to understand the approach
to equilibrium as the evolution from an unlikely macro-state to a
more likely macro-state and finally to the most likely macro-state.
If the system evolves from less to more likely macro-states
\emph{most of the time} then we have justified BL. Whether we have
any reasons to believe that this is indeed the case will be
discussed in Unit \ref{Probabilities}.

\emph{Micro-Probabilities}:\index{Probability!in statistical
mechanics!of the micro-state} A different approach assigns
probabilities to sets of micro-states (rather than to macro-states)
on the basis of the so-called statistical postulate
(SP).\footnote{It is not clear where this postulate originates. It
has recently---with some qualifications, as we shall see---been
advocated by Albert (2000), and also Bricmont (1996) uses arguments
based on probabilities introduced in this way; see also Earman
(2006, 405), where this postulate is discussed, but not endorsed.
Principles very similar to this one have also been suggested by
various writers within the Gibbsian tradition; see Section \ref{The
Gibbs Approach}.}

\begin{quote}
Statistical Postulate:\index{Statistical Postulate} Let $M$ be the
macro-state of a system at time $t$. Then the probability at $t$
that the fine-grained micro-state of the system lies in a subset $A$
of $\Gamma_{M}$ is
\begin{equation}\label{Statistical Postulate Equation}
    \mu_{_{L, \,E}}(A)/\mu_{_{L, \,E}}(\Gamma_{M}).
\end{equation}

\end{quote}

\noindent With this assumption the truth of BL depends on the
dynamics of the system, because now BL states that the overwhelming
majority of fine-grained micro-states in any $\Gamma_{M_{i}}$
(except the equilibrium macro-region) are such that they evolve
under the dynamics of the system towards some other region
$\Gamma_{M_{j}}$ of higher entropy. Hence, the truth of BL depends
on the features of the dynamics. The question is whether the systems
we are interested in have this property. I come back to this issue
in Subsection \ref{Micro Probabilities Revisited}.


\subsection{The Combinatorial Argument}\label{The Combinatorial
Argument}

An important element in most presentations of the Boltzmann approach
is what is now know as the `combinatorial argument'.
\index{Combinatorial argument} However, depending on how one
understands the approach, this argument is put to different uses---a
fact that is unfortunately hardly ever made explicit in the
literature on the topic. I will first present the argument and then
explain what these different uses are.

Consider the same system of $n$ identical particles as above, but
now focus on the 6 dimensional phase space of \emph{one} of these
particles, the so-called $\mu$-space $\Gamma_{\mu}$, rather than the
($6n$ dimensional) phase space $\Gamma_{\gamma}$ of the entire
system.\footnote{The use of the symbol $\mu$ both in `$\mu$-space'
and to refer to the measure on the phase space is somewhat
unfortunate as they have nothing to do with each other. However, as
this terminology is widely used I stick to it. The distinction
between $\mu$-space and $\gamma$-space goes back to Ehrenfest \&
Ehrenfest (1912); it is a pragmatic and not a mathematical
distinction in that it indicates how we \emph{use} these spaces
(namely to describe a single particle's or an entire system's
state). From a mathematical point of view both $\mu$-space and
$\gamma$-space are classical phase spaces (usually denoted by
$\Gamma$). This explains choice of the the seemingly unwieldy
symbols $\Gamma_{\mu}$ and $\Gamma_{\gamma}$.} A point in
$\Gamma_{\mu}$ denotes the particle's fine-grained micro-state. It
necessarily lies within a finite sub-region $\Gamma_{\mu, \, a}$ of
$\Gamma_{\mu}$, the accessible region of $\Gamma_{\mu}$. This region
is determined by the constraints that the motion of the particles is
confined to volume $V$ and that the system as a whole has constant
energy $E$. Now we choose a partition $\omega$ of $\Gamma_{\mu, \,
a}$; that is, we divide $\Gamma_{\mu, \, a}$ into a finite number
$l$ of disjoint cells $\omega_{j}$, which jointly cover the
accessible region of the phase space. The introduction of a
partition on a phase space is also referred to as
`coarse-graining'.\index{Coarse-graining} The cells are taken to be
rectangular with respect to the position and momentum coordinates
and of equal volume $\delta \omega$ (this is illustrated in Figure
2).

\vspace{1cm}

\noindent [Insert Figure 2 (appended at the end of the document).]

\vspace{1cm}

\noindent {\footnotesize{Figure 2: Partitioning (or coarse-graining)
of the phase space.}}

\vspace{1cm}

The so-called coarse-grained micro-state of a particle is given by
specifying in which cell $\omega_{j}$ its fine-grained micro-state
lies.\footnote{There is a question about what cell a fine-grained
micro-state belongs to if it lies exactly on the boundary between
two cells. One could resolve this problem by adopting suitable
conventions. However, it turns out later on that sets of measure
zero (such as boundaries) can be disregarded and so there is no need
to settle this issue.}

The micro-state of the entire system is a specification of the
micro-state of every particle in the system, and hence the
fine-grained micro-state of the system is determined by $n$ labelled
points in $\Gamma_{\mu}$.\footnote{The points are labelled in the
sense that it is specified which point represents the state of which
particle.} The so-called coarse-grained micro-state is a
specification of which particle's state lies in which cell of the
partition $\omega$ of $\Gamma_{\mu, \, a}$; for this reason the
coarse-grained micro-state of a system is also referred to as an
`arrangement'.\index{Statistical mechanics!arrangement}

The crucial observation now is that a number of arrangements
correspond to the same macro-state because a system's
macro-properties are determined solely by the number of particles in
each cell, while it is irrelevant exactly which particle is in which
cell. For instance, whether particle number 5 and particle number 7
are in cells $\omega_{1}$ and $\omega_{2}$ respectively, or vice
versa, makes no difference to the macro-properties of the system as
a whole because these do not depend on which particle is in which
cell. Hence, all we need in order to determine a system's
macro-properties is a specification of how many particles there are
in each cell of the coarse-grained $\mu$-space. Such a specification
is called a `distribution'.\index{Statistical
mechanics!distribution} Symbolically we can write it as a tuple $D =
(n_{1}, \ldots, n_{l})$, meaning the distribution comprising $n_{1}$
particles in cell $\omega_{1}$, etc. The $n_{j}$ are referred to as
`occupation numbers' \index{Occupation numbers} and they satisfy the
condition $\sum_{j=1}^{l} n_{j}=n$.

For what follows it is convenient to label the different
distributions with a discrete index $i$ (which is not a problem
since for any given partition $\omega$ and particle number $n$ there
are only a finite number of distributions) and denote the
$i^{\footnotesize{\textnormal{th}}}$ tuple by $D_{i}$. The beginning
of such labelling could be, for instance, $D_{1} = (n, 0, ...., 0)$,
$D_{2} = (n-1, 1, 0, ..., 0)$, $D_{3} = (n-2, 1, 1, 0, ..., 0)$,
etc.

How many arrangements are compatible with a given distribution $D$?
Some elementary combinatorial considerations show that
\begin{equation}\label{G(D)}
    G(D) := \frac{n!}{n_{1}! \ldots n_{l}!}
\end{equation}
arrangements are compatible with a given distribution $D$ (where `!'
denotes factorials, i.e. $k! := k (k-1) \, ... \,1$, for any natural
number $k$ and $0! :=1$). For this reason a distribution conveys
much less information than an arrangement.

Each distribution corresponds to well-defined region of
$\Gamma_{\gamma}$, which can be seen as follows. A partition of
$\Gamma_{\gamma}$ is introduced in exactly the same way as above. In
fact, the choice of a partition of $\Gamma_{\mu}$ induces a
partition of $\Gamma_{\gamma}$ because $\Gamma_{\gamma}$ is just the
Cartesian product of $n$ copies of $\Gamma_{\mu}$. The
coarse-grained state of the system is then given by specifying in
which cell of the partition its fine-grained state lies. This is
illustrated in Figure 3 for the fictitious case of a two particle
system, where each particle's $\mu$-space is one dimensional and
endowed with a partition consisting of four cells $\omega_{1}, ...,
\omega_{4}$. (This case is fictitious because in classical mechanics
there is no $\Gamma_{\mu}$ with less than two dimensions. I consider
this example for ease of illustration; the main idea carries over to
higher dimensional spaces without difficulties.)

\vspace{1cm}

\noindent [Insert Figure 3 (appended at the end of the document).]

\vspace{1cm}

\noindent {\footnotesize{Figure 3: Specification  of coarse-grained
state of a system.}}

\vspace{1cm}

\noindent This illustration shows that each distribution $D$
corresponds to a particular part of $\Gamma_{\gamma,\, a}$; and it
also shows the important fact that parts corresponding to different
distributions do not overlap. In Figure 3, the hatched areas (which
differ by which particle is in which cell) correspond to the
distribution $(1, 0, 0, 1)$ and the dotted area (where both
particles are in the same cell) correspond to $(0, 2, 0, 0)$.
Furthermore, we see that the hatched area is twice as large as the
dotted area, which illustrates an important fact about distributions
to which we now turn.

From the above it becomes clear that each point $x$ in
$\Gamma_{\gamma, \, a}$ corresponds to exactly one distribution;
call this distribution $D(x)$. The converse of this, of course,
fails since in general many points in $\Gamma_{\gamma, \, a}$
correspond to the same distribution $D_{i}$. These states together
form the set $\Gamma_{D_{i}}$:

\begin{equation}\label{Gamma M i}
    \Gamma_{D_{i}}\, := \, \{x\, \in \Gamma_{\gamma} \, | \, D(x) \, =
    \, D_{i} \}.
\end{equation}

From equations (\ref{G(D)}) and (\ref{Gamma M i}), together with the
assumption that all cells have the same size $\delta\omega$ (in the
6 dimensional $\mu$-space), it follows that

\begin{equation}\label{Omega(Gamma_D)=G(D)omega^n}
     \mu_{_{L}}(\Gamma_{D_{i}}) = G(D_{i})\,  (\delta \omega)^{n},
\end{equation}

Next we want to know the distribution $D$ for which $G(D)$, and with
it $\mu_{_{L}}(\Gamma_{D_{i}})$, assume their maximum. To solve this
problem we make two crucial sets of assumptions, one concerning the
energy of the system and one concerning the system's size (their
implications will be discussed in Subsection \ref{Limitations}).

First, we assume that the energy of a particle only depends on which
cell $\omega_{j}$ it is in, but \emph{not} on the states of the
other particles; that is, we neglect the contribution to the energy
of the system that stems from interactions between the particles. We
then also assume that the energy $E_{j}$ of a particle whose
fine-grained state lies in cell $\omega_{j}$ only depends on the
index $j$, i.e. on the cell in which the state is, and not on its
exact location within the cell. This can be achieved, for instance,
by taking $E_{j}$ to be the average energy in $\omega_{j}$. Under
these assumptions, the total energy of the system is given by
$\sum_{j=1}^{l} n_{j} E_{j}$.

Second, we assume that the system as a whole is large, and that
there are many particles in each individual cell: ($n_{j} \gg 1$ for
all $j$). These assumptions allows us to use Stirling's formula to
approximate factorials:

\begin{equation}
    n! \approx \sqrt{2\pi n} \bigg{(}\frac{n}{e}\bigg{)}^{n}
\end{equation}

Now we have to maximise $G(D)$ under the `boundary conditions' that
the number $n$ of particles is constant ($n=\sum_{j}n_{j}$) and that
the total energy $E$ of the system is constant
($E=\sum_{j}n_{j}E_{j}$). Under these assumptions one can then prove
(using Stirling's approximation and the Lagrange multiplier method)
that $G(D)$ reaches its maximum for

\begin{equation}\label{Maxwell-Boltzmann distribution}
    n_{j}\, = \,   \alpha \exp(-\beta E_{j}),
\end{equation}

\noindent which is the (discrete) \emph{Maxwell-Boltzmann
distribution}, \index{Maxwell-Boltzmann distribution} where $\alpha$
and $\beta$ are constants depending on the nature of the system
(Ehrenfest \& Ehrenfest 1912, 30; Tolman 1938, Ch. 4).

Before we turn to a discussion of the significance of these
calculations, something needs to be said about observable
quantities. It is obvious from what has been said so far that
observable quantities are averages of the form

\begin{equation}\label{average values}
    \langle f \rangle \, := \, \sum_{j=1}^{n} n_{j} f_{\omega_{j}},
\end{equation}
\noindent where $f$ is a function of position and momentum of a
particle, and $f_{\omega_{j}}$ is the value of the function in cell
$\omega_{j}$ (where, as in the case of the energy, it is assumed
that the values of $f$ depend only on the cell $\omega_{j}$ but not
of the particle's location within the cell; i.e. it is assumed that
$f$ does not fluctuate on a scale of $\delta\omega$). In particular
one can calculate the pressure of a gas in equilibrium in this
way.\footnote{In practice this is not straightforward. To derive the
desired results, one first has to express the Maxwell-Boltzmann
distribution in differential form, transform it to position and
momentum coordinates and take a suitable continuum limit. For
details see, for instance, Tolman (1938, Ch. 4).}

What is the relevance of these considerations for the Boltzmann
approach? The answer to this question is not immediately obvious.
Intuitively one would like to associate the $\Gamma_{D_{i}}$ with
the system's macro-macro regions $\Gamma_{M_{i}}$. However, such an
association is undercut by the fact that the $\Gamma_{D_{i}}$ are
$6n$ dimensional objects, while the $\Gamma_{M_{i}}$, as defined by
Equation (\ref{macro regions}), are subsets of the $6n-1$
dimensional energy hypersurface $\Gamma_{E}$.

Two responses to this problem are possible. The first is to replace
the definition of a macro-region given in the previous subsection by
one that associates macro-states with $6n$ dimensional rather than
$6n-1$ dimensional parts of $\Gamma_{\gamma}$, which amounts to
replacing Equation (\ref{macro regions}) by $\Gamma'_{M_{i}}\, := \,
\{x\, \in \Gamma_{\gamma} \, | \, M_{i} \, = \, M(x)\}$ for all
$i=1, ..., m$. Macro-states thus defined can then be identified with
the regions of $\Gamma_{\gamma}$ corresponding to a given
distribution: $\Gamma'_{M_{i}} = \Gamma_{D_{i}}$ for all $i=1, ...,
m$, where $m$ now is the number of different distributions.

This requires various adjustments in the apparatus developed in
Subsection \ref{The Framework}, most notably in the definition of
the Boltzmann entropy. Taking the lead from the idea that the
Boltzmann entropy is the logarithm of the hypervolume of the part of
the phase space associated with a macro-state we have

\begin{equation}\label{Def S = k ln Omega}
    S'_{_{B}}(M_{i}) \, := \, k_{_{B}} \log
    [\mu_{_{L}}(\Gamma'_{M_{i}})],
\end{equation}

\noindent and with Equation (\ref{Omega(Gamma_D)=G(D)omega^n}) we
get

\begin{equation}\label{S = k ln Omega}
    S'_{_{B}}(M_{i}) \, = \, k_{_{B}} \log [G(D_{i})]
    + k_{_{B}} n \log (\delta \omega).
\end{equation}

\noindent Since the last term is just an additive constant it can be
dropped (provided we keep the partition fixed), because ultimately
we are interested in entropy differences rather than in absolute
values. We then obtain $S'_{_{B}}(M_{i}) \, = \, k_{_{B}} \log
[G(D_{i})]$, which is the definition of the Boltzmann entropy we
find in Albert (2000, 50).

In passing it is worth mentioning that $S'_{_{B}}$ can be expressed
in alternative ways. If we plug Equation (\ref{G(D)}) into Equation
(\ref{S = k ln Omega}) and take into account the above assumption
that all $n_{j}$ are large (which allows us to use Stirling's
approximation) we obtain (Tolman 1938, Ch. 4):

\begin{equation} \label{Boltzmann Entropy 2}
    S'_{_{B}}(M_{i}) \, \approx - k_{_{B}} \, \sum_{j}n_{j}\log n_{j}
    + c(n, \, \delta \omega),
\end{equation}
where the $n_{j}$ are the occupation numbers of distribution $D_{i}$
and $c(n, \, \delta \omega)$ is a constant depending on $n$ and
$\delta \omega$. Introducing the quotients $p_{j}:=n_{j}/n$ and
plugging them into the above formula we find
\begin{equation}\label{Boltzmann Entropy 3}
    S'_{_{B}}(M_{i})\, \approx  - n \, k_{_{B}} \, \sum_{j}p_{j}\log p_{j}
    + \tilde{c}(n, \, \delta \omega),
\end{equation}
where, again, $\tilde{c}(n, \, \delta \omega)$ is a constant
depending on $n$ and $\delta \omega$.\footnote{This expression for
the Boltzmann entropy is particularly useful because, as we shall
see in Subsection \ref{The Shannon Entropy}, $\sum_{j}p_{j}\log
p_{j}$ is a good measure for the `flatness' of the distribution
$p_{j}$.} The quotients $p_{i}$ are often said to be the probability
of finding a randomly chosen particle in cell $\omega_{i}$. This is
correct, but is important not to confuse this probability, which is
simply a finite relative frequency, with the probabilities that
occur in BL. In fact, the two have nothing to do with each other.

What are the pros and the cons of this first response? The obvious
advantage is that it provides an explicit construction of the
macro-regions $\Gamma'_{M_{i}}$, and that this construction gives
rise to a definition of the Boltzmann entropy which allows for easy
calculation of its values.

The downside of this `$6n$ dimensional approach' is that the
macro-regions $\Gamma'_{M_{i}}$ almost entirely consist of
micro-states which the system never visits (remember that the motion
of the system's micro-state is confined to the $6n-1$ dimensional
energy hypersurface). This is a problem because it is not clear what
relevance considerations based on the hypervolume of certain parts
of the phase space have if we know that the system's actual
micro-state only ever visits a subset of these parts which is of
measure zero. Most notably, of what relevance is the observation
that the equilibrium macro-region has the largest ($6n$ dimensional)
hypervolume if the system can only ever access a subset of measure
zero of this macro-region? Unless there is a relation between the
$6n-1$ dimensional hypervolume of relevant parts of the energy
hypersurface and the $6n$ dimensional hypervolume of the parts of
$\Gamma_{\gamma}$ in which they lie, considerations based on the
$6n$ dimensional hypervolume are inconsequential.\footnote{Moreover,
the `$6n$ dimensional approach' renders the `orthodox' account of SM
probability, the time average interpretation (see Subsection
\ref{The Ergodicity Programme}), impossible. This interpretation is
based on the assumption that the system is ergodic on the union of
the macro-regions, which is impossible if macro regions are $6n$
dimensional.}

The second response to the above problem leaves the definition of
macro-regions as subsets of the $6n-1$ dimensional energy
hypersurface unaltered and endeavours to `translate' the results of
the combinatorial argument back into the original framework (as
presented in Subsection \ref{The Framework}). This, as we shall see,
is possible, but only at the cost of introducing a further
hypothesis postulating a relation between the values of the $6n$ and
the $6n-1$ dimensional hypervolumes of relevant parts of
$\Gamma_{\gamma}$.

The most important achievement of the combinatorial argument is the
construction of the $\Gamma_{D_{i}}$, the regions in phase space
occupied by micro-states with the same macro-properties. Given that
the original framework does not provide a recipe of how to construct
the macro-regions, we want to make use of the $\Gamma_{D_{i}}$ to
define the $\Gamma_{M_{i}}$. A straightforward way to obtain the
$\Gamma_{M_{i}}$ from the $\Gamma_{D_{i}}$ is intersect the
$\Gamma_{D_{i}}$ with$\Gamma_{E}$:\footnote{This construction
implicitly assumes that there is a one-to-one correspondence between
distributions and macro-states. This assumption is too simplistic in
at least two ways. First, $\Gamma_{D_{i}} \cap \Gamma_{E}$ may be
empty for some $i$. Second, characteristically several distributions
correspond to the same macro-state in that the macroscopic
parameters defining the macro-state assume the same values for all
of them. These problems can easily be overcome. The first can be
solved by simply deleting empty $M_{i}$ from the list of
macro-regions; the second can be overcome by intersecting
$\Gamma_{E}$ not with each individual $\Gamma_{D_{i}}$, but instead
with the union of all $\Gamma_{D_{i}}$ that correspond to the same
macro-state. Since this would not alter any of the considerations to
follow, I disregard this issue henceforth.}

\begin{equation}
    \Gamma_{M_{i}} \, := \, \Gamma_{D_{i}} \cap \Gamma_{E}.
\end{equation}

How can we calculate the Boltzmann entropy of the macro-states
corresponding to macro-regions thus defined? The problem is that in
order to calculate the Boltzmann entropy of these states we need the
$6n-1$ dimensional hypervolume of the $\Gamma_{M_{i}}$, but what we
are given (via Equation (\ref{Omega(Gamma_D)=G(D)omega^n})) is the
the $6n$ dimensional hypervolume of the $\Gamma_{D_{i}}$, and there
is no way to compute the former on the basis of the latter.

The way out of this impasse is to introduce a new postulate, namely
that the  $6n-1$ dimensional hypervolume of the $\Gamma_{M_{i}}$ is
proportional to $6n$ dimensional hypervolume of the
$\Gamma_{D_{i}}$: $\mu_{_{L, \, E}}(\Gamma_{M_{i}}) = k_{_{v}}
\mu_{_{L}}(\Gamma_{D_{i}})$, where $k_{_{v}}$ is a proportionality
constant. It is plausible to assume that this postulate is at least
approximately correct because the energy hypersurface of
characteristic SM systems is smooth and does not oscillate on the
scale of $\delta \omega$. Given this, we have

\begin{equation}\label{new S = k ln Omega}
    S_{_{B}}(M_{i}) \, = \, S'_{_{B}}(M_{i}) \, + \, k_{_{B}}
    \log(k_{_{v}});
\end{equation}

\noindent that is, $S_{_{B}}$ and $S'_{_{B}}$ differ only by an
additive constant, and so Equation (\ref{S = k ln Omega}) as well as
Equations (\ref{Boltzmann Entropy 2}) and (\ref{Boltzmann Entropy
3}) can be used to determine the values of $S_{_{B}}$.

In what follows I assume that this `proportionality assumption'
holds water and that the Boltzmann entropy of a macro-state can be
calculated using Equation (\ref{new S = k ln Omega}).


\subsection{Problems and Tasks}

In this subsection I point out what the issues are that the
Boltzmannian needs to address in order to develop the approach
introduced so-far into a full-fledged account of SM. Needless to
say, these issues are not independent of each other and the response
to one bears on the responses to others.


\subsubsection{Issue 1: The Connection with Dynamics}\label{Issue 1}

The Boltzmannian account as developed so far does not make reference
to dynamical properties of the system other than the conservation of
energy, which is a consequence of Hamilton's equations of motion.
But not every dynamical system---not even if it consists of a large
number of particles---behaves thermodynamically in that the
Boltzmann entropy increases most of the time.\footnote{Lavis (2005,
254-61) criticises the standard preoccupation with `local' entropy
increase as misplaced and suggests that what SM should aim to
explain is so-called thermodynamic-like behaviour, namely that the
Boltzmann entropy be close to its maximum most of the time.} For
such behaviour to take place it must be the case that a system,
which is initially prepared in any low entropy state, eventually
moves towards the region of $\Gamma_{\gamma}$ associated with
equilibrium. This is illustrated in Figure 4 (which is adapted from
Penrose 1989, 401 and 407).

\vspace{1cm}

\noindent [Insert Figure 4 (appended at the end of the document).]

\vspace{1cm}

\noindent {\footnotesize{Figure 4: Trajectory from a low entropy
state to a region associated with equilibrium.}}

\vspace{1cm}

\noindent But this need not be so. If, for instance, the initial low
entropy macro-region is separated from the equilibrium region by an
invariant surface, then no approach to equilibrium takes place.
Hence, the question is: of what kind the dynamics has to be for the
system to behave thermodynamically.

A common response begins by pointing out that equilibrium is not
only associated with the largest part of $\Gamma_{\gamma}$; in fact,
the equilibrium macro-region is \emph{enormously} larger than
\emph{any} other macro-region (Ehrenfest \& Ehrenfest 1912, 30).
Numerical considerations show that the ratio
$\Gamma_{M_{eq}}/\Gamma_{M}$, where $M$ is a typical non-equilibrium
distribution, is of the magnitude of $10^{n}$ (Goldstein 2001, 43;
Penrose 1989, 403). If we now assume that the system's state drifts
around more or less `randomly' on $\Gamma_{\gamma, \, a}$ then,
because $\Gamma_{M_{eq}}$ is vastly larger than any other macro
region, sooner or later the system will reach equilibrium and stay
there for at least a very long time. The qualification `more or less
randomly' is essential. If the motion is too regular, it is possible
that the system successfully avoids equilibrium positions. But if
the state wanders around on the energy hypersurface randomly, then,
the idea is, it simply cannot avoid moving into the region
associated with equilibrium sooner or later.

Plausible as it may seem, this argument has at best heuristic value.
What does it mean for a system to drift around randomly? In
particular in the context of Hamiltonian mechanics, a deterministic
theory, the notion of drifting around randomly is in need of
explanation: what conditions does a classical system have to satisfy
in order to possess `random properties' sufficient to warrant the
approach to equilibrium?\index{Randomness}


\subsubsection{Issue 2: Introducing and Interpreting
Probabilities}\label{Probabilities}
\index{Probability!interpretations}
There are several different (albeit interrelated) issues that must
be addressed in order to understand the origin and meaning of
probabilities in SM, and all of them are intimately connected to
issue 1. The first of these is the problem of interpretation.

\emph{The interpretation of SM probabilities}. How are SM
probabilities to be understood? Approaches to probability can be
divided into two broad groups.\footnote{What follows is only the
briefest of sketches. Those options that have been seriously pursued
within SM will be discussed in more detail below. For an in-depth
discussion of all these approaches see, for instance, Howson (1995),
Gillies (2000), Galavotti (2005) and Mellor (2005).} First,
\textit{epistemic approaches} take probabilities to be measures for
degrees of belief. Those who subscribe to an \textit{objective
epistemic theory} take probabilities to be degrees of rational
belief, whereby `rational' is understood to imply that given the
same evidence all rational agents have the same degree of belief in
any proposition. This is denied by those who hold a
\textit{subjective epistemic theory}, regarding probabilities as
subjective degrees of belief that can differ between persons even if
they are presented with the same body of
evidence.\footnote{`Subjective probability' is often used as a
synonym for `epistemic probability'. This is misleading because not
all epistemic probabilities are also subjective. Jaynes's approach
to probabilities, to which I turn below, is a case in point.}
Second, \textit{ontic approaches} take probabilities to be part of
the `furniture of the world'. On the \textit{frequency approach},
probabilities are long run frequencies of certain events. On the
\textit{propensity theory}, probabilities are tendencies or
dispositions inherent in objects or situations. The \textit{Humean
best systems approach}---introduced in Lewis (1986)---views
probability as defined by the probabilistic laws that are part of
that set of laws which strike the best balance between simplicity,
strength and fit.\footnote{Sometimes ontic probabilities are
referred to as `objective probabilities'. This is misleading because
epistemic probabilities can be objective as well.} To which of these
groups do the probabilities introduced in the Boltzmannian scheme
belong?

We have introduced two different kinds of probabilities (micro and
macro), which, prima facie, need not be interpreted in the same way.
But before delving into the issue of interpretation, we need to
discuss whether these probabilities can, as they should, explain
Boltzmann's law. In fact, serious problems arise for both kinds of
probabilities.

\emph{Macro-Probabilities}. Boltzmann suggested that
macro-probabilities explain the approach to equilibrium: if the
system is initially prepared in an improbable macro-state (i.e. one
far away from equilibrium), it will from then on evolve towards more
likely states until it reaches, at equilibrium, the most likely
state (1877, 165). This happens because Boltzmann takes it as a
given that the system `always evolves from an improbable to a more
probable state' (\emph{ibid}., 166).

This assumption is unwarranted. Equation (\ref{Proportionality
Postulate}) assigns \emph{unconditional}
probabilities\index{Probability!unconditional} to macro-states, and
as such they do not imply anything about the succession of states,
let alone that ones of low probability are followed by ones of
higher probability. As an example consider a brassed die; the
probability to get a `six' is 0.25 and all other numbers of spots
have probability 0.15. Can we then infer that after, say, a `three'
we have to get a `six' because the six is the most likely event? Of
course not; in fact, we are much more likely not to get a `six' (the
probability for non-six is 0.75, while the one for six is 0.25). A
further (yet related) problem is that BL makes a statement about a
\emph{conditional} probability,\index{Probability!conditional}
namely the probability of the system's macro-state at $t_{2}$ being
such that $S_{_{B}}(t_{2})
> S_{_{B}}(t_{1})$, \emph{given} that the system's macro-state at
the earlier time $t_{1}$ was such that its Boltzmann entropy was
$S_{_{B}}(t_{1})$. The probabilities of PP (see Equation
(\ref{Proportionality Postulate})) are not of this kind, and they
cannot be turned into probabilities of this kind by using the
elementary definition of conditional probabilities, $p(B | A) = p(B
\& A)/p(A)$, for reasons pointed out by Frigg (2007a). For this
reason non-equilibrium SM cannot be based on the PP, no matter how
the probabilities in it are interpreted.

However, PP \emph{does} play a r\^{o}le in equilibrium SM. It posits
that the equilibrium state is the most likely of all states and
hence that the system is most likely to be in equilibrium. This
squares well with an intuitive understanding of equilibrium as the
state that the system reaches after a (usually short) transient
phase, after which it stays there (remember the spreading gas in the
introduction). Granting that, what notion of probability is at work
in PP? And why, if at all, is this postulate true? That is, what
facts about the system make it the case that the equilibrium state
is indeed the most likely state? These are the questions that
Boltzmannian equilibrium SM has to answer, and I will turn to these
in Subsection \ref{The Ergodicity Programme}.

\emph{Micro-Probabilities}. The conditional probabilities needed to
explain BL can be calculated on the basis of SP (see Equation
(\ref{Statistical Postulate Equation})).\footnote{To keep things
simple I assume that there corresponds only one macro-state to a
given entropy value. If this is not the case the exactly same
calculations can be made using the union of the macro-regions of all
macro-states with the same entropy.} Let $M$ be the macro-state of a
system at time $t$. For every point $x \in \Gamma_{M}$ there is a
matter of fact (determined by the Hamiltonian of the system) about
whether $x$ evolves into a region of higher or lower entropy or
stays at the same level of entropy. Call $\Gamma_{M_{+}}$,
$\Gamma_{M_{-}}$, and $\Gamma_{M_{0}}$ the sets of those points of
$\Gamma_{M}$ that evolve towards a region of higher, lower, or same
entropy respectively (hence $\Gamma_{M} = \Gamma_{M_{+}} \cup
\Gamma_{M_{-}} \cup \Gamma_{M_{0}}$). The probability for the
system's entropy to either stay the same or increase as time evolves
is $\mu(\Gamma_{M_{+}} \cup \Gamma_{M_{0}})/\mu(\Gamma_{M})$. Hence,
it is a necessary and sufficient condition for BL to be true that
$\mu(\Gamma_{M_{+}} \cup \Gamma_{M_{0}}) \gg \mu(\Gamma_{M_{-}})$
for all micro-states $M$ except the equilibrium state itself (for
which, trivially, $\mu(\Gamma_{M_{+}})=0$). BL then translates into
the statement that the overwhelming majority of micro-states in
every macro-region $\Gamma_{M}$ except $\Gamma_{M_{eq}}$ evolve
under the dynamics of the system towards regions of higher entropy.

This proposal is seriously flawed. It turns out that \emph{if} the
system, in macro-state $M$, is very likely to evolve towards a
macro-state of higher entropy in the future (which we want to be the
case), \emph{then}, because of the time reversal invariance of the
underlying dynamics, the system is also very likely to have evolved
into the current macro-state $M$ from another macro-state $M'$ of
\emph{higher} entropy than $M$ (see Appendix A for a discussion of
time reversal invariance). So whenever the system is very likely to
have a high entropy future it is also very likely to have a high
entropy past; see Albert (2000, Ch. 4) for a discussion of this
point. This stands in stark contradiction with both common sense
experience and BL itself. If we have a lukewarm cup of coffee on the
desk, SP makes the radically wrong retrodiction that is
overwhelmingly likely that 5 minutes ago the coffee was cold (and
the air in the room warmer), but then fluctuated away from
equilibrium to become lukewarm and five minutes from now will be
cold again. However, in fact the coffee was hot five minutes ago,
cooled down a bit and will have further cooled down five minutes
from now.

This point is usually attributed to the Ehrenfests. It is indeed
true that the Ehrenfests (1912, 32-34) \index{Ehrenfests} discuss
transitions between different entropy levels and state that
higher-lower-higher transitions of the kind that I just mentioned
are overwhelmingly likely. However, they base their statement on
calculations about a \emph{probabilistic} model, their famous
urn-model, \index{Ehrenfests!urn model}  and hence it is not clear
what bearing, if any, their considerations have on deterministic
dynamical systems; in fact, some of the claims they make are not in
general true in conservative deterministic systems.

Nor is it true that the objection to the proposal follows directly
from the time reversal invariance of the underlying dynamics on the
simple grounds that everything that can happen in one direction of
time can also happen in the other direction. However, one can indeed
prove that the statement made in the last paragraph about entropic
behaviour is true in conservative deterministic dynamical systems,
if SP is assumed (Frigg 2007b). Hence there is a serious problem,
because the micro-dynamics and (SP) lead us to expect the system to
behave in a way that is entirely different from how the system
actually behaves and from what the laws of thermodynamics lead us to
expect. The upshot is that the dynamics at the micro-level and SP
\emph{by itself} do not underwrite the asymmetrical behaviour that
we find at the macro level, and which is captured by BL. Hence the
question is: where then does the irreversibility at the macro level
come from, if not from the dynamical laws governing the micro
constituents of a system? I turn to a discussion of this question in
Subsection \ref{The Past Hypothesis}.


\subsubsection{Issue 3: Loschmidt's Reversibility Objection}\label{Loschmidt's
Reversibility Objection}

As we observed in the introduction, the world is rife with
irreversible processes; that is, processes that happen in one
temporal direction but not in the other. This asymmetry is built
into the Second Law of thermodynamics.\index{Second law of
thermodynamics}. As Loschmidt pointed out in controversy with
Boltzmann (in the 1870s), this does not sit well with the fact that
classical mechanics is time-reversal invariant.\index{Loschmidt's
Objection} The argument goes as follows:\footnote{The following is a
more detailed version of the presentation of the argument in
Ehrenfest \& Ehrenfest (1907, 311).}

\emph{Premise 1}: It follows from the time reversal invariance of
classical mechanics that if a transition from state $x_{_{i}}$ to
state $x_{_{f}}$ (`$i$' for `initial' and `$f$' for `final') in time
span $\Delta$ is possible (in the sense that there is a Hamiltonian
that generates it), then the transition from state $Rx_{_{f}}$ to
state $Rx_{_{i}}$ in time span $\Delta$ is possible as well, where
$R$ reverses the momentum of the instantaneous state of the system
(see Appendix A for details).

\emph{Premise 2}: Consider a system in macro-state $M$ with
Boltzmann entropy $S_{_{B}}(M)$. Let $RM$  be the reversed
macro-state, i.e. the one with macro-region $\Gamma_{RM} := \{x \in
\Gamma | Rx \in \Gamma_{M} \}$ (basically we obtain $RM$ by
reversing the momenta of all particles at all points in
$\Gamma_{M}$). Then we have $S_{_{B}} (M)= S_{_{B}}(RM)$; that is,
the Boltzmann entropy is invariant under $R$.

Now consider a system that assumes macro-states $M_{i}$ and $M_{f}$,
at $t_{_{i}}$ and $t_{_{f}}$ respectively, where $S_{i} :=
S_{_{B}}(M_{i}) < S_{_{B}}(M_{f}) =: S_{f}$ and $t_{_{i}}<
t_{_{f}}$. Furthermore assume that the system's fine grained state
is $x_{_{i}} \in \Gamma_{M_{i}}$ at $t_{_{i}}$ and is $x_{_{f}} \in
\Gamma_{M_{f}}$ at $t_{_{f}}$, and that the transition from
$x_{_{i}}$ to $x_{_{f}}$ during the interval $\Delta := t_{_{f}} -
t_{_{i}}$ is allowed by the underlying dynamics. Now, by Premise 1
the system is time reversal invariant and hence the transition from
$Rx_{_{f}}$ to $Rx_{_{i}}$ during $\Delta$ is possible as well.
Because, by Premise 2, $S_{_{B}}$ is invariant under $R$, we have to
conclude that the transition from $S_{f}$ to $S_{i}$ is possible as
well. This contradicts the Second Law of thermodynamics which says
that high to low entropy transitions cannot occur. So we are in the
awkward position that a transition that is ruled out by the macro
theory is allowed by the micro theory which is supposed to account
for why the macro laws are the way they are.


\subsubsection{Issue 4: Zermelo's Recurrence Objection}\label{Zermelo}

Poincar\'{e}'s recurrence theorem says, roughly speaking, that for
the systems\index{Poincar\'{e}'s Recurrence Theorem} at stake in SM,
almost every initial condition will, after some \emph{finite} time
(the Poincar\'{e} recurrence time), return to a state that is
arbitrarily close to its initial state (see Appendix A for details).
As Zermelo pointed out in 1896, this has the
unwelcome\index{Zermelo's recurrence objection} consequence that
entropy cannot keep increasing all the time; sooner or later there
will be a period of time during which the entropy of the system
decreases. For instance, if we consider again the initial example of
the gas (Figure 1), it follows from Poincar\'{e}'s recurrence
theorem that there is a future instant of time at which the gas
returns \emph{by itself} to the left half of the container. This
stands in contradiction to the second law.

A first attempt to dismiss this objection points to the fact that
the time needed for this to happen in a realistic system is several
times the age of the universe. In fact Boltzmann himself estimated
that the time needed for a recurrence to occur for a system
consisting of a cubic centimeter of air was about $10^{10^{19}}$
seconds (Uffink 2007, 984). Hence, we never observe such a
recurrence, which renders the objection irrelevant.

This response misses the point. The objection is not concerned with
whether we ever \emph{experience} a decrease of entropy; the
objection points to the fact that there is an \emph{in-principle}
incompatibility between the Second Law and the behaviour of
classical mechanical systems. This is, of course, compatible with
saying that there need not be any conflict with actual experience.

Another response is to let the number of particles in the system
tend towards infinity (which is the basic idea of the so-called
thermodynamic limit; see Subsection \ref{Khinchin's Programme and
the Thermodynamic Limit}). In this case the Poincar\'{e} recurrence
time becomes infinite as well. However, actual systems are not
infinite and whether such limiting behaviour explains the behaviour
of actual systems is at least an open question.


\subsubsection{Issue 5: The Justification of Coarse-Graining}\label{Issue 5}

The introduction of a finite partition on the system's $\mu$-space
is crucial to the combinatorial argument. Only with respect to such
a partition can the notion of a distribution be introduced and
thereby the Maxwell-Boltzmann equilibrium distribution be derived.
Hence the use of a partition is essential. However, there is nothing
in classical mechanics itself that either suggests or even justifies
the introduction of a partition. How, then, can coarse-graining be
justified?

This question is further aggravated by the fact that the success of
the combinatorial argument crucially depends on the choice of the
\emph{right} partition. The Maxwell-Boltzmann distribution is
derived under the assumption that $n$ is large and $n_{i} \gg 1$,
$i=1, ..., l$. This assumption is false if we choose too fine a
partition (for instance one for which $l \geq n$), in which case
most $n_{i}$ are small.

There are also restrictions on what kinds of partitions one can
choose. It turns out that the combinatorial argument only works if
one partitions phase space. Boltzmann first develops the argument by
partitioning the particles' energy levels and shows that in this
case the argument fails to reproduce the Maxwell-Boltzmann
distribution (1877, 168-190). The argument yields the correct result
only if the phase space is partitioned along the position and
momentum axes into cells of equal size.\footnote{Strictly speaking
this requirement is a bit too stringent. One can choose a different
(but constant) cell size along each axis and still get the right
results (Boltzmann 1877, 190).} But why is the method sensitive to
such choices and what distinguishes `good' from `bad' partitions
(other than the ability to reproduce the correct distribution law)?


\subsubsection{Issue 6: Limitations of the Formalism}\label{Issue 6}

When deriving the Maxwell-Boltzmann
distribution\index{Maxwell-Boltzmann distribution} in Subsection
\ref{The Combinatorial Argument}, we made the assumption that the
energy of a particle depends only on its coarse-grained micro-state,
i.e. on the cell in which its fine-grained micro-state comes to lie,
which (trivially) implies that a particle's energy does not depend
on the other particles' states. This assumption occupies centre
stage in the combinatorial argument because the derivation of the
Maxwell-Boltzmann distribution depends on it. However, this is true
only if there is no interaction between the particles; wherever
there is an interaction potential between the particles of a system
the argument is inapplicable. Hence, the only system satisfying the
assumptions of the argument is the ideal gas (which, by definition,
consists of non-interacting particles).

This restriction is severe. Although some real gases approximately
behave like ideal gases under certain circumstances (basically: if
the density is low), most systems of interest in statistical
mechanics cannot be regarded as ideal gases. The behaviour both of
solids and liquids (and even of dense gases) essentially depends on
the interaction between the micro-constituents of the system, and a
theory that is forced to idealise these interactions away (should
this be possible at all) is bound to miss out on what is essential
to how real systems behave.\footnote{A discussion of this point can
be found, for instance, in Schr\"{o}dinger (1952, Ch. 1).}

A further limitation is that the argument assumes that all the
particles of the system have the same phase space, which essentially
amounts to assuming that we are dealing with a system of identical
objects. A paradigm example for such a system is a monoatomic gas
(e.g. helium). But many systems are not of that kind; most solids,
for instance, contain constituents of different types.

Finally, the formalism remains silent about what happens when
systems interact with their environments. In practice many systems
are not completely isolated and one would like the formalism to
cover at least selected kinds of interactions.

Hence, the question is whether the approach can be generalised so
that it applies to the cases that, as it stands, are not within its
scope.


\subsubsection{Issue 7: Reductionism}\label{Issue 7}

There is a consensus that the principal aim of SM is to account for
the thermodynamic behaviour of a macroscopic system in terms of the
dynamical laws governing its micro constituents; and it is a measure
of the success of SM how much of TD it is able to reproduce (see
Subsection \ref{trailer}). In philosophical parlance, the aim of SM
is to \emph{reduce} TD to mechanics plus probabilistic assumptions.

What does such a reduction involve?\index{Reduction!of
thermodynamics to statistical mechanics} How do the micro and the
macro level have to relate to one another in order for it to be the
case that the former reduces to the latter? The term `reduction' has
been used in many different senses and there is no consensus over
what exactly it involves to reduce one domain to another. So we need
to specify what exactly is asserted when SM is claimed to reduce TD
and to discuss to what extent this assertion is true.

A particular problem for reductionism is that idealisations play a
constitutive r\^{o}le in SM (Sklar 2000, 740).\index{Reduction!and
idealisation} Depending on which approach we favour, we work with,
for example: non-interacting particles or hard spheres in a box
instead of `realistic' interactions; or systems of infinite volume
and an infinite number of particles; or vanishing densities; or
processes that take infinitely long. These idealisations are more
than the `usual' inexactness that is unavoidable when applying a
general theory to the messy world; they are essential to SM since
the desired results usually cannot be derived without them. What is
the status of results that only hold in highly idealised systems
(and often are know to fail in more realistic systems) and what
r\^{o}le can they play in a reduction of TD to SM?


\subsubsection{Plan}

Subsection \ref{The Ergodicity Programme} presents and discusses the
`orthodox' response to Issues 1 and 2, which is based on ergodic
theory and the use of macro-probabilities. In Subsections \ref{The
Past Hypothesis} and \label{Micro Probabilities Revisited} I discuss
the currently most influential alternative answer to these issues,
which invokes the so-called Past Hypothesis and uses
micro-probabilities. Issues 3 and 4 are addressed in Unit
\ref{Loschmidt's and Zermelo's Objections}. In Subsection
\ref{Limitations} I deal with Issues 5 and 6, and Issue 7 is
discussed in Subsection \ref{Reductionism - Boltzmann}.


\subsection{The Ergodicity Programme}\label{The Ergodicity Programme}

The best-known response to Issues 1 and 2, if macro probabilites are
considered, is based on the notion of ergodicity. For this reason
this subsection begins with an introduction to ergodic theory, then
details how ergodicity is supposed to address the problems at stake,
and finally explains what difficulties this approach faces.


\subsubsection{Ergodic Theory}\label{Ergodic Theory}
\index{Ergodic theory} Modern ergodic theory is developed within the
setting of dynamical systems theory.\footnote{The presentation of
ergodic theory in this subection follows by and large Arnold and
Avez (1968) and Cornfeld \emph{et al.} (1982). For accounts of the
long and intertwined history of ergodic theory see Sklar (1993, Chs.
2 and 5) and von Plato (1991, 1994, Ch. 3).} A dynamical system is a
triplet $(X, \lambda, \phi_{t})$, where $X$ is a state space endowed
with a normalised measure $\lambda$ (i.e. $\lambda(X)=1$) and
$\phi_{t}: X \rightarrow X$, where $t$ is a real number, is a
one-parameter family of measure preserving automorphisms (i.e.
$\lambda(\phi_{t}(A)) = \lambda(A)$ for all measurable $A\subseteq
X$ and for all $t$); the parameter $t$ is interpreted as time. The
Hamiltonian systems considered so far are dynamical systems in this
sense if the following associations are made: $X$ is the accessible
part of the energy hypersurface; $\lambda$ is the standard measure
on the energy hypersurface, renormalised so that the measure of the
accessible part is one; $\phi_{t}$ is the Hamiltonian flow.

Now let $f(x)$ be any complex-valued and Lebesgue-integrable
function defined on $X$. Its space mean $\bar f$ (sometimes also
`phase space average' or simply `phase average') is defined as
\begin{equation}
    \bar f := \int_{X}f(x)d\lambda,
    \label{eq:¥}
\end{equation}¥
and its time mean $f^{*}$ (sometimes also `time average') at $x_{0}
\in X$ is defined as
\begin{equation}
    f^{*}(x_{0})= \lim_{\tau \rightarrow
\infty}(1/\tau) \int_{t_{0}}^{t_{0}+\tau}f[\phi_{t}(x_{0})] dt.
    \label{eq:¥}
\end{equation}¥

\noindent The question is whether the time mean exists; the Birkhoff
theorem asserts that it does:\index{Birkhoff Theorem}

\begin{quote}
    \textit{Birkhoff Theorem}. Let $(X, \lambda, \phi_{t})$ be a
    dynamical system and $f$ a complex-valued, $\lambda$-integrable
    function on $X$. Then the time average $f^{*}(x_{0})$\newline
    \noindent (i) exists almost everywhere (i.e. everywhere except,
    perhaps, on a set of measure zero);\newline
    \noindent (ii) is invariant (i.e. does not depend on the
    initial time $t_{0}$): $f^{*}(x_{0})=f^{*}(\phi_{t}(x_{0}))$ for all $t$;\newline
    \noindent (iii) is integrable: $\int_{X} f^{*}(x_{0})d\lambda =
    \int_{X} f(x)d\lambda$.
\end{quote}

\noindent We can now state the central definition:\index{Ergodic theorem!definition of ergodicity}

\begin{quote}
    \textit{Ergodicity}. A dynamical system is ergodic iff
    for every complex-valued, $\lambda$-integrable function $f$ on
    $X$ we have $f^{*}(x_{0}) = \bar f$ almost everywhere; that is,
    everywhere except, perhaps, on a set of measure zero.
\end{quote}

\noindent Two consequences of ergodicity are worth emphasising. First, if a
system is ergodic, then for almost all trajectories, the fraction of
time a trajectory spends in a region $R$ equals the fraction of the
area of $X$ that is occupied by $R$. This can easily be seen by
considering $f(x) = \chi_{_{R}} (x)$, where $\chi_{_{R}} (x)$ is the
characteristic function of the region $R$: $\chi_{_{R}} (x) = 1$ if
$x \in R$ and $\chi_{_{R}} (x) = 0$ otherwise. We then have $\bar f
= \lambda(R) = f^{*}(x)$, meaning that the fraction of time the
system spends in $R$ equals $\lambda(R)$, which is the fraction of
the area of $X$ that is occupied by $R$.

Second, almost all trajectories (i.e. trajectories through almost
all initial conditions) come arbitrarily close to any point on the
energy hypersurface infinitely many times; or to put it another way,
almost all trajectories pass through every subset of $X$ that has
positive measure infinitely many times. This follows from the fact
that the time mean equals the space mean, which implies that the
time mean cannot depend on the initial condition $x_{0}$. Hence a
system can be ergodic only if its trajectory may access all parts of
the energy hypersurface.

The latter point is also closely related to the decomposition
theorem. We first define:

\begin{quote}
     \textit{Decomposability}. A system is decomposable (sometimes also
     `metrically decomposable' or `metrically intransitive')
     iff there exist
     two regions $X_{1}$ and $X_{2}$ of non-zero measure
     such that $X_{1} \cap X_{2} = \emptyset$ and $X_{1} \cup X_{2} =
     X$, which are invariant under the dynamics of the system:
     $\phi_{t}(X_{1}) \subseteq X_{1}$ and $\phi_{t}(X_{2}) \subseteq X_{2}$ for all
     $t$. A system that is not decomposable is
     indecomposable (`metrically indecomposable' or `metrically
     transitive').
\end{quote}

\noindent Then we have:

\begin{quote}
    \textit{Decomposition Theorem}. A dynamical system
    is ergodic if and only if it is indecomposable; i.e. if every
    invariant measurable set has either measure zero or one.
\end{quote}

The ergodic measure is unique, up to a continuity requirement, in
the sense that there is only one measure invariant under the
dynamics. We first define:

\begin{quote}
     \textit{Absolute Continuity}. A measure $\lambda'$ is absolutely continuous with
     respect to $\lambda$ iff for any measurable region $A \subseteq X$: if
     $\lambda(A)=0$ then $\lambda'(A)=0$.
\end{quote}

\noindent We then have:

\begin{quote}
    \textit{Uniqueness Theorem}. Assume that $(X, \lambda, \phi_{t})$ is
    ergodic and $\lambda$ is normalised. Let $\lambda'$ be another
    measure on $X$ which is normalised, invariant under $\phi_{t}$, and
    absolutely continuous with respect to $\lambda$. Then $\lambda=\lambda'$.
\end{quote}

For what follows it will also be important to introduce the notion
of mixing.

\begin{quote}
    \textit{Mixing}. A system is mixing\footnote{Strictly speaking
    this property is called `strong mixing'
    since there is a similar condition called `weak mixing'. The
    differences between these need not occupy us here. For details see
    Arnold and Avez (1968, Ch. 2).} if and only if for all measurable subsets $A$ and $B$
    of $X$: $\lim_{t \rightarrow \infty} \mu(\phi_{t}B \cap A) \, =
    \,\mu(A)\mu(B)$.
\end{quote}

The meaning of this concept can be visualised as follows. Think of
the phase space as a glass of water to which a shot of scotch has
been added. The volume of the cocktail $X$ (scotch + water) is
$\mu(X)$ and the volume of scotch is $\mu(B)$; hence the
concentration of scotch in $X$ is $\mu(B) / \mu(X)$. Now stir.
Mathematically, the time evolution operator $\phi_{t}$ represents
the stirring, meaning that $\phi_{t}(B)$ is the region occupied by
the scotch after time $t$. The cocktail is mixed, if the
concentration of scotch equals $\mu(B) / \mu(X)$ not only with
respect to the \textit{entire} `glass' $X$, but with respect to any
arbitrary (but non-zero measure) region $A$ in that volume; that is,
it is mixed if $\mu(\phi_{t}(B) \cap A) / \mu(A)\, = \, \mu(B) /
\mu(X)$ for any finite volume $A$. This condition reduces to
$\mu(\phi_{t}(B) \cap A) / \mu(A)\, = \, \mu(B)$ for any region $B$
because, by assumption, $\mu(X)=1$. If we now assume that mixing is
achieved only for $t\rightarrow \infty$ we obtain the above
condition.

One can then prove the following two theorems.

\begin{quote}
    \textit{Implication Theorem}. Every dynamical system that is mixing is
    also ergodic, but not vice versa.
\end{quote}

\begin{quote}
    \textit{Convergence Theorem}. Let $(X, \lambda, \phi_{t})$ be
    a dynamical system and let $\rho$ be a measure on $X$ that is
    absolutely continuous with $\lambda$ (but otherwise arbitrary).
    Define $\rho_{t}(A):= \rho(\phi_{t}(A))$ for all measurable $A
    \subseteq X$. Let $f(x)$ be a bounded measurable function on $X$.
    If the system is mixing, then $\rho_{t} \rightarrow \lambda$ as
    $t \rightarrow \infty$ in the sense that for all such $f$:
    \begin{equation}\label{convergence theorem}
        \lim_{t \rightarrow \infty} \int f(x) d\rho_{t} \, = \, \int
        f(x) d\lambda.
    \end{equation}

\end{quote}


\subsubsection{Promises}\label{Promises}

Assuming that the system in question is ergodic seems to provide us
with neat responses to Issues 1 and 2, if macro-probabilities are
considered.

Thus, let us ask the question of how are we to understand statements
about the probability of a macro-state? That is, how are we to
interpret the probabilities introduced in Equation
(\ref{Proportionality Postulate})? A natural suggestion is that
probabilities should be understood as time averages. More
specifically, the suggestion is that the probability of a
macro-state $M$ is the fraction of time that the system's state
spends in $\Gamma_{M}$ (the so-called sojourn time):

\begin{equation}\label{p(R)}
    p(M) = \frac{1}{\tau} \int_{t_{0}}^{t_{0}+\tau}
    \chi_{_{\Gamma_{M}}} [\phi_{t}(x)] dt,
\end{equation}¥

\noindent where $\chi_{_{\Gamma_{M}}}$ is the characteristic
function (as defined above) and $[t_{0}, t_{0}+\tau]$ is some
suitable interval of time.

This definition faces some prima facie problems. First, what is the
suitable interval of time? Second, does this time average exist?
Third, as defined in Equation (\ref{p(R)}), $p(M)$ exhibits an
awkward dependence on the initial condition $x$. These difficulties
can be overcome by assuming that the system is ergodic. In this case
the relevant time interval is infinity; the existence question is
resolved by Birkhoff's theorem, which states that the infinite time
limit exists almost everywhere; and the awkward dependence on the
initial condition vanishes because in an ergodic system the infinite
time means equals the space means for almost all initial conditions,
and hence a fortiori the time mean, does not depend on the initial
condition $x$ (for almost all $x$).

This puts the time average interpretation of SM probabilities on a
solid foundation and at the same time also offers a response to the
problem of the mechanical foundation of the PP. The combinatorial
considerations in the last subsection have shown that the
equilibrium state occupies by far the largest part of
$\Gamma_{\gamma}$. Combining this with the fact that the time a
system spends in a given region of $\Gamma_{\gamma}$ is proportional
to its measure provides a faultless mechanical justification of
PP.\footnote{For a further discussion of ergodicity and issues in
the interpretation of probability in the Boltzmann approach see von
Plato (1981, 1982, 1988, 1989), Gutmann (1999), van Lith (2003),
Emch (2005).}

In sum, if the system is ergodic, we seem to have a neat mechanical
explanation of the system's behaviour as well as a clear
interpretation of the probabilities that occur in the PP.


\subsubsection{Problems}\label{Problems with Ergodicity}

The Ergodic programme faces serious difficulties. To begin with, it
turns out to be extremely difficult to prove that the systems of
interest really are ergodic. Contrary to what is sometimes asserted,
not even a system of $n$ elastic hard balls moving in a cubic box
with hard reflecting walls has  been proven to be ergodic for
arbitrary $n$; it has been proven to be ergodic only for $n\leq 4$.
Moreover, hard ball systems are highly idealised (molecules do not
behave like hard balls) and it is still an open question whether
systems with more realistic interaction potentials (e.g.
Lenard-Jones potentials) are ergodic.\footnote{For further
discussions of this issue see Sklar (1993, Ch. 5), Earman \& Redei
(1996, Sec. 4), Uffink (2007, Sec. 6),  Emch \& Liu (2002, Chs.
7-9), and Berkovitz \emph{et al.} (2006, Sec. 4).}

What is worse than the absence of proofs that the systems of
interest are ergodic is that there are systems that show the
appropriate behaviour and yet are known not to be ergodic. For
instance, in a solid the molecules oscillate around fixed positions
in a lattice, and as a result the phase point of the system can only
access a small part of the energy hypersurface (Uffink 2007, 1017).
Bricmont (2001) investigates the Kac Ring Model (Kac 1959) and a
system of $n$ uncoupled anharmonic oscillators of identical mass,
and points out that both systems exhibit thermodynamic behaviour---
and yet they fail to be ergodic. And most notably, a system of
non-interacting point particles is known not be ergodic; yet
ironically it is exactly this system on which the combinatorial
argument is based (Uffink 1996b, 381). Hence, ergodicity is not
necessary for thermodynamic behaviour.\footnote{It has been argued
that ergodicity is not sufficient either because there are systems
that are ergodic but don't show an approach to equilibrium, for
instance two hard spheres in a box (Sklar 1973, 209). This is, of
course, correct. But this problem is easily fixed by adding the
qualifying clause that if we consider a system \emph{of interest in
the context of SM}---i.e. one consisting of something like $10^{23}$
particles---then if the system is ergodic it shows SM behaviour.}
But, as Earman \& Redei (1996, 70) and van Lith (2001a, 585) point
out, if ergodicity is not necessary for thermodynamic behaviour,
then ergodicity cannot provide a satisfactory explanation for this
behaviour. Either there must be properties other than ergodicity
that explain thermodynamic behaviour in cases in which the system is
not ergodic, or there must be an altogether different explanation
for the approach to equilibrium even for systems which are
ergodic.\footnote{The term `explanation' here is used in a
non-technical sense; for a discussion of how the use of ergodicity
ties in with certain influential philosophical views about
explanation see Sklar (1973) and Quay (1978).}

But even if a system turns out to be ergodic, further problems
arise. All results and definitions of ergodic theory come with the
qualification `almost everywhere': the Birkhoff theorem ensures that
$f^{*}$ exists \textit{almost everywhere} and a dynamical system is
said to be ergodic iff for every complex-valued, Lebesgue-integrable
function $f$ the time mean equals the space mean \textit{almost
everywhere}. This qualification is usually understood as suggesting
that sets of measure zero can be neglected or ignored. This,
however, is neither trivial nor evidently true. What justifies the
neglect of these sets? This has become known as the `measure zero
problem'.\index{Measure zero problem} The idea seems to be that
points falling in a set of measure zero are `sparse' and this is why
they can be neglected. This view receives a further boost from an
application of the Statistical Postulate, which assigns probability
zero to events associated with such sets. Hence, so goes the
conclusion, what has measure zero simply doesn't
happen.\footnote{This piece of `received wisdom' is clearly
explained but not endorsed in Sklar (2000a, 265-6).}

This is problematic for various reasons. First, sets of measure zero
can be rather `big'; for instance, the rational numbers have measure
zero within the real numbers. Moreover, a set of measure zero need
not be (or even appear) negligible if sets are compared with respect
to properties other than their measures. For instance, we can judge
the `size' of a set by its cardinality or Baire category rather than
by its measure, which leads us to different conclusions about the
set's size (Sklar 1993, 182-188).

Furthermore it is a mistake to assume that an event with measure
zero cannot occur. In fact, having measure zero  and being
impossible are distinct notions. Whether or not the system at some
point was in one of the special initial conditions for which the
space and time mean fail to be equal is a factual question that
cannot be settled by appeal to measures; pointing out that such
points are scarce in the sense of measure theory does not do much,
because it does not imply that they are scarce in the world as
well.\footnote{Sklar (1973, 210-211) makes a very similar point when
discussing the Gibbs approach.} All we can do is find out what was
the case, and if the system indeed was in one of these initial
conditions then considerations based on this equality break down.
The fact that SM works in so many cases suggests that they indeed
are scarce, but this is a matter of fact about the world and not a
corollary of measure theory.\footnote{For a further discussion of
this issue see Friedman (1976).} Hence, an explanation of SM
behaviour would have to consist of the observation that the system
is ergodic and that it additionally started in an initial condition
which is such that space and time means are equal.

For these reasons a time average interpretation of
macro-probabilities is problematic. However, alternative
interpretations do not fare better. Frequentism is ruled out by the
fact that the relevant events in SM do not satisfy the requirement
of von Mises' theory (van Lith 2001, 587), and a propensity
interpretation (Popper 1959) fails because the existence of
propensities is ultimately incompatible with a deterministic
underlying micro theory (Clark 2001).\footnote{For a further
discussion of this issue see Butterfield (1987) and Clark (1987;
1989; 1995).}

A peculiar way around the problem of interpreting probabilities is
to avoid probabilities altogether. This is the strategy pursued,
among others, by Goldstein (2001),  Lebowitz (1993b), Goldstein \&
Lebowitz (2004) and Zangh\`{i} (2005) in their presentation of the
Boltzmannian account. The leading idea of this approach is that
equilibrium states are `typical' while non-equilibrium states are
`atypical', and that the approach to equilibrium can be understood
as a transition from atypical to typical states. For a discussion of
this approach to SM see Frigg (2007b).


\subsection{The Past Hypothesis}\label{The Past Hypothesis}


\subsubsection{The Past Hypothesis Introduced}

Let us now turn to Issues 2 and 3, and base our discussion on
micro-probabilities. The two problems we have to solve are (a) that
high to low entropy transitions are allowed by the dynamics (by the
reversibility objection) and (b) that most trajectories compatible
with a given non-equilibrium state are ones that have evolved into
that state from a state of \emph{higher} entropy (which is a
consequence of SP and the time reversal invariance of the micro
dynamics).

There is a common and now widely accepted solution to these problems
which  relies on the fact that a system's actual behaviour is
determined by its dynamical laws \emph{and} its initial condition.
Hence there need not be a contradiction between time reversal
invariant laws and the fact that high to low entropy transitions do
not (or only very rarely) occur \emph{in our world}. All we have to
do is to assume that the relevant systems in our world have initial
conditions which are such that the system's history is indeed one
that is characterised by low to high entropy transitions. That
initial conditions of this kind are scarce is irrelevant; all that
matters is that the system \emph{de facto} started off in one of
them. If this is the case, we find the irreversible behaviour that
we expect. However, this behaviour is now a consequence not only of
the laws governing the system, but also of its special initial
condition.

The question is at what point in time the relevant low entropy
initial condition is assumed to hold. A natural answer would be that
the beginning of an experiment is the relevant instant; we prepare
the gas such that it sits in the left half of the container before
we open the shutter and this is the low entropy initial condition
that we need. The problem with this answer is that the original
problem recurs if we draw an entropy curve for the system we find
that the low entropy state at the beginning of the experiment
evolved another high entropy state. The problem is obvious by now:
whichever point in time we chose to be the point for the low entropy
initial condition to hold, it follows that the overwhelming majority
of trajectories compatible with this state are such that their
entropy was higher in the past.\index{The past hypothesis} An
infinite regress looms large. This regress can be undercut by
assuming that there is an instant that simply has no past, in which
case it simply does not make sense to say that the system has
evolved into that state from another state. In other words, we have
to assume that the low entropy condition holds at the beginning of
the universe.

At this point modern cosmology enters the scene: proponents of
Boltzmannian SM take cosmology to inform us that the universe was
created in the big bang a long but finite time ago and that it then
was in a low entropy state. Hence, modern cosmology seems to provide
us with exactly what we were looking for. This is a remarkable
coincidence, so remarkable that Price sees in it `the most important
achievement of late-twentieth-century physics' (2004, 228). The
posit that the universe has come into existence in a low entropy
state is now (following Albert 2000) commonly referred to as the
`Past Hypothesis' (PH); let us call the state that it posits the
`Past State'. In Albert's formulation PH is the claim

\begin{quote}
    `[...] that the world first came into being in whatever particular
    low-entropy highly condensed big-bang sort of macrocondition it is
    that the normal inferential procedures of cosmology will eventually
    present to us' (2000, 96).
\end{quote}

\noindent This idea can be traced back to Boltzmann (see Uffink
2007, 990) and has since been advocated, among others, by Feynman
(1965, Ch. 5), Penrose (1989, Ch. 7; 2006, Ch. 27), Price (1996,
2004, 2006), Lebowitz (1993a, 1993b, 1999), Albert (2000), Goldstein
(2001), Callender (2004a, 2004b), and Wald (2006).

There is a remarkable consensus on the formulation and content of
PH; different authors diverge in what status they attribute to it.
For Feynman, Goldstein, and Penrose PH, seems to have the status of
a law, which we simply add to the laws we already have. Whether such
a position is plausible depends on one's philosophical commitments
as regards laws of nature. A discussion of this issue would take us
too far afield; surveys of the philosophical controversies
surrounding the concept of a law of nature can be found in, among
others, Armstrong (1983), Earman (1984) and Cartwright and
Alexandrova (2006). Albert regards PH as something like a Kantian
regulative principle in that its truth has to be assumed in order to
make knowledge of the past possible at all. On the other hand,
Callender, Price, and Wald agree that PH is not a law, but just a
contingent matter of fact; but they have conflicting opinions about
whether this fact is in need of explanation.\footnote{Notice that
this view has the consequence that the Second Law of thermodynamics,
or rather its `statistical cousin', Boltzmann's Law, becomes a de
facto regularity and is thus deprived it of its status as a law
properly speaking.} Thus for Price (1996, 2004) the crucial question
in the foundation of SM is not so much why entropy increases, but
rather why it ever got to be so low in the first place.\index{Price,
H.} Hence, what really needs to be explained is why the universe
shortly after the big bang was in the low entropy state that PH
posits. Callender (1998, 2004a, 2004b) argues that this quest is
wrong. PH simply specifies initial conditions of a process because
initial conditions, irrespective of whether they are special or not,
are not the kind of thing that is in need of
explanation.\index{Callender, C.} Similar concerns have also been
raised by Sklar (1993, 309-318).


\subsubsection{Problems and Criticisms}

PH has recently come under attack. Earman (2006) argues that what
at\index{Earman, J.} first glance looks like a great
discovery---that modern cosmology posits exactly the kind of Past
State that the Boltzmannian account requires---turns out to be `not
even false' (400). Earman first investigates a particular
Friedman-Robertson-Walker model of cosmology suggested by Hawking
and Page and shows that in this model probabilities are typically
ill-defined or meaningless, and he then argues that this result is
not an artefact of the idealisations of the models and would crop up
equally in more realistic models (417-418). Hence, for the
cosmologies described in general relativity there is no well-defined
sense in which the Boltzmann entropy has a low value. And worse,
even if quantum gravity or some other yet to be discovered theory
came to rescue and made it possible to give a well-defined
expression for the Boltzmann entropy at the beginning of the
universe, this would be of little help because the dynamics of the
cosmological models does not warrant the claim that there will be
monotonic increase in entropy (418-420). For these two reasons,
Earman concludes, the past hypothesis is untenable.

Whatever the eventual verdict of Earman's critique of PH, there is a
further problem is that the Boltzmann entropy is a global quantity
characterising the macro-state of an entire system, in this case the
entire universe. The fact that this quantity is low does not imply
that the entropy of a particular small subsystem of interest is also
low. And what is worse, just because the overall entropy of the
universe increases it need not be the case that the entropy in a
small subsystem also increases. A decrease in the entropy in one
part of the universe may be balanced by an increase in entropy in
some other part of the universe and hence is compatible with an
increase in the overall entropy. Hence, SM cannot explain the
behaviour of small systems like gases in laboratories. Winsberg
(2004a, 499-504) addresses this problem and argues that the only way
to avoid it is to make a further conjecture about the theory (he
calls it `Principle 3'), which in effect rules out local `entropic
misbehaviour'. However, as he points out, this principle is clearly
false and hence there is no way for the Boltzmannian to rule out
behaviour of this kind.

it is not the time to notice that a radical shift has occurred at
the beginning of this subsection. We started with a pledge to
explain the behaviour of homely systems like a vessel full of gas
and ended up talking about the Big Bang and the universe as a whole.
At least to some, this looks like using a sledgehammer to crack
nuts, and not a very wise move because most of the problems that it
faces are caused by the move to the cosmological scale. The natural
reaction to this is to downsize again and talk about laboratory
scale systems. This is what happens in the so-called `branch systems
approach', which is inspired by Reichenbach's (1956) discussion of
the direction of time, and is fully articulated in Davies (1974) and
discussed in Sklar (1993, 318-332).

The leading idea is that the isolated systems relevant to SM have
neither been in existence forever, nor continue to exist forever
after the thermodynamic processes took place. Rather, they separate
off from the environment at some point (they `branch') then exist as
energetically isolated systems for a while and then usually merge
again with the environment. Such systems are referred to as `branch
systems'. For instance, the system consisting of a glass and an ice
cube comes into existence when someone puts the ice cube into the
water, and it ceases to exist when someone pours it into the sink.
So the question becomes why a branch system like the water with the
ice cube behaves in the way it does. An explanation can be given
along the lines of the past hypothesis, with the essential
difference that the initial low entropy state has to be postulated
not for the beginning of the universe but only for the state of the
system immediately after the branching. Since the system, by
stipulation, did not exist before that moment, there is also no
question of whether the system has evolved into the current state
from a higher entropy state. This way of looking at things is in
line with how working physicists think about these matters for the
simple reason that low entropy states are routinely prepared in
laboratories---hence Lebowitz's (1993b, 11) remark that the origin
of low entropy initial states is no problem in laboratory
situations.

Albert dismisses\index{Albert, D. Z.} this idea as `sheer madness'
(2000, 89) for three reasons. First, it is impossible to specify the
precise moment at which a particular system comes into being; that
is, we cannot specify the precise branching point. Second, there is
no unambiguous way to individuate the system. Why does the system in
question consist of the glass with ice, rather than the glass with
ice and the table on which the glass stands, or the glass and ice
and the table and the person watching it, or ... And this matters
because what we regard as a relevant low entropy state depends on
what we take the system to be. Third, it is questionable whether we
have any reason to assume, or whether it is even consistent to
claim, that SP holds for the initial state of the branch
system.\footnote{As we shall see in the next subsection, it is
necessary to assume that SP holds for the initial state. Proponents
of the past hypothesis and of the branch systems approach differ in
what they regard as the beginning.}

The first and the second criticism do not seem to be successful. Why
should the system's behaviour have anything to do with our inability
to \emph{decide} at what instant the system becomes energetically
isolated? So Albert's complaint must be that there is no matter of
the fact about when a system becomes isolated. If this was true, it
would indeed be a problem. But there does not seem to be a reason
why this should be so. If we grant that there is such a thing as
being isolated from one's environment (an assumption not challenged
in the first criticism), then there does not seem to be a reason to
claim that becoming isolated at some point in time should be more
problematic than the lights going off at some point in time, or the
game beginning at some point in time, or any other event happening
at some instant. The second criticism does not cut any ice either
(see Winsberg 2004b, 715). Being energetically isolated from the
rest of the universe is an objective feature of certain things and
not others. The glass and its contents are isolated from the rest of
the universe and this is what makes them a branch system; the table,
the observer, the room, the house, etc. are not, and this is why
they are not branch systems. There is nothing subjective or
arbitrary about this division. One can, of course, question whether
systems ever really are isolated (we come to this in Subsection
\ref{Interventionism}). But this is a different point. If one goes
down that road, then there simply are no branch systems; but then
there is no individuation problem either.

The third criticism leads us into deep waters. Why would we want to
deny that SP applies to the branch system at the instance of its
creation? Although Albert does not dwell on this point, his
reasoning seems to be something like the following (see Winsberg
2004b, 715-717). Take the universe at some particular time. Now
things happen: someone opens the freezer, takes an ice cube and puts
it into a glass of lukewarm water. These are physical processes
governed by the laws of mechanics; after all, at the micro level all
that happens is that swarms of particles move around in some
specific way. But then the micro-state of the glass with ice is
determined by the laws of mechanics and the micro-condition at the
earlier point of time and we can't simply `reset' the glass' state
and postulate that it is now such that SP, or any other condition
for that matter, holds. In brief, the glass' state at some point is
dictated by the laws of the theory and is not subject to
stipulations of any kind.

Whether or not one finds this criticism convincing depends on one's
philosophical commitments as regards the nature of laws. The above
argument assumes that laws are universal and valid all the time; it
assumes that not only the behaviour of the water and the ice, but
also of the table, the room, the fridge and, last but not least, the
person putting the ice into the water and everything else in the
universe are governed by the laws of mechanics. If one shares this
view, then Albert's third criticism is valid. However, this view of
laws is not uncontroversial. It has been argued that the domain of
applicability of laws is restricted: we are making a mistake if we
assume them to be universal. To someone of the latter persuasion the
above argument  has no force at all against branch systems. This
conflict surfaces again when we discuss the interventionist approach
to SM in Subsection \ref{Interventionism} and for this reason I
postpone till then a more detailed discussion of the issue of the
scope of laws.


\subsection{Micro-Probabilities Revisited}\label{Micro Probabilities Revisited}

As we have seen above, SP gives us wrong retrodictions and this
needs to be fixed. PH, as introduced in the last subsection, seems
to provide us with the means to reformulate SP so that this problem
no longer arises (Unit \ref{Conditionalising on PH}). Once we have a
rule that assigns correct probabilities to past states, we come back
to the question of how to interpret these probabilities
(\ref{Interpreting Micro Probabilities}) and then address the
reversibility and recurrence objections (Unit \ref{Loschmidt's and
Zermelo's Objections}).


\subsubsection{Conditionalising on PH}\label{Conditionalising on PH}

PH, if true, ensures that the system indeed starts in the desired
low entropy state. But, as we have seen in Unit \ref{Probabilities},
our probabilistic machinery tells us that this is overwhelmingly
unlikely. Albert (2000, Ch. 4) argues that this is unacceptable
since it just cannot be that the \emph{actual} past is
overwhelmingly unlikely for this would lead us to believe wrong
things about the past.\footnote{In fact, Albert (2000, Chs. 4 and 6)
even sees this as a fundamental problem threatening the very notion
of having knowledge of the past. Leeds (2003) takes the opposite
stance and points out that this conclusion is not inevitable since
it depends on the view that we explain an event by its having a high
probability to occur. Explaining the past, then, involves showing
that the actual past has high probability. However, if we deny that
we are in the business of explaining the past on the basis of the
present and the future, then this problem looks far less dramatic.
For a further discussion of Albert's view on past knowledge and
intervention see Frisch (2005) and Parker (2005).} The source of
this problem is that we have (tacitly) assumed that SP is valid at
all times. Hence this assumption must be renounced and a postulate
other than SP must be true at some times.

Albert (2000, 94-96) suggests the following remedy: SP is valid only
for the Past State (the state of universe just after the Big Bang);
for all later states the correct probability distribution is the one
that is uniform (with respect to the Lebesgue measure) over the set
of those conditions that are compatible with the current macro-state
\emph{and} the fact that the original macro-state of the system (at
the very beginning) was the Past State. In brief, the suggestion is
that we conditionalise on the Past Hypothesis and the current
macro-state.

More precisely, let $M_{P}$ be the macro-state of the system just
after the big bang (the Past State) and assume (without loss of
generality) that this state obtains at time $t=0$; let $M_{t}$ be
the system's macro-state at time $t$ and let $\Gamma_{t}:=
\Gamma_{M_{t}}$ be the parts of $\Gamma_{\gamma, \, a}$ that
corresponds to $M_{t}$. Then we have:

\begin{quote}
Past Hypothesis Statistical Postulate (PHSP):\index{Past hypothesis statistical postulate} SP is valid for the
Past State. For all times $t>0$, the probability at $t$ that the
fine-grained micro-state of the system lies in a subset $A$ of
$\Gamma_{t}$ is
\begin{equation}\label{PHPS}
    \mu_{_{L,\, t}}(A):= \frac{\mu_{_{L}}(A \cap R_{t})}{\mu_{_{L}}(R_{t})}
\end{equation}
\noindent whenever $R_{t} \neq 0$, where $R_{t}:= \Gamma_{t} \cap
\phi_{t}(\Gamma_{P})$ and $\phi_{t}(\Gamma_{P})$, as above, is the
image of $\Gamma_{P}$ under the dynamics of the system after time
$t$ has elapsed.

\end{quote}

\noindent This is illustrated in Figure 5. Now, by construction,
those fine-grained micro-states in $\Gamma_{t}$ which have a high
entropy past have probability zero, which is what we needed.

\vspace{1cm}

\noindent [Insert Figure 5 (appended at the end of the document).]

\vspace{1cm}

\noindent {\footnotesize{Figure 5: Illustration of the past
hypothesis statistical postulate.}}

\vspace{1cm}

\noindent However, PHSP needs to be further qualified. There might
be a `conspiracy' in the system to the effect that states with a low
entropy past and ones with a low entropy future are clumped
together. Let $\Gamma_{t, \, f}$ be the subregions of $\Gamma_{t}$
occupied by states with a low entropy future. If it now happens that
these lie close to those states compatible with PH, then PHSP---
wrongly---predicts that a low entropy future is very likely despite
the fact that the fraction of $\Gamma_{t}$ occupied by $\Gamma_{t,
\, f}$ is tiny and that SP---correctly---predicts that a low entropy
future is very unlikely (see Figure 6).

\vspace{1cm}

\noindent [Insert Figure 6 (appended at the end of the document).]

\vspace{1cm}

\noindent {\footnotesize{Figure 6: Illustration of a conspiracy
involving clumping together of low entropy past and low entropy
future states.}}

\vspace{1cm}

\noindent This problem can be avoided by requiring that $\Gamma_{t,
\, f}$ is scattered in tiny clusters all over $\Gamma_{t}$ (see
Albert 2000, 67, 81-85) so that the fraction of $\Gamma_{t, \, f}$
that comes to lie in $R_{t}$ is exactly the same as the fraction of
$\Gamma_{t}$ taken up by $\Gamma_{t, \, f}$, i.e.
$\mu_{_{L}}(\Gamma_{t, \, f})/\mu_{_{L}}(\Gamma_{t}) =
\mu_{_{L}}(\Gamma_{t, \, f} \cap R_{t})/\mu_{_{L}}(R_{t})$ (see
Figure 7).

\vspace{1cm}

\noindent [Insert Figure 7 (appended at the end of the document).]

\vspace{1cm}

\noindent {\footnotesize{Figure 7: Scattering condition as solution
to the conspiracy problem.}}

\vspace{1cm}

\noindent Let us call this the
`scattering condition'. If this condition falls in place, then the
predictions of PHSP and SP coincide and the problem is solved.

In sum, replacing SP by PHSP and requiring that the scattering
condition holds for all times $t$ is sufficient to get both
predictions and retrodictions right.

The remaining question is, of course, whether the scattering
condition holds. Albert simply claims that it is plausible to assume
that it holds, but he does so without presenting, or even
mentioning, a proof. Since this condition concerns mathematical
facts about the system, we need a proof, or a least a plausibility
argument, that it holds. Such a proof is not easy to get because the
truth of this condition depends on the dynamics of the system.


\subsubsection{Interpreting Micro-Probabilities}\label{Interpreting Micro Probabilities}

How are we to interpret the probabilities defined by PHSP?
Frequentism,\index{Probability!interpretations} time averages and
the propensity interpretation are unworkable for the same reasons as
in the context of macro-probabilities. Loewer (2001, 2004) suggested
that the way out of the impasse is to interpret PHSP probabilities
as Humean chances in Lewis' (1994) sense. \index{Lewis, D.} Consider
all deductive systems that make true assertions about what happens
in the world and also specify probabilities of certain events. The
\emph{best} system is the one that strikes the best balance between
simplicity, strength and fit, where the fit of a system is measured
by how likely the system regards it that things go the way they
actually do. Lewis then proposes as an analysis of the concept of a
law of nature that laws are the regularities of the best system and
chances are whatever the system asserts them to be. Loewer suggests
that the package of classical mechanics, PH and PHSP is a putative
best system of the world and that therefore the chances that occur
in this system can be understood as chances in Lewis' sense.

Frigg (2006, 2007a) argues that this suggestion faces serious
difficulties. First, Lewis' notion of fit is modeled on the
frequentist notion of a sequence and cannot be carried over to a
theory with continuous time. Second, even when discretising time in
order to be able to calculate fit, it turns out that Loewer's
putative best system is not the best system because there are
distributions over the initial conditions that lead to a better fit
of the system than the distribution posited in PHSP. The details of
these arguments suggest that PHSP probabilities are best understood
as epistemic probabilities of some sort.


\subsubsection{Loschmidt's and Zermelo's Objections}\label{Loschmidt's and Zermelo's Objections}

We now return to Loschmidt's and Zermelo's objections and
discuss\index{Loschmidt's Objection} in what way the micro
probability approach can address them.

\emph{Reversal Objection}: Consider the same scenario as in Unit
\ref{Loschmidt's Reversibility Objection}. Denote by $\Gamma_{if}$
the subset of $\Gamma_{M_{i}}$ consisting of all points that evolve
into $\Gamma_{M_{f}}$ during the interval $\Delta$, and likewise let
$\Gamma_{fi}$ be set of all points in $\Gamma_{M_{f}}$ that evolve
into $\Gamma_{M_{i}}$ during $\Delta$. We then have $\Gamma_{fi} = R
(\phi_{\Delta}(\Gamma_{if}))$, where $\phi_{\Delta}$ is the time
evolution of the system during time span $\Delta$. Therefore
$\mu(\Gamma_{fi}) / \mu(\Gamma_{M_{f}})$ $= \mu(R
(\phi_{\Delta}(\Gamma_{if}))) / \mu(\Gamma_{M_{f}})$ $=
\mu(\Gamma_{if}) / \mu(\Gamma_{M_{f}})$, because $\mu(RA)=\mu(A)$
for all sets $A$. By assumption $\mu(\Gamma_{M_{f}}) >
\mu(\Gamma_{M_{i}})$ (because $M_{f}$ has higher entropy than
$M_{i}$), hence $\mu(\Gamma_{if}) / \mu(\Gamma_{M_{f}}) <
\mu(\Gamma_{if}) / \mu(\Gamma_{M_{i}})$. Assuming that
conditionalising on PH would not upset these proportions, it follows
that the system is more likely to evolve from low to high entropy
than it is to evolve from high to low entropy. Now take $M_{i}$ and
$M_{f}$ to be, respectively, the state of a gas confined to the left
half of container and the state of the gas spread out evenly over
the entire available space. In this case $\mu(\Gamma_{M_{f}}) /
\mu(\Gamma_{M_{i}}) \approx 10^{n}$ ($n$ being the number of
particles in the system), and hence the system is $10^{n}$ times
more likely to evolve from low to high entropy than vice versa. This
is what BL asserts.\footnote{See Bricmont (1996, Sec. 3) for a more
detailed discussion.}

\emph{Recurrence Objection}: \index{Zermelo's recurrence objection}
Roughly speaking, the recurrence objection (see Subsection
\ref{Zermelo}) states that entropy cannot always increase because
every mechanical system returns arbitrarily close to its initial
state after some \emph{finite} time ({Poincar\'{e}'s Recurrence
Theorem). The common response (Callender 1999, 370; Bricmont 1996,
Sec. 4) to the recurrence objection has a somewhat empiricist
flavour and points out that, according to the Past Hypothesis, the
universe is still today in a low entropy state far away from
equilibrium and recurrence will therefore presumably not occur
within all relevant observation times. This, of course, is
compatible with there being periods of decreasing entropy at some
later point in the history of the universe. Hence, we should not
view BL as valid at all times.


\subsection{Limitations}\label{Limitations}

There are serious questions about the use of coarse graining, i.e.
partitions, in the combinatorial argument (issue 5) and the scope of
the theory (issue 6). I now discuss these problems one at a time.

How can coarse-graining be justified? The standard answer is an
appeal to knowledge: we can never observe the precise value of a
physical quantity because measuring instruments invariably have a
finite resolution (just as do human observation capabilities); all
we can assert is that the result lies within a certain range. This,
so the argument goes, should be accounted for in what we assert
about the system's state and the most natural way to do this is to
choose a partition whose cell size reflects what we can reasonably
hope to know about the system.

This argument is problematic because the appeal to observation
introduces a kind of subjectivity into the theory that does not
belong there. Systems approach equilibrium irrespective of what we
happen to know about them. Hence, so the objection concludes, any
reference to incomplete knowledge is out of place.\footnote{Many
authors have criticised approaches to SM that invoke limited
knowledge as deficient. Since these criticisms have mainly been put
forward against Gibbsian approaches to SM, I will come back to this
point in more detail below.}

Another line of argument is that there exists an objective
separation of relevant scales---in that context referred to as
`micro' and `macro'\footnote{Notice that this use of the terms
`micro' and `macro' does not line up with how these terms have been
used above, where both fine-grained and coarse-grained states were
situated at the `micro' level (see Subsection \ref{The
Framework}).}---and that this justifies
coarse-graining.\footnote{This point of view is often alluded to by
physicists but rarely explained, let alone defended. It also seems
to be what Goldstein has in mind when he advises us to `partition
the 1-particle phase space (the $q,p$-space) into macroscopically
small but microscopically large cells $\Delta_{\alpha}$' (2001,
42).} The distinction between the two scales is considered objective
in much the same way as, say, the distinction between dark and
bright: it may not be clear where exactly to draw the line, but
there is no question that there is a distinction between dark and
bright. From a technical point of view, the separation of scales
means that a macro description is bound to use a finite partition
(whose cell size depends on where exactly one draws the line between
the micro and macro scales). This justifies Boltzmannian
coarse-graining.

The question is whether there really is an objective micro-macro
distinction of this kind. At least within the context of classical
mechanics this is not evidently the case. In quantum mechanics
Planck's constant gives a natural limit to how confined a state can
be in both position and momentum, but classical mechanics by itself
does not provide any such limit. So the burden of proof seems to be
on the side of those who wish to uphold that there is an objective
separation between micro and macro scales.

And this is not yet the end of the difficulties . Even if the above
arguments were successful, they would remain silent about the
questions surrounding the choice of the `right' partition. Nothing
in either the appeal to the limits of observation or the existence
of an objective separation of scales explains why coarse-graining
energy is `bad' while coarse-graining position and momentum is
`good'.

These problems are not easily overcome. In fact, they seem so
serious that they lead Penrose to think that `entropy has the status
of a ``convenience'', in present day theory, rather than being
``fundamental'''(2006, 692) and that it only would acquire a `more
fundamental status' in the light of advances in quantum theory, in
particular quantum gravity, as only quantum mechanics provides the
means to compartmentalise phase space (\emph{ibid}.).

In the light of these difficulties the safe strategy seems to be to
renounce commitment to coarse-graining by downgrading it to the
status of a mere expedient, which, though instrumentally useful, is
ultimately superfluous. For this strategy to be successful the
results of the theory would have to be robust in the limit $\delta
\omega \rightarrow 0$.

But this is not the case. The terms on the right hand side of
Equation (\ref{S = k ln Omega}) diverge in the limit $\delta \omega
\rightarrow 0$. And this is not simply a `technical accident' that
one can get straight given enough mathematical ingenuity. On the
contrary, the divergence of the Boltzmann entropy is indicative of
the fact that the whole argument is intimately tied to there being
finitely many cells which serve as the starting point for a
combinatorial argument. Using combinatorics simply does not make
sense when dealing with a continuum; so it is only natural that the
argument breaks down in the continuum limit.

Let us now turn to the limitations of the formalism, which are
intimately connected to the Boltzmannian conception of equilibrium.
The equilibrium macro-state, by definition, is the one for which
$S_{_{B}}$ is maximal. \emph{Per se} this is just a definition and
its physical relevance needs to be shown. This is done in two steps.
First, we use the combinatorial argument to explicitly construct the
macro-regions as those parts of the energy hypersurface that
correspond to a certain distribution, and then show that the largest
macro-region is the one that corresponds to the Maxwell-Boltzmann
distribution. But why is this the equilibrium distribution of a
physical system? This is so, and this is the second step, because
(a) predictions made on the basis of this distribution bear out in
experiments, and (b) Maxwell showed in 1860 that this distribution
can be derived from symmetry considerations that are entirely
independent of the use of a partition (see Uffink (2007, 943-948)
for a discussion of Maxwell's argument). This provides the
sought-after justification of the proposed definition of
equilibrium.

The problem is that this justification is based on the assumption
that there is no interaction between the particles in the system and
that therefore the total energy of the system is the sum of the
`individual' particle energies. While not a bad characterisation of
the situation in dilute gases, this assumption is radically false
when we consider systems with non-negligible interactions such as
liquids, solids, or gravitating systems. Hence, the above
justification for regarding the macro-state for which $S_{_{B}}$ is
maximal as the equilibrium state is restricted to dilute gases, and
it is not clear whether the equilibrium macro-state can be defined
in the same way in systems that are not of this kind.

There is a heuristic argument for the conclusion that this is
problematic. Consider a system of gravitating particles. These
particles attract each other and hence have the tendency to clump
together. So if it happens that a large amount of these are
distributed evenly over a bounded space, then they will move
together and eventually form a lump. However, the phase volume
corresponding to a lump is much smaller than the one corresponding
to the original spread out state, and hence it has lower Boltzmann
entropy.\footnote{A possible reply to this is that the loss in
volume in configuration space is compensated by an increase in
volume in momentum space. Whether this argument is in general
correct is an open question; there at least seem to be scenarios in
which it is not, namely ones in which all particles end up moving
around with almost the same velocity and hence only occupy a small
volume of momentum space.} So we have here a system that evolves
from a high to a low entropy state. This problem is usually `solved'
by declaring that things are different in a gravitating system and
that we should, in such cases, regard the spread out state as one of
low entropy and the lump as one of high entropy. Whether or not this
\emph{ad hoc} move is convincing may well be a matter of contention.
But even if it is, it is of no avail to the Boltzmannian. Even if
one redefines entropy such that the lump has high and the spread out
state low entropy, it is still a fact that the phase volume
corresponding to the spread out state is substantially larger than
the one corresponding to the lump, and Boltzmannian explanations of
thermodynamic behaviour typically make essential use of the fact
that the equilibrium macro-region is the largest of all macro
regions.

Hence macro-states need to be defined differently in the context of
interacting systems. Goldstein \& Lebowitz (2004, 60-63) discuss the
problem of defining macro-states for particles interacting with a
two-body potential $\phi(q_{i} - q_{j})$, where $q_{i}$ and $q_{j}$
are the position coordinates of two particles, and they develop a
formalism for calculating the Boltzmann entropy for systems
consisting of a large number of such particles. However, the
formalism yields analytical results only for the special case of a
system of hard balls. Numerical considerations also provide results
for (two-dimensional) particles interacting with a cutoff
Lennard-Jones potential, i.e. a potential that has the Lennard-Jones
form for $|q_{i} - q_{j}| \leq r_{c}$ and is zero for all $|q_{i} -
q_{j}| > r_{c}$, where $r_{c}$ is a cutoff distance (Garrido,
Goldstein \& Lebowitz 2004, 2).

These results are interesting, but they do not yet provide the
sought-after generalisation of the Boltzmann approach to more
realistic systems. Hard ball systems are like ideal gases in that
the interaction of the particles do not contribute to the energy of
the system; the only difference between the two is that hard balls
are extended while the `atoms' of an ideal gas are point particles.
Similarly, the cutoff Lennard-Jones potential also represents only a
small departure from the idea of the ideal gas as the cutoff
distance ensures that no long range interactions contribute to the
energy of the system. However, typical realistic interactions such
as gravity and electrostatic attraction/repulsion are long range
interactions. Hence, it is still an open question whether the
Boltzmann formalism can be extended to systems with realistic
interactions.


\subsection{Reductionism}\label{Reductionism - Boltzmann}

Over the past decades,\index{Reduction!Reductionism} the issue of
reductionism has attracted the attention of many philosophers and a
vast body of literature on the topic has grown; Kim (1998) presents
a brief survey; for a detailed discussion of the different positions
see Hooker (1981) and Batterman (2002, 2003); Dupr\'{e} (1993)
expounds a radically sceptical perspective on reduction. This
enthusiasm did not resonate with those writing on the foundations of
SM and the philosophical debates over the nature (and even
desirability) of reduction had rather little impact on work done on
the foundations of SM (this is true for both the Boltzmannian and
Gibbsian traditions). This is not the place to make up for this lack
of interaction between two communities, but it should be pointed out
that it might be beneficial to both those interested in reduction as
well as those working on the foundations of SM to investigate
whether, and if so how, philosophical accounts of reduction relate
to SM and what consequences certain philosophical perspectives on
reduction would have on how we think about the aims and problems of
SM.

One can only speculate about what the reasons for this mutual
disinterest are. A plausible explanation seems to be that
reductionism has not been perceived as problematic by those working
on SM and hence there did not seem to be a need to turn to the
philosophical literature. A look at how reductionism is dealt with
in the literature on SM confirms this suspicion: by and large there
is agreement that the aim of SM is to derive, fully and rigorously,
the laws of TD from the underlying micro theory. This has a familiar
ring to it for those who know the philosophical debates over
reductionism. In fact, it is precisely what Nagel (1961, Ch. 11)
declared to be the aim of reduction. So one can say that the
Nagelian model of reduction is the (usually unquestioned and
unacknowledged) `background philosophy' of SM. This sets the agenda.
I will first introduce Nagel's account of reduction, discuss some of
its problems, mention a possible ramification, and then examine how
well the achievements of SM square with this conception of
reduction. At the end I will mention some further issues in
connection with reduction.

The core idea of Nagel's theory of reduction is that \index{Nagel,
E.} a theory $T_{1}$ reduces a theory $T_{2}$ (or $T_{2}$ is reduced
to $T_{1}$) only if the laws of $T_{2}$ are derivable from those of
$T_{1}$; $T_{1}$ is then referred to as the `reducing theory' and
$T_{2}$ as the `reduced theory'. In the case of a so-called
homogeneous reduction both theories contain the same descriptive
terms and use them with (at least approximately) the same meaning.
The derivation of Kepler's laws of planetary motion and Galileo's
law of free fall from Newton's mechanics are proposed as paradigm
cases of reductions of this kind. Things get more involved in the
case of so-called `heterogeneous' reductions, when the two theories
do not share the same descriptive vocabulary. The reduction of TD
belongs to this category because both TD and SM contain concepts
that do not form part of the other theory (e.g. temperature is a TD
concept that does not appear in the core of SM, while trajectories
and phase functions are foreign to TD), and others are used with
very different meanings (entropy is defined in totally dissimilar
ways in TD and in SM). In this case so-called `bridge laws' need to
be introduced, which connect the vocabulary of both theories. More
specifically, Nagel requires that for every concept $C$ of $T_{2}$
that does not appear in $T_{1}$ there be a bridge law connecting
$C$\index{Reduction!bridge laws} to concepts of $T_{1}$ (this is the
so-called `requirement of connectability'). The standard example of
a bridge law is the equipartition relation $\langle E \rangle = 3/2
k_{_{B}} T$, connecting temperature $T$ with the mean kinetic energy
$\langle E \rangle$.

Bridge laws carry with them a host of interpretative problems. What
status do they have? Are they linguistic conventions? Or are they
factual statements? If so, of what sort? Are they statements of
constant conjunction (correlation) or do they express nomic
necessities or even identities? And depending on which option one
chooses the question arises of how a bridge law is established. Is
it a factual discovery? By which methods is it established?
Moreover, in what sense has $T_{1}$ reduced $T_{2}$ if the reduction
can only be carried out with the aid of bridge laws which, by
definition, do not belong to $T_{1}$? Much of the philosophical
discussions on Nagelian reduction has centred around these issues.

Another problem is that strict derivability often is too stringent a
requirement because only approximate versions of the $T_{2}$-laws
can be obtained. For instance, it is not possible to derive strict
universal laws from a statistical theory. To make room for a certain
mismatch between the two theories, Schaffner (1976) introduced the
idea that concepts of $T_{2}$ often need to be modified before they
can be reduced to $T_{1}$. More specifically, Schaffner holds that
$T_{1}$ reduces $T_{2}$ only if there is a corrected version
$T_{2}^{*}$ of $T_{2}$ such that $T_{2}^{*}$ is derivable from
$T_{1}$ given that (1) the primitive terms of $T_{2}^{*}$ are
associated via bridge laws with various terms of $T_{1}$, (2)
$T_{2}^{*}$ corrects $T_{2}$ in the sense that $T_{2}^{*}$ makes
more accurate predictions than $T_{2}$  and (3) $T_{2}^{*}$ and
$T_{2}$ are strongly analogous.

With this notion of reduction in place we can now ask whether
Boltzmannian SM reduces TD in this sense. This problem is
usually\index{Reduction!of thermodynamics to statistical mechanics}
narrowed down to the question of whether the Second Law of TD can be
deduced from SM. This is of course an important question, but it is
by no means the only one; I come back to other issues below. From
what has been said so far it is obvious that the Second
Law\index{Second law of thermodynamics} cannot be derived from SM.
The time reversal invariance\index{Time reversal invariance} of the
dynamics and Poincar\'{e} recurrence imply that the Boltzmann
entropy does not increase monotonically at all times. In fact, when
an SM system has reached equilibrium it fluctuates away from
equilibrium every now and then. Hence, a strict Nagelian reduction
of TD to SM is not possible. However, following Schaffner, this is
anyway too much to ask for; what we should look for is a corrected
version TD* of TD, which satisfies the above-mentioned conditions
and which can be reduced to SM. Callender (2001, 542-45) argues that
this is precisely what we should do because trying to derive the
\emph{exact} Second Law would amount to `taking thermodynamics too
seriously'; in fact, what we need to derive from SM is an `analogue'
of the Second Law.\footnote{The same problem crops up when reducing
the notions of equilibrium (Callender 2001, 545-547) and the
distinction between intensive and extensive TD variables (Yi 2003,
1031-2) to SM: a reduction can only take place if we first present a
suitably revised version of TD.} One such analogue is BL, although
there may be other candidates.

The same move helps us to reduce thermodynamic irreversibility.
Callender (1999, 359, 364-67) argues that it is a mistake to try to
deduce strict irreversibility from SM. All we need is an explanation
of how phenomena that are irreversible on an appropriate time scale
emerge from SM, where what is appropriate is dictated by the
conditions of observation. In other words, what we need to recover
from SM is the phenomena supporting TD, not a strict reading of the
TD laws.

Given this, the suggestion is that $S_{_{B}}$ can plausibly be
regarded as the SM counterpart of the entropy of TD*. This is a
plausible suggestion, but it seems that more needs to be said by way
of justification. Associating $S_{_{B}}$ with the entropy of TD*
effectively amounts to introducing a bridge law that defines the TD*
entropy in terms of the logarithm of the phase volume of macro
regions. This brings back all the above questions about the nature
of bridge laws. What justifies the association of TD* entropy with
its SM counterpart? Of what kind is this association? The discussion
of the relation between the two entropies is usually limited to
pointing out that the values of the two coincide in relevant
situations. This certainly is an important point, but it does not
answer the deeper questions about the relationship between the two
concepts.

Although the second law occupies centre stage in TD, it is not the
only law that needs to be reduced; in particular, we need to account
for how the First Law of TD reduces to SM. And in this context a
further problem crops up (Sklar 1999, 194). To explain how systems
of very different kinds can transfer energy to one another, we need
to assume that these systems have temperatures. This, in turn,
implies that temperature can be realised in radically different
ways; in other words, temperature is multiply realisable. How can
that be? How do the various `realisers' of temperature relate to one
another? What exactly makes them realisers of this concept and why
can we give them a uniform treatment in the theory?\footnote{For a
further discussion of temperature see Sklar (1993, 351-354), Uffink
(1996, 383-386) and Yi (2003, 1032-1036).}

Similar problems also appear when we reduce more `local' laws and
properties to SM. For instance, the relation between pressure,
volume and temperature of an ideal gas is given by the equation
$pV=nk_{_{B}}T$, the so called `ideal gas law'. In order to derive
this law we need to make associations, for instance between pressure
and mechanical properties like mass and momentum transfer, that have
the character of bridge laws. How are these justified? Sklar (1993,
349-350) points out how complex even this seemingly straightforward
case is.

And then there are those TD concepts that SM apparently remains
silent about. Most importantly the concept of a quasi-static
transformation (or process), which lies at the heart of TD. The laws
of TD only apply to equilibrium situations and therefore changes in
a system have to be effected in a way that never pushes the system
out of equilibrium , i.e. by so-called quasi-static transformations
(see Uffink (2001) for discussion of this concept). But what does it
mean in SM to perform a quasi-static transformation on a
system?\footnote{Thanks to Wolfgang Pietsch for drawing my attention
to this point.}

Furthermore, one of the alleged payoffs of a successful reduction is
explanation, i.e. the reduction is supposed to explain the reduced
theory. Does SM explain TD and if so in what sense? This question is
clearly stated by Sklar (1993, 148-154; 2000, 740) Callender (1999,
372-373) and Hellman (1999, 210), but it still awaits an in-depth
discussion.


\section{The Gibbs Approach}\label{The Gibbs Approach}

At the beginning of the Gibbs approach stands a radical rupture with\index{Gibbs, J. W.!approach to statistical mechanics}
the Boltzmann programme. The object of study for the Boltzmannians
is an individual system, consisting of a large but finite number of
micro constituents. By contrast, within the Gibbs framework the
object of study is a so-called \emph{ensemble}\index{Ensemble}, an uncountably infinite
collection of independent systems that are all governed by the same
Hamiltonian but distributed over different states. Gibbs introduces
the concept as follows:

\begin{quote}

We may imagine a great number of systems of the same nature, but
differing in the configurations and velocities which they have at a
given instant, and differing not only infinitesimally, but it may be
so as to embrace every conceivable combination of configuration and
velocities. And here we may set the problem, not to follow a
particular system through its succession of configurations, but to
determine how the whole number of systems will be distributed among
the various conceivable configurations and velocities at any
required time, when the distribution has been given for some one
time. (Gibbs 1902, v)

\end{quote}

\noindent Ensembles are fictions, or `mental copies of the one
system under consideration' (Schr\"{o}dinger 1952, 3); they do not
interact with each other, each system has its own dynamics, and they
are not located in space and time.

Ensembles should not be confused with collections of micro-objects
such as the molecules of a gas. The ensemble corresponding to a gas
made up of $n$ molecules, say, consists of an infinite number of
copies of the \emph{entire} gas; the phase space of each system in
the ensemble is the $6n$-dimensional $\gamma$-space of the gas as a
whole.


\subsection{The Gibbs Formalism}\label{Gibbs Formalism}

Consider an ensemble of systems. The instantaneous state of one
system of the ensemble is specified by one point in its
$\gamma$-space, also referred to as the system's
micro-state.\footnote{To be more precise, the system's fine-grained
micro-state. However, within the Gibbs approach coarse-graining
enters the stage only much later (in Subsection \ref{The Approach to
Equilibrium}) and so the difference between coarse-grained and
fine-grained micro-states need not be emphasised at this point.} The
state of the ensemble is therefore specified by an everywhere
positive density function $\rho(q, p, t)$ on the system's
$\gamma$-space.\footnote{That is, $\rho(q, p, t) \geq 0$ for all
$(q, p) \in \Gamma_{\gamma}$ and all instants of time $t$.} The time
evolution of the ensemble is then associated with changes in the
density function in time.

Within the Gibbs formalism $\rho(q, p, t)$ is regarded as a
probability density, reflecting the probability of finding the state
of a system chosen at random from the entire ensemble in region $R
\subseteq \Gamma$\footnote{The $\mu$-space of a system does not play
any r\^{o}le in the Gibbs formalism. For this reason I from now on drop
the subscript `$\gamma$' and only write `$\Gamma$' instead of
`$\Gamma_{\gamma}$' when referring to a system's $\gamma$-space.} at
time $t$:

\begin{equation}\label{Gibbs probability}
    p_{_{t}}(R) \, = \, \int_{R}\rho(q, p, t) d\Gamma
\end{equation}

\noindent For this reason the distribution has to be normalised:

\begin{equation}\label{Normalising rho}
    \int_{\Gamma}\rho (q, p, t) \, d\Gamma = 1.
\end{equation}

Now consider a real valued function $f: \, \Gamma \times t \,
\rightarrow \, R$. The phase average (some times also `ensemble
average') of this function is given by:

\begin{equation}\label{Phase average}
    \bar{f}(t) = \int_{\Gamma} f(q, p, t)
    \rho(q, p, t) \, d\Gamma.
\end{equation}
\noindent Phase averages occupy centre stage in the Gibbs formalism
because it is these that, according to the formalism, we observe in
experiments. More specifically, the Gibbs formalism postulates that
to every experimentally observable quantity $F(t)$ (with the
exception of absolute temperature and entropy) there corresponds a
phase function $f(q, p, t)$ such that Equation (\ref{Phase average})
yields the value that we should expect to find in an experiment:
$F(t) = \bar{f}(t)$.

Using the principles of Hamiltonian mechanics one can then prove
that the total derivative of the density function equals zero,

\begin{equation}
     \frac{d\rho}{dt} = 0,
\end{equation}
\noindent which is commonly referred to as `Liouville's
theorem'\index{Liouville's theorem} in this context. Intuitively,
this theorem says that $\rho$ moves in phase space like an
incompressible fluid. With Equation (\ref{f-dot}) in the Appendix it
follows that the time evolution of $\rho$ is given by Liouville's
equation:

\begin{equation}\label{Liouville's equation}
     \frac{\partial \rho}{\partial t} \, = \, -\{\rho, H
     \},
\end{equation}
\noindent where $\{\, \cdot \, , \, \cdot \, \}$ is the Poisson
bracket and $H$ the Hamiltonian governing the system's dynamics. By
definition, a distribution is stationary iff $\partial \rho /
\partial t  \, = 0$ for all $t$.

Given that observable quantities are associated with phase averages
and that equilibrium is defined in terms of the constancy of the
macroscopic parameters characterising the system, it is natural to
regard the stationarity of the distribution as defining equilibrium
because a stationary distribution yields constant
averages.\footnote{Provided that the observable $f$ itself is not
explicitly time dependent, in which case one would not require
equilibrium expectation values to be constant.} For this reason
Gibbs refers to stationarity as the `condition of statistical
equilibrium'.

Among all stationary distributions\footnote{As Gibbs notes, every
distribution that can be written as a function of the Hamiltonian is
stationary.} those satisfying a further requirement, the Gibbsian
maximum entropy principle, play a special r\^{o}le. The fine-grained
Gibbs entropy (sometimes also `ensemble entropy') is defined as

\begin{equation}\label{Gibbs Entropy}
      S_{_{G}}(\rho)\, := \, - k_{_{B}} \int_{\Gamma} \rho
      \log(\rho)
      d\Gamma.
\end{equation}
\noindent The Gibbsian maximum entropy principle then requires that
$S_{_{G}}(\rho)$ be maximal, given the constraints that are imposed
on the system.

The last clause is essential because different constraints single
out different distributions. A common choice is to keep both the
energy and the particle number in the system fixed: $E$=const and
$n$=const (while also assuming that the spatial extension of the
system is finite). One can prove that under these circumstances
$S_{_{G}}(\rho)$ is maximal for what is called the `microcanonical
distribution' (or `microcanonical ensemble'), the distribution which
is uniform on the energy hypersurface $H(q,p)=E$ and zero elsewhere:

\begin{equation}\label{microcanonical distribution}
    \rho (q,p)=C \, \delta[E-H(q,p)],
\end{equation}

\noindent where $C$ is some suitable normalisation constant and
$\delta$ is Dirac's delta function.\footnote{This distribution is
sometimes referred to as the `super microcanonical distribution'
while the term `microcanonical distribution' is used to refer to a
slightly different distribution, namely one that is constant on a
thin but finite `sheet' around the accessible parts of the energy
hypersurfce and zero elsewhere. It turns out that the latter
distribution is mathematically more manageable.}

If we choose to hold the number of particles constant while allowing
for energy fluctuations around a given mean value we obtain the
so-called canonical distribution; if we also allow the particle
number to fluctuate around a given mean value we find the so-called
grand-canonical distribution (for details see, for instance, Tolman
1938, Chs. 3 and 4).

Consider now a system out of equilibrium, characterised by a density
$\rho(t)$. This density is not stationary and its entropy is not
maximal. Given the laws of TD we would expect this density to
approach the equilibrium density as time evolves (e.g. in the case
of a system with constant energy and constant particle number we
would expect $\rho(t)$ to approach the microcanonical distribution),
which would also be reflected in an increase in entropy.

This expectation is frustrated. Using Liouville's equation one can
prove that $\rho(t)$ does \emph{not} approach the microcanonical
distribution (I will come back to this point below); and what seems
worse, the entropy does not increase at all. In fact, it is
straightforward to see that $S_{G}$ is a constant of the motion (Zeh
2001, 48-9); that is, $dS_{G}(\rho(t)) / dt = 0$, and hence
$S_{G}(\rho(t)) = S_{G}(\rho(0))$ for all times $t$.

This precludes an explanation of the approach to equilibrium along
the lines explored so far. Hence the main challenge for Gibbsian
non-equilibrium theory is either to get $S_{G}$ moving somehow, or
to explain why this is not necessary after all.


\subsection{Problems and Tasks}

In this subsection I list the issues that need to be addressed in
the Gibbs programme and make some remarks about how they differ from
the problems that arise in the Boltzmann framework. Again, these
issues are not independent of each other and the response to one
bears on the responses to the others.


\subsubsection{Issue 1: Ensembles and Systems}

The most obvious problem concerns the use of ensembles. The
probability distribution in the Gibbs approach is defined over an
ensemble, the formalism provides ensemble averages, and equilibrium
is regarded as a property of an ensemble. But what we are really
interested in is the behaviour of a single system. What can the
properties of an ensemble, a fictional entity consisting of
infinitely many copies of a system, tell us about the one real
system that we investigate? And how are we to reconcile the fact
that the Gibbs formalism treats equilibrium as a property of an
ensemble with physical common sense and thermodynamics, both of
which regard an individual system as the bearer of this property?

These difficulties raise the question of whether the commitment to
ensembles could be renounced. Are ensembles really an irreducible
part of the Gibbsian scheme or are they just an expedient, or even a
pedagogical ploy, of no fundamental significance? If so, how can the
theory be reformulated without appeal to ensembles?

These questions are of fundamental significance, not least because
it is the use of ensembles that frees the Gibbs approach from some
of the most pressing problems of the Boltzmann approach, namely the
reversal and the recurrence objections. These arise exactly because
we are focussing on what happens in an individual system; in an
ensemble recurrence and reverse behaviour are no problem because it
can be accepted that some systems in the ensemble will behave
non-thermodynamically, provided that their contribution to the
properties of the ensemble as a whole is taken into account when
calculating ensemble averages. So some systems behaving strangely is
no objection as this does not imply that the ensemble as a whole
behaves in a strange way too.


\subsubsection{Issue 2: The Connection with Dynamics and the
Interpretation of Probability}

The microcanonical distribution has been derived from the Gibbsian
maximum entropy principle and the requirement that the equilibrium
distribution be stationary. Neither of these requirements make
reference to the dynamics of the system. However, as in the case of
the combinatorial argument, it seems odd that equilibrium conditions
can be specified without any appeal to the dynamics of the systems
involved. That equilibrium can be characterised by a microcanonical
distribution must, or so it seems, have something to do with facts
about the system in question. Understanding the connection between
the properties of a system and the Gibbsian probability distribution
is complicated by the fact that the distribution is one pertaining
to an ensemble rather than an individual system. What, if anything,
in the dynamics gives rise to, or justifies, the use of the
microcanonical distribution? And if there is no such justification,
what is the reason for this?

Closely related to the question of how the probability distribution
relates to the system's dynamics is the problem of interpreting
these probabilities. The options are the same as in Unit
\ref{Probabilities} and need not be repeated here. What is worth
emphasising is that, as we shall see, different interpretations of
probability lead to very different justifications of the maximum
entropy requirement and its connection to the dynamics of the
system; in fact, in non-equilibrium theory they lead to very
different formalisms. Thus, this is a case where philosophical
commitments shape scientific research programmes.


\subsubsection{Issue 3: Why does Gibbs phase averaging work?}

The Gibbs formalism posits that what we observe in actual
experiments are phase averages. Practically speaking this method
works just fine. But why does it work? Why do averages over an
ensemble coincide with the values found in measurements performed on
an actual physical system in equilibrium? There is no obvious
connection between the two and if Gibbsian phase averaging is to be
more than a black-box technique then we have to explain what the
connection between phase averages and measurement values is.


\subsubsection{Issue 4: The Approach to Equilibrium}\label{Gibbs
Issue 4}

Phase averaging only applies to equilibrium systems and even if we
have a satisfactory explanation of why this procedure works, we are
still left with the question of why and how the system reaches
equilibrium at all if, as it often happens, it starts off far from
equilibrium.

Gibbsian non-equilibrium theory faces two serious problems. The
first is that, as we have seen above, the Gibbs entropy is constant.
This precludes a characterisation of the approach to equilibrium in
terms of increasing Gibbs entropy. Hence, either such a
characterisation has to be given up (at the cost of being
fundamentally at odds with thermodynamics), or the formalism has to
be modified in a way that makes room for entropy increase.

The second problem is the characterisation of equilibrium in terms
of a stationary distribution. The Hamiltonian equations of motion,
which govern the system, preclude an evolution from a non-stationary
to a stationary distribution: if, at some point in time, the
distribution is non-stationary, then it will remain non-stationary
for all times  and, conversely, if it is stationary at some time,
then it must have been stationary all along (van Lith 2001a, 591-2).
Hence, if a system is governed by Hamilton's equation, then a
characterisation of equilibrium in terms of stationary distributions
contradicts the fact that an approach to equilibrium takes place in
systems that are not initially in equilibrium.

Clearly, this is a \emph{reductio} of a characterisation of
equilibrium in terms of stationary distributions. The reasoning that
led to this characterisation was that an equilibrium state is one
that remains unchanged through time, which, at the mechanical level,
amounts to postulating an unchanging, i.e. stationary, distribution.
This was too quick. Thermodynamic equilibrium is defined as a state
in which all macro-parameters describing the system are constant. So
all that is needed for equilibrium is that the distribution be such
that mean values of the functions associated with thermodynamic
quantities are constant in time (Sklar 1978, 191). This is a much
weaker requirement because it can be met by distributions that are
not stationary. Hence we have to come to a more `liberal'
characterisation of equilibrium; the question is what this
characterisation is.\footnote{Leeds (1989, 328-30) also challenges
as too strong the assumption that a physical system in an
equilibrium state has a precise probability distribution associated
with it. Although this may well be true, this seems to be just
another instance of the time-honoured problem of how a precise
mathematical description is matched up with a piece of physical
reality that is not intrinsically mathematical. This issue is beyond
the scope of this review.}


\subsubsection{Issue 5: Reductionism}

Both the Boltzmannian and the Gibbsian approach to SM eventually aim
to account for the TD behaviour of the systems under investigation.
Hence the questions for the Gibbs approach are exactly the same as
the ones mentioned in Unit \ref{Issue 7}, and the starting point
will also be Nagel's model of reduction (introduced in Subsection
\ref{Reductionism - Boltzmann}).


\subsubsection{Plan}

As mentioned above, the methods devised to justify the use of the
microcanonical distribution and the legitimacy of phase averaging,
as well as attempts to formulate a coherent non-equilibrium theory
are radically different depending on whether probabilities are
understood ontically or epistemically. For this reason it is best to
discuss these two families of approaches separately. Subsection
\ref{Why Does Gibbs Phase Averaging Work?} presents arguments
justifying Gibbs phase averaging on the basis of an ontic
understanding of the probabilities involved. What this understanding
might be is discussed in Subsection \ref{Ontic Probabilities}. I
turn to this question only after a discussion of different
justifications of phase averaging because although an ontic
understanding of probabilities is clearly assumed, most writers in
this tradition do not discuss this assumption explicitly and one can
only speculate about what interpretation of probability they might
endorse. Subsection \ref{The Approach to Equilibrium} is concerned
with different approaches to non-equilibrium that are based on this
interpretation of probabilities. In Subsection \ref{The Epistemic
Approach} I discuss the epistemic approach to the Gibbs formalism. I
close this Section with a discussion of reductionism in the Gibbs
approach (Subsection \ref{Reductionism - Gibbs}).


\subsection{Why Does Gibbs Phase Averaging Work?}\label{Why Does Gibbs Phase Averaging Work?}
\index{Gibbs phase averaging} Why do phase averages coincide with
values measured in actual physical systems? There are two families
of answers to this question, one based on ergodic theory (using
ideas we have seen in the Boltzmann Section), the other building on
the notion of a thermodynamic limit. For reasons of space we will
treat this second approach much more briefly.


\subsubsection{Time Averages and Ergodicity}

Common wisdom justifies the use of phase averages as
follows.\footnote{This view is discussed but not endorsed, for
instance, in Malament \& Zabell (1980, 342), Bricmont (1996, 145-6),
Earman \& Redei (1996, 67-69), and van Lith (2001a, 581-583).} The
Gibbs formalism associates physical quantities with functions on the
system's phase space. Making an experiment to measure one of these
quantities takes some time. So what measurement devices register is
not the instantaneous value of the function in question, but rather
its time average over the duration of the measurement; hence it is
time averages that are empirically accessible. Then, so the argument
continues, although measurements take an amount of time that is
short by human standards, it is long compared to microscopic time
scales on which typical molecular processes take place (sometimes
also referred to as `microscopic relaxation time'). For this reason
the actually measured value is approximately equal to the
\emph{in}finite time average of the measured function. This by
itself is not yet a solution to the initial problem because the
Gibbs formalism does not provide us with time averages and
calculating these would require an integration of the equations of
motion, which is unfeasible. This difficulty can be circumvented by
assuming that the system is ergodic. In this case time averages
equal phase averages, and the latter can easily be obtained from the
formalism. Hence we have found the sought-after connection: the
Gibbs formalism provides phase averages which, by ergodicity, are
equal to infinite time averages, and these are, to a good
approximation, equal to the finite time averages obtained from
measurements.

This argument is problematic for at least two reasons (Malament \&
Zabell 1980, 342-343; Sklar 1973, 211). First, from the fact that
measurements take some time it does not follow that what is actually
measured are time averages. Why do measurements produce time
averages and in what way does this depend on how much time
measurements take?

Second, even if we take it for granted that measurements do produce
finite time averages, then equating these with infinite time
averages is problematic. Even if the duration of the measurement is
very long (which is often not the case as actual measurement may not
that take much time), finite and infinite averages may assume very
different values. And the infinity is crucial: if we replace
infinite time averages by finite ones (no matter how long the
relevant period is taken to be), then the ergodic theorem does not
hold any more and the explanation is false.

Besides, there is another problem once we try to apply the Gibbs
formalism to non-equilibrium situations. It is a simple fact that we
do observe how systems approach equilibrium, i.e. how macroscopic
parameter values change, and this would be impossible if the values
we observed were infinite time averages.

These criticisms seem decisive and call for a different strategy in
addressing Issue 3. Malament \& Zabell (1980) respond to this
challenge by suggesting a new way of explaining the success of
equilibrium theory, at least for the microcanonical ensemble. Their
method still invokes ergodicity but avoids altogether appeal to time
averages and only invokes the uniqueness of the measure (see
Subection \ref{The Ergodicity Programme}). Their explanation is
based on two Assumptions (\emph{ibid}., 343).

\emph{Assumption 1}. The phase function $f$ associated with a
macroscopic parameter of the system exhibits small dispersion with
respect to the microcanonical measure; that is, the set of points on
the energy hypersurface $\Gamma_{E}$ at which $f$ assumes values
that differ significantly from its phase average has vanishingly
small microcanonical measure. Formally, for any `reasonably small'
$\varepsilon \, > 0$ we have

\begin{equation}\label{small dispersion}
    \lambda \Big{(}\, \Big{\{}\, x \in \, \Gamma_{E} \, : \,
    \mid f(x) - \int_{\Gamma_{E}} f\, d\lambda \mid \, \geq \varepsilon \,
    \Big{\}}\, \Big{)} \approx 0,
\end{equation}
\noindent where $\lambda$ is the microcanonical measure (i.e. the
measure that is constant on the accessible part of the energy
hypersurface and normalised).

\emph{Assumption 2}. At any given time, the microcanonical measure\index{Microcanonical measure}
represents the probability of finding the system in a particular
subset of the phase space: $p(A) = \lambda(A)$, where $A$ is a
measurable but otherwise arbitrary subset of $\Gamma_{E}$. These two
assumptions jointly imply that, at any given time, it is
overwhelmingly likely that the system's micro-state is one for which
the value of $f$ coincides with, or is very close to, the phase
average.

The question is how these assumptions can be justified. In the case
of Assumption 1 Malament \&  Zabell refer to a research programme
that originated with the work of Khinchin. The central insight of
this programme is that phase functions which are associated with
macroscopic parameters satisfy strong symmetry requirements and as a
consequence turn out to have small dispersion on the energy surface
for systems with a large number of constituents. This is just what
is needed to justify Assumption 1. This programme will be discussed
in the next subsection; let us assume for now that it provides a
satisfactory justification of Assumption 1.

To justify Assumption 2 Malament \& Zabell introduce a new
postulate: the equilibrium probability measure $p(\, \cdot \,)$ of
finding a system's state in a particular subset of $\Gamma_{E}$ must
be absolutely continuous with respect to the microcanonical measure
$\lambda$ (see Subsection \ref{The Ergodicity Programme}). Let us
refer to this as the `Absolute Continuity Postulate'
(ACP).\footnote{I formulate ACP in terms of $\lambda$ because this
simplifies the argument to follow. Malament \& Zabell require that
$p(\, \cdot \,)$ be absolutely continuous with $\mu_{E}$, the
Lebesgue measure $\mu$ on $\Gamma$ restricted to $\Gamma_{E}$.
However, on $\Gamma_{E}$ the restricted Lebesgue measure and the
microcanonical measure only differ by a constant: $\lambda = c \,
\mu_{E}$, where $c:=1/\mu_{E}(\Gamma_{E})$ and hence whenever a
measure is absolutely continuous with $\mu_{E}$ it is also with
$\lambda$ and vice versa.}

Now consider the dynamical system $(X, \phi, \lambda)$, where $X$ is
$\Gamma_{E}$, $\phi$ is the flow on $\Gamma_{E}$ induced by the
equations of motion governing the system, and $\lambda$ is the
microcanonical measure on $\Gamma_{E}$. Given this, one can present
the following argument in support of Assumption 2 (\emph{ibid}.
345):

\begin{list}{}{%
    \setlength{\labelwidth}{0pt}
    \setlength{\labelsep}{0pt}
    \setlength{\leftmargin}{34pt}
    \setlength{\itemindent}{0pt}
  }

\item (P1) $(X, \phi, \lambda)$ is ergodic.
\item (P2) $p(\, \cdot \,)$ is invariant in time because this is the defining
        feature of equilibrium probabilities.
\item (P3) by ACP, $p(\, \cdot \,)$ is absolutely continuous with $\lambda$.
\item (P4) According to the uniqueness theorem (see Subsection \ref{Ergodic Theory}),
           $\lambda$ is the only measure invariant in time.
\item Conclusion: $p(\, \cdot \,) = \lambda$.
\end{list}

\noindent Hence the microcanonical measure is singled out as the one
and only correct measure for the probability of finding a system's
micro-state in a certain part of phase space.

The new and crucial assumption is ACP and the question is how this
principle can be justified. What reason is there to restrict the
class of measures that we take into consideration as acceptable
equilibrium measures to those that are absolutely continuous with
respect to the microcanonical measure? Malament \& Zabell respond to
this problem by introducing yet another principle, the `displacement
principle' (DP). This principle posits that if of two measurable
sets in $\Gamma_{E}$ one is but a small displacement of the other,
then it is plausible to believe that the probability of finding the
system's micro-state in one set should be close to that of finding
it in the other (\emph{ibid}., 346). This principle is interesting
because one can show that it is equivalent to the claim that
probability distributions are absolutely continuous with respect to
the Lebesgue measure, and hence the microcanonical measure, on
$\Gamma_{E}$ (\emph{ibid}., 348-9).\footnote{As Leeds (1989, 327)
points out, Malament \& Zabell's proof is for $R^{n}$ and they do
not indicate how the proof could be modified to apply to the energy
hypersurface, where translations can take one off the surface.}

To sum up, the advantages of this method over the `standard account'
are that it does not appeal to measurement, that it takes into
account that SM systems are `large' (via Assumption 1), and that it
does not make reference to time averages at all. In fact, ergodicity
is used only to justify the uniqueness of the microcanonical
measure.

The remaining question is what reasons there are to believe in DP.
Malament \& Zabell offer little by way of justification: they just
make some elusive appeal to the `method by which the system is
prepared or brought to equilibrium' (\emph{ibid}., 347.) So it is
not clear how one gets from some notion of state preparation to DP.
But even if it was clear, why should the success of equilibrium SM
depend on the system being \emph{prepared} in a particular way? This
seems to add an anthropocentric element to SM which, at least if one
is not a proponent of the ontic approach (referred to in Subsection
\ref{Probabilities}), seems to be foreign to it.

The argument in support of Assumption 2 makes two further
problematic assumptions. First, it assumes equilibrium to be defined
in terms of a stationary distribution, which, as we have seen above,
is problematic because it undercuts a dynamical explanation of the
approach to equilibrium (variants of this criticism can be found in
Sklar (1978) and Leeds (1989)).

Second, it is based on the premise that the system in question is
ergodic. As we have seen above, many systems that are successfully
dealt with by the formalism of SM are not ergodic and hence the
uniqueness theorem, on which the argument in support of Assumption 2
is based, does not apply.

To circumvent this difficulty Vranas (1998) has suggested replacing
ergodicity with what he calls $\varepsilon$-ergodicity. The leading
idea behind this move is to challenge the commonly held belief that
even if a system is just a `little bit' non-ergodic, then the
uniqueness theorem fails completely (Earman \& Redei 1996, 71).
Vranas points out that there is a middle ground between holding and
failing completely and then argues that this middle ground actually
provides us with everything we need.

Two measures $\lambda_{1}$ and $\lambda_{2}$ are $\varepsilon$-close
iff for every measurable set $A$: $| \lambda_{1}(A) -
\lambda_{2}(A)| \leq \varepsilon$, where $\varepsilon$ is a small
but finite number. The starting point of Vranas' argument is the
observation that we do not need $p(\, \cdot \,) = \lambda$: to
justify Gibbsian phase averaging along the lines suggested by
Malament \& Zabell all we need is that $p(\, \cdot \,)$ and
$\lambda$ are $\varepsilon$-close, as we cannot tell the difference
between a probability measure that is exactly equal to $\lambda$ and
one that is just a little bit different from it. Hence we should
replace Assumption 2 by Assumption 2', that statement that $p(\,
\cdot \,)$ and $\lambda$ are $\varepsilon$-close. The question now
is: how do we justify Assumption 2'?

If a system is non-ergodic then its phase space $X$ is decomposable;
that is, there exist two sets, $A \subseteq X$ and $B := X \setminus
A$, with measure greater than zero which are invariant under the
flow. Intuitively, if the system is `just a little bit non-ergodic',
then the system is ergodic on $B$ and $\lambda(A)\ll \lambda(B)$
(where, again, $\lambda$ is the microcanonical measure). This
motivates the following definition: A dynamical system $(X, \lambda,
\phi)$ is $\varepsilon$-ergodic iff the system's dynamics is ergodic
on a subset $Y$ of $X$ with $\lambda(Y) =
1-\varepsilon$.\footnote{Vranas (1998, 695) distinguishes between
`$\varepsilon$-ergodic' and `epsilon-ergodic', where a system is
epsilon-ergodic if it is $\varepsilon$-ergodic with $\varepsilon$
tiny or zero. In what follows I always assume $\varepsilon$ to be
tiny and hence do not distinguish between the two.} Strict
ergodicity then is the limiting case of $\varepsilon=0$.
Furthermore, given two small but finite numbers $\varepsilon_{1}$
and $\varepsilon_{2}$, Vranas defines $\lambda_{2}$ to be
`$\varepsilon_{1}/\varepsilon_{2}$-continuous' with $\lambda_{1}$
iff for every measurable set $A$: $\lambda_{2}(A) \leq
\varepsilon_{2}$ if $\lambda_{1}(A) \leq \varepsilon_{1}$ (Vranas
1998, 695).

Vranas then proves an `$\varepsilon$-version' of the uniqueness
theorem, the $\varepsilon$-equivalence theorem (\emph{ibid.},
703-5): if $\lambda_{1}$ is $\varepsilon_{1}$-ergodic and
$\lambda_{2}$ is $\varepsilon_{1} / \varepsilon_{2}$-continuous with
respect to $\lambda_{1}$ and invariant, then $\lambda_{1}$ and
$\lambda_{2}$ are $\varepsilon_{3}$-close with $\varepsilon_{3} =
2\varepsilon_{2} + \varepsilon_{1}(1-\varepsilon_{1})^{-1}$.

Given this, the Malament \& Zabell argument can be rephrased as
follows:

\begin{list}{}{%
    \setlength{\labelwidth}{0pt}
    \setlength{\labelsep}{0pt}
    \setlength{\leftmargin}{34pt}
    \setlength{\itemindent}{0pt}
  }

\item (P1') $(X, \phi, \lambda)$ is $\varepsilon$-ergodic.
\item (P2) $p(\, \cdot \,)$ is invariant in time because this is the defining
        feature of equilibrium probabilities.
\item (P3') $\varepsilon$-ACP: $p(\, \cdot \,)$ is $\varepsilon/\varepsilon_{2}$-continuous
        with respect to $\lambda$.
\item (P4') The $\varepsilon$-equivalence theorem.

\item Conclusion': $p(\, \cdot \,)$ and $\lambda$ are $\varepsilon_{3}$-close
with $\varepsilon_{3} = 2\varepsilon_{2} +
\varepsilon(1-\varepsilon)^{-1}$.

\end{list}

The assessment of this argument depends on what can be said in
favour of (P1') and (P3'), since (P4') is a mathematical theorem and
(P2) has not been altered. In support of (P1') Vranas (\emph{ibid}.,
695-98) reviews computational evidence showing that systems of
interest are indeed $\varepsilon$-ergodic. In particluar, he
mentions the following cases. A one-dimensional system of $n$
self-gravitating plane parallel sheets of uniform density was found
to be strictly ergodic as $n$ increases (it reaches strict
ergodicity for $n=11$). The Fermi-Pasta-Ulam system (a one
dimensional chain of $n$ particles with weakly nonlinear
nearest-neighbor interaction) is $\varepsilon$-ergodic for large
$n$. There is good evidence that a Lennard-Jones gas is
$\varepsilon$-ergodic for large $n$ and in the relevant energy
range, i.e. for energies large enough so that quantum effects do not
matter. From these Vranas draws the tentative conclusion that the
dynamical systems of interest in SM are indeed
$\varepsilon$-ergodic. But he is clear about the fact that this is
only a tentative conclusion and that it would be desirable to have
theoretical results.

The justification of (P3') is more difficult. This does not come as
a surprise because Malament \& Zabell did not present a
justification for ACP either. Vranas (\emph{ibid}., 700-2) presents
some arguments based on the limited precision of measurement but
admits that this argument invokes premises that he cannot justify.

To sum up, this argument enjoys the advantage over previous
arguments that it does not have to invoke strict ergodicity.
However, it is still based on the assumption that equilibrium is
characterised by a stationary distribution, which, as we have seen,
is an obstacle when it comes to formulating a workable Gibbsian
non-equilibrium theory. In sum, it is still an open question whether
the ergodic programme can eventally explain in a satisfactory way
why Gibbsian SM works.


\subsubsection{Khinchin's Programme and the Thermodynamic
Limit}\label{Khinchin's Programme and the Thermodynamic Limit}

Ergodic theory works at a general level in that it makes no
assumptions about the number of  degrees of freedom of the system
under study and does not restrict the allowable phase functions
beyond the requirement that they be integrable. Khinchin
(1949)\index{Khinchin, A. Y.} points out that this generality is not
only unnecessary; actually it is the source of the problems that
this programme encounters. Rather than studying dynamical systems at
a general level, we should focus on those cases that are relevant in
statistical mechanics. This involves two restrictions. First, we
only have to consider systems with a large number of degrees of
freedom; second, we only need to take into account a special class
of phase function, so-called sum functions. A function is a sum
function if it can be written as a sum over one-particle functions:

\begin{equation}\label{sum function}
    f(x) = \sum_{i=1}^{n} f_{i}(x_{i}),
\end{equation}

\noindent where $x_{i}$ is the vector containing the position and
momentum coordinates of particle $i$ (that is, $x_{i} \in
{\mathbf{R}}^{6}$ while $x \in {\mathbf{R}}^{6n}$) Under the
assumption that the Hamiltonian of the system is a sum function as
well, Khinchin can prove the following theorem:

\begin{quote}
    \emph{Khinchin's Theorem}. For all sum functions $f$ there are
    positive constants $k_{1}$ and $k_{2}$ such that
\begin{equation}\label{Khinchin's Theorem}
    \lambda \Big{(}\, \Big{\{}\, x \in \, \Gamma_{E} \, : \, \Big{|}
    \frac{f^{*}(x) - \bar{f}}{\bar{f}}\Big{|} \, \geq k_{1} \, n^{-1/4}
    \Big{\}}\, \Big{)} \, \, \leq \, \, k_{2} \, n^{-1/4},
\end{equation}

\noindent where $\lambda$ is the microcanonical measure.

\end{quote}

\noindent This theorem is sometimes also referred to as `Khinchin's
ergodic theorem'; let us say that a system satisfying the condition
specified in Khinchin's theorem is
`K-ergodic'.\footnote{K-ergodicity should not be conflated with the
property of being a K-system; i.e. being a system having the
Kolmogorov property.} For a summary and a discussion of the proof
see Batterman (1998, 190-198), van Lith (2001b, 83-90) and Badino
(2006). Basically the theorem  says that as $n$ becomes larger, the
measure of those regions on the energy hypersurface where the time
and the space means differ by more than a small amount tends towards
zero. For any finite $n$, K-ergodicity is weaker than ergodicity in
the sense that the region where time and phase average do not
coincide can have a finite measure, while it is of measure zero if
the system is ergodic; this discrepancy vanishes for $n \rightarrow
\infty$. However, even in the limit for $n \rightarrow \infty$ there
is an important difference between ergodicity and K-ergodicity: if
K-ergodicity holds, it only holds for a very special class of phase
functions, namely sum-functions; ergodicity, by contrast, holds for
any $\lambda$-integrable function.

A number of problems facing an explanation of equilibrium SM based
on K-ergodicity need to be mentioned. First, like the
afore-mentioned approaches based on ergodicity, Khinchin's programme
associates the outcomes of measurements with infinite time averages
and is therefore vulnerable to the same objections. Second,
ergodicity's measure zero problem turns into a `measure $k_{2} \,
n^{-1/4}$ problem', which is worse because now we have to justify
that a part of the energy hypersurface of \emph{finite} measure
(rather than measure zero) can be disregarded. Third, the main
motivation for focussing attention on sum-functions is the claim
that all relevant functions, i.e. the ones that correspond to
thermodynamic quantities, are of that kind. Batterman (1998, 191)
points out that this is too narrow as there are functions of
interest that do not have this form.

A further serious difficulty is what Khinchin himself called the
`methodological paradox' (Khinchin 1949, 41-43). The proof of the
above theorem assumes the Hamiltonian to be a sum function (and this
assumption plays a crucial r\^{o}le in the derivation of the
theorem). However, for an equilibrium state to arise to begin with,
the particles have to interact (collide), which cannot happen if the
Hamiltonian is a sum function. Khinchin's response is to assume that
there are only short range interactions between the molecules (which
is the case, for instance, in a hard ball gas). If this is the case,
Khinchin argues, the interactions are effective only on a tiny part
of the phase space and hence have no significant effect on averages.

This response has struck many as unsatisfactory and ad hoc, and so
the methodological paradox became the starting point for a research
programme now known as the `thermodynamic limit', investigating the
question of whether one can still prove `Khinchin-like' results in
the case of Hamiltonians \emph{with} interaction terms. Results of
this kind can be proven in the limit for $n \rightarrow \infty$, if
also the volume $V$ of the system tends towards infinity in such a
way that the numberdensity $n/V$ remains constant. This programme,
championed among others by Lanford, Mazur, Ruelle, and van der
Linden, has reached a tremendous degree of mathematical
sophistication and defies summary in simple terms. Classic
statements are Ruelle (1969, 2004); surveys and further references
can be found in Compagner (1989), van Lith (2001b, 93-101) and
Uffink (2007, 1020-28).

A further problem is that for finite $n$ K-ergodic systems need not
be metrically transitive. This calls into question the ability of an
approach based on K-ergodicity to provide an answer to the question
of why measured values coincide with microcanonical averages.
Suppose there is some global constant of motion other than $H$, and
as a result the motion of the system remains confined to some part
of the energy hypersurface. In this case there is no reason to
assume that microcanonical averages with respect to the
\emph{entire} energy hypersurface coincide with measured values.
Faced with this problem one could argue that each system of that
kind has a decomposition of its energy hypersurface into different
regions of non-zero measure, some ergodic and others not, and that,
as $n$ and $V$ get large, the average values of relevant phase
functions get insensitive towards the non-ergodic parts.

Earman \& R\'{e}dei (1996, 72) argue against this strategy on the
grounds that it is straightforward to construct an infinity of
normalised invariant measures that assign different weights to these
regions than does the microcanonical measure. However, phase
averages with respect to these other measures can deviate
substantially form microcanonical averages, and it is to be expected
that these predictions turn out wrong. But why? In a non-ergodic
system there is no reason to grant the microcanonical measure a
special status and Khinchin's approach does not provide a reason to
expect microcanonical averages rather than any other average value
to correspond to measurable quantities.

Batterman (1998) grants this point but argues that there is another
reason to expect correspondence with observed values; but this
reason comes from a careful analysis of renormalisation group
techniques and their application to the case at hand, rather than
any feature of either Khinchin's approach or the thermodynamic
limit. A discussion of these techniques is beyond the scope of this
review; the details of the case at hand are considered in Batterman
(1998), and a general discussion of renormalisation and its relation
to issues in connection with reductionism and explanation can be
found in Batterman (2002).


\subsection{Ontic Probabilities in Gibbs' Theory}\label{Ontic
Probabilities}

Two ontic interpretations of Gibbsian probabilities have been
suggested in the literature: frequentism and time averages. Let us
discuss them in turn. \index{Probability!interpretations!ontic}


\subsubsection{Frequentism}

A common way of looking at ensembles is to think about them in
analogy with urns, but rather than containing balls of different
colours they contain systems in different micro-states. This way of
thinking about $\rho$ was first suggested in a notorious remark by
Gibbs (1902, 163), in which he observes that `[w]hat we know about a
body can generally be described most accurately and most simply by
saying that it is one taken at random from a great number (ensemble)
of bodies which are completely described'. Although Gibbs himself
remained non-committal as regards an interpretation of probability,
this point of view naturally lends itself to a frequentist analysis
of probabilities. In this vein Malament \& Zabell (1980, 345)
observe that one can regard Gibbsian probabilities as representing
limiting relative frequencies within an infinite ensemble of
identical systems.

First appearances notwithstanding, this is problematic. The strength
of frequentism is that it grounds probabilities in facts about the
world. There are some legitimate questions for those who associate
probabilities with infinite limiting frequencies, as these are not
experimental facts. However, frequentists of any stripe agree that
one single outcome is not enough to ground a probability claim. But
this is the best we can ever get in Gibbsian SM. The ensemble is a
fictitious entity; what is real is only the one system in the
laboratory and so we can make at most one draw from this ensemble.
All the other draws would be hypothetical. But on what grounds do we
decide what the result of these draws would be? It is obvious that
these hypothetical draws do not provide a basis for a frequentist
interpretation of probabilities.

Another way of trying to ground a frequency interpretation is to
understand frequencies as given by consecutive measurements made on
the actual system. This move successfully avoids the appeal to
hypothetical draws. Unfortunately this comes at the price of another
serious problem. Von Mises' theory requires that successive trials
whose outcomes make up the sequence on which the relative
frequencies are defined (the collective) be independent. This, as
von Mises himself pointed out, is generally not the case if the
sequence is generated by one and the same system.\footnote{Von Mises
discussed this problem in connection with diffusion processes and
suggested getting around this difficulty by reconstructing the
sequence in question, which is not a collective, as a combination of
two sequences that are collectives (von Mises 1939, Ch. 6). Whether
this is a viable solution in the context at hand is an open
question.} So making successive measurements on the same system does
not give us the kind of sequences needed to define frequentist
probabilities.


\subsubsection{Time Averages}

Another interpretation regards Gibbsian probabilities as time
averages of the same kind as the ones we discussed in Subsection
\ref{The Ergodicity Programme}. On this view, $p_{_{t}}(R)$ in
Equation (\ref{Gibbs probability}) is the average time that the
system spends in region $R$. As in the case of Boltzmannian
probabilities, this is in need of qualification as a relevant
interval over which the time average is taken has to be specified
and the dependence on initial conditions has to vanish. If, again,
we assume that the system is ergodic on the energy hypersurface we
obtain neat answers to these questions (just as in the Boltzmann
case).

Assuming the system to be ergodic solves two problems at once. For
one, it puts the time average interpretation on solid grounds (for
the reasons discussed in Subsection \ref{Promises} in the context of
the Boltzmannian approach). For another, it offers an explanation of
why the microcanonical distribution is indeed the right
distribution; i.e. it solves the uniqueness problem. This is
important because even if all interpretative issues were settled, we
would still be left with the question of which among the infinitely
many possible distribution would be the correct one to work with.
The uniqueness theorem of ergodic theory answers this question in an
elegant way by stating that the microcanonical distribution is the
only distribution absolutely continuous with respect to the Lebesgue
measure (although some argument still would have to be provided to
establish that every acceptable distribution has to be absolutely
continuous with respect to the Lebesgue measure).

However, this proposal suffers from all the difficulties mentioned
in Subsection \ref{Problems with Ergodicity}, which, as we saw, are
not easily overcome. A further problem is that it undercuts an
extension of the approach to non-equilibrium situations.
Interpreting probabilities as infinite time averages yields
stationary probabilities. As a result, phase averages are constant.
This is what we expect in equilibrium, but it is at odds with the
fact that we witness change and observe systems approaching
equilibrium departing from a non-equilibrium state. This evolution
has to be reflected in a change of the probability distribution,
which is impossible if it is stationary by definition. Hence the
time average interpretation of probability together with the
assumption that the system is ergodic make it impossible to account
for non-equilibrium behaviour (Sklar 1973, 211; Jaynes 1983, 106;
Dougherty 1993, 846; van Lith 2001a, 586).

One could try to circumvent this problem by giving up the assumption
that the system is ergodic and define $p_{_{t}}(R)$ as a finite time
average. However, the problem with this suggestion is that it is not
clear what the relevant time interval should be, and the dependence
of the time average on the initial condition would persist. These
problems make this suggestion rather unattractive. Another
suggestion is to be a pluralist about the interpretation of
probability and hold that probabilities in equilibrium have to be
interpreted differently than probabilities in non-equilibrium.
Whatever support one might muster for pluralism about the
interpretation of probability in other contexts, it seems out of
place when the equilibrium versus non-equilibrium distinction is at
stake. At least in this case one needs an interpretation that
applies to both cases alike (van Lith 2001a, 588).


\subsection{The Approach to Equilibrium}\label{The Approach to
Equilibrium}

The main challenge for Gibbsian non-equilibrium theory is to find a
way to get the Gibbs entropy moving. Before discussing different
solutions to this problem, let me again illustrate what the problem
is. Consider the by now familiar gas that is confined to the left
half of a container ($V_{\tiny{\textnormal{left}}}$). Then remove
the separating wall. As a result the gas will spread and soon evenly
fill the entire volume ($V_{\tiny{\textnormal{total}}}$). From a
Gibbsian point of view, what \textit{seems} to happen is that the
equilibrium distribution with respect to the left half evolves into
the equilibrium distribution with respect to the entire container;
more specifically, what \textit{seems} to happen is that the
microcanonical distribution over all micro-states compatible with
the gas being in $V_{\tiny{\textnormal{left}}}$,
$\Gamma_{\tiny{\textnormal{left}}}$, evolves into the microcanonical
distribution over all states compatible with the gas being in
$V_{\tiny{\textnormal{total}}}$,
$\Gamma_{\tiny{\textnormal{total}}}$. The problem is, that this
development is ruled out by the laws of mechanics for an isolated
system. The time evolution of an ensemble density is subject to
Liouville's equation (\ref{Liouville's equation}), according to
which the density moves in phase space like an incompressible
liquid, and therefore it is not possible that a density that was
uniform over $\Gamma_{\tiny{\textnormal{left}}}$ at some time can be
uniform over $\Gamma_{\tiny{\textnormal{total}}}$ at some later
point. Hence, as it stands, the Gibbs approach cannot explain the
approach to equilibrium.


\subsubsection{Coarse-Graining}

The `official' Gibbsian proposal is that this problem is best
addressed by coarse-graining the phase space; the idea is introduced
in Chapter XII of Gibbs (1902) and has since been endorsed, among
others, by Penrose (1970), Farquhar (1964), and all supporters of
the programme of stochastic dynamics discussed below. The procedure
is exactly the same as in the Boltzmann case (Subsection \ref{The
Combinatorial Argument}), with the exception that we now
coarse-grain the system's $\gamma$-space rather than its
$\mu$-space.

The so-called coarse-grained density $\bar{\rho}$ is defined as the
density that is uniform within each cell, taking as its value the
average value in this cell of the original continuous density
$\rho$:

\begin{equation}\label{coarse-grained density}
     \bar{\rho}_{\omega}(q,p,t) := \frac{1}{\delta\omega} \int_{\omega(q,p)} \rho
     (q', p', t) d\Gamma',
\end{equation}
where $\omega(q,p)$ is the cell in which the point $(q,p)$ lies and
$\delta\omega$ is the Lebesgue measure of a cell. Whether we work
with $\bar{\rho}_{\omega}$ or $\rho$ is of little importance to the
practitioner because for any phase function that does not fluctuate
on the scale of $\delta\omega$ (which is true of most physically
relevant phase functions) the phase average with respect to
$\bar{\rho}_{\omega}$ and $\rho$ are approximately the same.

We can now define the coarse-grained entropy $S_{\omega}$:

\begin{equation}
      S_{\omega} (\rho) \, := S_{G} (\bar{\rho}_{\omega})\, = - k_{_{B}} \int_{\Gamma}
      \bar{\rho}_{\omega} \log(\bar{\rho}_{\omega})
      d\Gamma
\end{equation}

\noindent One can prove that the coarse-grained entropy is always
greater or equal to the fine-grained entropy: $S_{\omega} (\rho)
\geq S_{G}(\rho)$; the equality holds only if the fine grained
distribution is uniform over the cells of the coarse-graining (see
Uffink 1995b, 155; Wehrl 1978, 229; Lavis 2004, 672).

What do we gain by working with $\bar{\rho}_{\omega}$ rather than
with $\rho$? The main point is that the coarse-grained density
$\bar{\rho}_{\omega}$ is not governed by Liouville's equation and
hence is not subject to the restrictions mentioned above. So it is,
at least in principle, possible for $\bar{\rho}_{\omega}$ to evolve
in such a way that it will be uniform over the portion of the phase
space available to the system in equilibrium. This state is referred
to as `coarse-grained equilibrium' (Ridderbos 2002, 69). The
approach to coarse-grained equilibrium happens if under the dynamics
of the system $\rho$ becomes so scrambled that an equal portion of
it is located in every cell of the partition. Because the averaged
density is `blind' to differences within each cell, the spread out
states of the initial equilibrium condition will, on the averaged
level, look like a homogenous distribution. This is illustrated in
Figure 8 for the example mentioned at the beginning of this
subsection, where the initial density is constant over
$\Gamma_{\tiny{\textnormal{left}}}$ while the final density is
expected to be constant over $\Gamma_{\tiny{\textnormal{total}}}$
(this figure is adapted from Uffink 1995b, 154).

\vspace{1cm}

\noindent [Insert Figure 8 (appended at the end of the document).]

\vspace{1cm}

\noindent {\footnotesize{Figure 8: Evolution into a
quasi-equilibrium distribution.}}

\vspace{1cm}

\noindent A fine-grained distribution which has evolved in this way, i.e.
which appears to be uniform at the coarse-grained level, is said to
be in a quasi-equilibrium (Blatt 1959, 749; Ridderbos 2002, 73). On\index{Quasi-equilibrium}
the coarse-graining view, then, all that is required to explain the
approach to equilibrium in the Gibbs approach is a demonstration
that an arbitrary initial distribution indeed evolves into a
quasi-equilibrium distribution (Ridderbos 2002, 73).

The question then is under what circumstances this happens. The
standard answer is that the system has to be mixing (see Subsection
\ref{Ergodic Theory} for a discussion of mixing). This suggestion
has some intuitive plausibility given the geometrical interpretation
of mixing, and it receives further support from the convergence
theorem (Equation (\ref{convergence theorem})). In sum, the proposal
is that we coarse-grain the system's phase space and then consider
the coarse-grained entropy, which indeed increases if the system is
mixing.

What can be said in support of this point of view? The main thrust
of arguments in favour of coarse-graining is that even if there are
differences between the fine-grained and the coarse-grained density,
we cannot empirically distinguish between them and hence there is no
reason to prefer one to the other. There are various facets to this
claim; these are discussed but not endorsed in Ridderbos (2002, 73).
First, measurements have finite precision and if $\delta \omega$ is
chosen so that it is below that precision, no measurement that we
can perform on the system will ever be able to tell us whether the
true distribution is $\rho$ or $\bar{\rho}_{\omega}$.

Second, as already observed above, the values of macroscopic
variables calculated using the coarse-grained density coincide with
those calculated using the fine-grained density (if the relevant
phase function does not fluctuate so violently as to fluctuate on
the scale of $\delta \omega$). This is all we need because
thermodynamic equilibrium is defined in terms of the values of
macroscopic parameters and as long as these coincide there is no
reason to prefer the fine-grained to a coarse-grained density.

This programme faces several serious difficulties. To begin with,
there is the problem that mixing is only defined, and so only
achieved, for $t \rightarrow \infty$, but thermodynamic systems seem
to reach equilibrium in finite time. One might try to mitigate the
force of this objection by saying that it is enough for a system to
reach an `almost mixed' state in the relevant finite time. The
problem with this suggestion is that from the fact that a system is
mixing nothing follows about how fast it reaches a mixed state and
hence it is not clear whether it becomes `almost mixed' over the
relevant observation times (see Berkovitz \emph{et al.} 2006, 687).
Moreover, mixing is too stringent a requirement for many realistic
systems. Mixing implies ergodicity and, a fortiori, if a system is
not ergodic it cannot be mixing (see Subsection \ref{Ergodic
Theory}). But there are relevant systems that fail to be ergodic,
and hence also fail to be mixing (as we have seen in Unit
\ref{Problems with Ergodicity}). This is a serious difficulty and
unless it can be argued---as Vranas did with regards to
ergodicity---that systems which fail to be mixing are `almost
mixing' in some relevant sense and reach some `almost mixed state'
in some finite time, an explanation of the approach to equilibrium
based on mixing is not viable.

Second, there is a consistency problem, because we now seem to have
two different definitions of equilibrium (Ridderbos 2002, 73). One
is based on the requirement that the equilibrium distribution be
stationary; the other on apparent uniformity. These two concepts of
equilibrium are not co-extensive and so we face the question of
which one we regard as the constitutive one. Similarly, we have two
notions of entropy for the same system. Which one really is the
system's entropy? However, it seems that this objection need not
really trouble the proponent of coarse-graining. There is nothing
sacrosanct about the formalism as first introduced above and, in
keeping with the revisionary spirit of the coarse-graining approach,
one can simply declare that equilibrium is defined by uniformity
relative to a partition and that $S_{\omega}$ is the `real' entropy
of the system.

Third, as in the case of Boltzmannian coarse-graining, there is a
question about the justification of the introduction of a partition.
The main justification is based on the finite accuracy of
observations, which can never reveal the precise location of a
system's micro-state in its $\gamma$-space. As the approach to
equilibrium only takes place on the coarse-grained level, we have to
conclude that the emergence of thermodynamic behaviour depends on
there being limits to the observer's measurement resolution. This,
so the objection continues, is misguided because thermodynamics does
not appeal to observers of any sort and thermodynamic systems
approach equilibrium irrespective of what those witnessing this
process can know about the system's micro-state.

This objection can be challenged on two grounds. First, one can
mitigate the force of this argument by pointing out that
micro-states have no counterpart in thermodynamics at all and hence
grouping some of them together on the basis of experimental
indistinguishability cannot possibly lead to a contradiction with
thermodynamics. All that matters from a thermodynamic point of view
is that the macroscopic quantities come out right, and this is the
case in the coarse-graining approach (Ridderbos 2002, 71). Second,
the above suggestion does not rely on there being actual observers,
or actual observations taking place. The claim simply is that the
fine-grained distribution has to reach quasi-equilibrium. The
concept is defined relative to a partition, but there is nothing
subjective about that. Whether or not a system reaches
quasi-equilibrium is an objective matter of fact that depends on the
dynamics of the system, but has nothing to do with the existence of
observers.

Those opposed to coarse-graining reply that this is besides the
point because the very justification for introducing a partition to
begin with is an appeal to limited observational capacities so that
whether or not quasi-equilibrium is an objective property
\emph{given} a particular partition is simply a non-issue. So, at
bottom, the disagreement seems to be over the question of whether
the notion of equilibrium is essentially a macroscopic one. That is,
does the notion of equilibrium make sense to creatures with
unlimited observational powers? Or less radically: do they need this
notion? It is at least conceivable that for them the gas indeed does
not approach equilibrium but moves around in some very complicated
but ever changing patterns, which only look stable and unchanging to
those who cannot (or simply do not) look too closely. Whether or not
one finds convincing a justification of coarse-graining by appeal to
limited observational powers depends on how one regards this
possibility.

Fourth, one can question the central premise of the argument for
regarding $\bar{\rho}_{\omega}$ as the relevant equilibrium
distribution, namely that $\bar{\rho}_{\omega}$ and $\rho$ are
empirically indistinguishable. Blatt (1959) and Ridderbos \& Redhead
(1998) argue that this is wrong because the spin-echo experiment
(Hahn 1950)\index{Spin-echo experiment} makes it possible to
empirically discriminate between $\rho$ or $\bar{\rho}$, even if the
size of the cells is chosen to be so small that no direct
measurement could distinguish between states within a cell. For this
reason, they conclude, replacing $\rho$ with $\bar{\rho}_{\omega}$
is illegitimate and an appeal to coarse-graining to explain the
approach to equilibrium has to be renounced.

In the spin-echo experiment, a collection of spins is placed in a
magnetic field $\vec{B}$ pointing along the $z$-axis and the spins
are initially aligned with this field (Figure 9.1).

\vspace{1cm}

\noindent [Insert Figure 9.1 (appended at the end of the document).]

\vspace{1cm}

\noindent {\footnotesize{Figure 9.1: Spins aligned with a magnetic
field.}}

\vspace{1cm}

\noindent Then the spins are subjected to a radio frequency pulse,
as a result of which they are tilted by 90 degrees so that they now
point in the $x$-direction (Figure 9.2).

\vspace{1cm}

\noindent [Insert Figure 9.2 (appended at the end of the document).]

\vspace{1cm}

\noindent {\footnotesize{Figure 9.2: Spins shifted $90^{\circ}$ by a
pulse.}}

\vspace{1cm}

\noindent Due to the presence of the magnetic field $\vec{B}$, the
spins start precessing around the $z$-axis and in doing so emit an
oscillating electromagnetic pulse, the `free induction decay signal'
(Figure 9.3}; the curved dotted arrows indicate the direction of
rotation).

\vspace{1cm}

\noindent [Insert Figure 9.3 (appended at the end of the document).]

\vspace{1cm}

\noindent {\footnotesize{Figure 9.3: Precession of spins.}}

\vspace{1cm}

\noindent This signal is the macroscopic evidence for the fact that
all spins are aligned and precess around the same axis. After some
time this signal decays, indicating that the spins are now no longer
aligned and point in `random' directions (Figure 9.4).

\vspace{1cm}

\noindent [Insert Figure 9.4 (appended at the end of the document).]

\vspace{1cm}

\noindent {\footnotesize{Figure 9.4: Signal decays and spins point
in random directions.}}

\vspace{1cm}

\noindent The reason for this is that the precession speed is a
function of the field strength of $\vec{B}$ and it is not possible
to create an exactly homogeneous magnetic field. Therefore the
precession frequencies of the spins differ slightly, resulting in
the spins pointing in different directions after some time $t=\tau$
has elapsed. At that point a second pulse is applied to the system,
tilting the spins in the $x-z$ plane by 180 degrees (Figure 9.5; the
straight dotted arrows indicate the direction of the spins before
the pulse).

\vspace{1cm}

\noindent [Insert Figure 9.5 (appended at the end of the document).]

\vspace{1cm}

\noindent {\footnotesize{Figure 9.5: Spins after a second pulse.}}

\vspace{1cm}

\noindent The result of this is a reversial of the order of the
spins in the sense that the faster spins that were ahead of the
slower ones are now behind the slower ones (Figure 9.6; $s_{1}$ and
$s_{2}$ are two spins, $s'_{1}$ and $s'_{2}$ their `tilted
versions').

However, those that were precessing faster before the second pulse
keep doing so after the pulse and hence `catch up' with the slower
ones. After time $t=2\tau$ all spins are aligned again and the free
induction decay signal reappears (the `echo pulse'). This is the
macroscopic evidence that the original order has been
restored.\footnote{It is often said that this experiment is the
empirical realisation of a Loschmidt velocity reversal (in which a
`Loschmidt demon' \index{Loschmidt demon} instantaneously transforms
the velocities $\vec{v}_{i}$ of all particles in the system into
$-\vec{v}_{i}$). This is incorrect. The directions of precession
(and hence the particles' velocities) are \emph{not} reversed in the
experiment. The reflection of the spins in the $x-z$ plane results
in a reversal of their ordering while leaving their velocities
unaltered. The grain of truth in the standard story is that a
reversal of the ordering with unaltered velocities is in a sense
`isomorphic' to a velocity reversal with unaltered ordering.}

\vspace{1cm}

\noindent [Insert Figure 9.6 (appended at the end of the document).]

\vspace{1cm}

\noindent {\footnotesize{Figure 9.6: Order of spins (in terms of
speed) is reversed.}}

\vspace{1cm}

\noindent At time $t=\tau$, when all spins point in random
directions, $\bar{\rho}$ is uniform and the system has reached its
coarse-grained equilibrium. From a coarse-grainer's point of view
this is sufficient to assert that the system \emph{is} in
equilibrium, as we cannot distinguish between true and
coarse-grained equilibrium. So according to Blatt and Redhead \&
Ridderbos the spin-echo experiment shows that this rationale is
wrong because we actually \emph{can} distinguish between true and
coarse-grained equilibrium. If the system was in true equilibrium at
$t=\tau$ the second radio pulse flipping the spins by 180 degrees
would not have the effect of aligning the spins again at $t=2\tau$;
this only happens because the system is merely in a coarse-grained
equilibrium. Hence equilibrium and quasi-equilibrium distributions
can be shown to display radically different behaviour. Moreover,
this difference is such that we can experimentally detect it without
measuring microdynamical variables: we simply check whether there is
an echo-pulse at $t=2\tau$. This pulls the rug from under the feet
of the coarse-grainer and we have to conclude that it is therefore
not permissible to base fundamental arguments in statistical
mechanics on coarse-graining (Blatt 1959, 746).

What is the weight of this argument? Ridderbos (2002, 75) thinks
that the fact that we can, after all, experimentally distinguish
between $\bar{\rho}$ and $\rho$, and hence between `real'
equilibrium and quasi-equilibrium, is by itself a sufficient reason
to dismiss the coarse-graining approach. Others are more hesitant.
Ainsworth (2005, 626-7) points out that, although valid, this
argument fails to establish its conclusion because it assumes that
for coarse-graining approach to be acceptable $\bar{\rho}$ and
$\rho$ must be are empirically indistinguishable. Instead, he
suggests appealing to the fact, proffered by some in support of
Boltzmannian coarse-graining, that there is an objective separation
of the micro and macro scales (see Subsection \ref{Limitations}). He
accepts this point of view as essentially correct and submits that
the same response is available to the Gibbsian: coarse-graining can
be justified by an appeal to the separation of scales rather than by
pointing to limitations of what we can observe. As the notion of
equilibrium is one that inherently belongs to the realm of the
macroscopic, coarse-grained equilibrium is the correct notion of
equilibrium, irrespective of what happens at the micro scale.
However, as I have indicated above, the premise of this argument is
controversial since it is not clear whether there is indeed an
objective separation of micro and macro scales.

Ridderbos \& Redhead make their case against coarse-graining by
putting forward two essentially independent arguments. Their first
argument is based on theoretical results. They introduce a
mathematical model of the experiment and then show that the
coarse-grained distribution behaves in a way that leads to false
predictions. They show that the system reaches a uniform
coarse-grained distribution over the entire phase space at $t=\tau$
(as one would expect), but then fails to evolve back into a
non-equilibrium distribution under reversal, so that, in
coarse-grained terms, the system is still described by a uniform
distribution at $t=2\tau$ (1998, 1250). Accordingly, the
coarse-grained entropy reaches its maximum at $t=\tau$ and does not
decrease as the spins evolve back to their initial positions. Hence,
the coarse-grained entropy is still maximal when the echo pulse
occurs and therefore the occurrence of the echo is, from a
coarse-grained perspective, completely miraculous (1998, 1251).

Their second argument is based on the assumption that we can,
somehow, experimentally observe the coarse-grained entropy (as
opposed to calculating it in the model). Then we face the problem
that observational results seem to tell us that the system has
reached equilibrium at time $t=\tau$ and that after the application
of the second pulse at that time evolves away from equilibrium; that
is, we are led to believe that the system behaves
anti-thermodynamically. This, Ridderbos and Redhead (1998, 1251)
conclude, is wrong because the experiments do not actually
contradict the Second Law.

So the experimental results would stand in contradiction both with
the theoretical results predicting that the coarse-grained entropy
assumes its maximum value at $t=2\tau$ and with the second law of
thermodynamics, which forbids high to low entropy transitions in
isolated systems (and the spin echo system is isolated after the
second pulse). This, according to Ridderbos \& Redhead, is a
\emph{reductio} of the coarse-graining approach.\footnote{Their own
view is that the fine grained entropy is the correct entropy and
that we were wrong to believe that the entropy ever increased.
Despite appearances, the thermodynamic entropy does not increase
between $t=0$ and $t=\tau$ and hence there is no need for it to
decrease after $t=\tau$ in order to resume its initial value at
$t=2\tau$; it is simply constant throughout the experiment. However,
this view is not uncontroversial (Sklar 1993, 253-4).}

These arguments have not gone unchallenged. The first argument has
been criticised by Lavis (2004) on the grounds that the behaviour of
$\bar{\rho}$ and the coarse-grained entropy predicted by Ridderbos
\& Redhead is an artifact of the way in which they calculated these
quantities. There are two methods for calculating $\bar{\rho}$. The
first involves a coarse-graining of the fine-grained distribution at
each instant of time; i.e. the coarse-grained distribution at time
$t$ is determined by first calculating the fine-grained distribution
at time $t$ (on the basis of the time evolution of the system and
the initial distribution) and then coarse-graining it. The second
method is based on re-coarse-graining as time progresses; i.e. the
coarse-grained distribution at time $t$ is calculated by evolving
the coarse-grained distribution at an earlier time and then
re-coarse-graining. Lavis points out that Ridderbos \& Redhead use
the second method: they calculate $\bar{\rho}$ at time $t=2\tau$ by
evolving $\bar{\rho}$ at time $t=\tau$ forward in time. For this
reason, the fact that they fail to find $\bar{\rho}$ returning to
its initial distribution is just a manifestion of the impossibility
of `un-coarse-graining' a coarse-grained distribution. Lavis then
suggests that we should determine that coarse-grained distribution
at some time $t$ by using the first method, which, as he shows,
yields the correct behaviour: the distribution returns to its
initial form and the entropy decreases in the second half of the
experiment, assuming its initial value at $t=2\tau$. Hence the
echo-pulse does not come as a surprise after all.

The question now is which of the two coarse-graining methods one
should use. Although he does not put it quite this way, Lavis'
conclusion seems to be that given that there are no physical laws
that favour one method over the other, the principle of charity
should lead us to choose the one that yields the correct results.
Hence Ridderbos \& Redhead's result has no force against the
coarse-graining.

As regards the second argument, both Lavis (2004) and Ainsworth
(2005) point out that the decrease in entropy during the second half
of the experiment need not trouble us too much. Ever since the work
of Maxwell and Boltzmann `entropy increase in an isolated system is
taken to be highly probable but not certain, and the spin-echo
model, along with simulations of other simple models, is a nice
example of the working of the law' (Lavis 2004, 686). On this view,
the spin-echo experiment simply affords us one of these rare
examples in which, due to skilful engineering, we can prepare a
system in one of these exceptional states which evolve from high to
low entropy.


\subsubsection{Interventionism}\label{Interventionism}

One of the crucial assumptions, made more or less tacitly so far, is
that the systems under consideration are isolated. This, needless to
say, is an idealising assumption that can never be realised in
practice. Real systems cannot be perfectly isolated from their
environment and are always subject to interactions; for instance, it
is impossible to shield a system from gravitation. Blatt (1959)
suggested that taking systems to be isolated not only fails to be
the harmless idealisation that it is generally believed to be; it
actually is the source of the problem. This recognition is the
starting point for the interventionist programme, at the heart of
which lies the idea that real systems are open in that they are
constantly subject to outside perturbations, and that it is exactly
these perturbations that drive the system into equilibrium.

In more detail, the leading idea is that every system interacts with
its environment (the gas, for instance, collides with the wall of
the container and the walls interact in many different ways with
their surroundings), and that these interactions are `\emph{in
principle} not amenable to a causal description, and \emph{must of
necessity} be described in statistical terms' (Blatt 1959, 751,
original emphasis). The perturbations from outside serve as a kind
of `stirring mechanism' or `source of randomness'  that drives the
system around randomly in the phase space, in much the same way as
it would be the case if the system was mixing. As a consequence, the
observable macroscopic quantities are soon driven towards their
equilibrium values. This includes the Gibbs entropy; in an open
system Liouville's theorem no longer holds and there is nothing to
prevent the Gibbs entropy from increasing.\footnote{It is a curious
fact about the literature on the subject that interventionism is
always discussed within the Gibbs framework. However, it is obvious
that interventionism, if true, would also explain the approach to
equilibrium in the Boltzmannian framework as it would explain why
the state of the system wanders around randomly on the energy
surface, which is needed for it to ultimately end up in the
equilibrium region (see Unit \ref{Issue 1}).}

Of course, from the fact that the Gibbs entropy can increase it does
not follow that it actually does increase; whether or not this is
the case depends on the system as well as the properties of the
outside perturbations. Blatt (1959) and Ridderbos \& Redhead (1998)
assure us that in realistic model systems one can prove this to be
the case. Granting this, we have an elegant explanation of why and
how systems approach equilibrium, which also enjoys the advantage
that no revision of the classical laws is needed.\footnote{For a
discussion of interventionism and time-reversal see Ridderbos \&
Redhead (1998, 1259-62) and references therein.}

A common objection against this suggestion points out that we are
always free to consider a larger system, consisting of our
`original' system \textit{and} its environment. For instance, we can
consider the `gas cum box' system,  which, provided that classical
mechanics is a universal theory, is also governed by classical
mechanics. So we are back to where we started. Interventionism,
then, seems wrong because it treats the environment as a kind of
\emph{deus ex machina} that is somehow `outside physics'; but the
environment is governed by the fundamental laws of physics just as
the system itself is and so it cannot do the job that the
interventionist has singled out for it to do.

The interventionist might now reply that the `gas cum box' system
has an environment as well and it is \textit{this} environment that
effects the desired perturbations. This answer does not resolve the
problems, of course. We can now consider an even larger system that
also encompasses the environment of the `gas cum box' system. And we
can keep expanding our system until the relevant system is the
entire universe, which, by assumption, has no environment any more
that might serve as a source of random perturbations.

Whether this constitutes a reductio of the interventionist programme
depends on one's philosophical commitments. The above argument
relies on the premise that classical mechanics (or quantum
mechanics, if we are working within quantum SM) is a universal
theory, i.e. one that applies to everything that there is without
restrictions. This assumption, although widely held among scientists
and philosophers alike, is not uncontroversial. Some have argued
that we cannot legitimately claim that laws apply universally. In
fact, laws are always tested in highly artificial laboratory
situations and claiming that they equally apply outside the
laboratory setting involves an inductive leap that is problematic.
Hence we have no reason to believe that classical mechanics applies
to the universe as a whole; see for instance Reichenbach (1956) and
Cartwright (1999) for a discussion of this view. This, if true,
successfully undercuts the above argument against
interventionism.\footnote{Interventionists are sometimes charged
with being committed to an instrumentalist take on laws, which, the
critics continue, is an unacceptable point of view. This is
mistaken. Whatever one's assessment of the pros and cons of
instrumentalism, all the interventionist needs is the denial that
the laws (or more specifically, the laws of mechanics) are universal
laws. This is compatible with realism about laws understood as
providing `local' descriptions of `parts' of the universe (a
position sometime referred to as `local realism').}

There is a way around the above objection even for those who do
believe in the generality of laws, namely to deny Blatt's assumption
that the environment needs to be genuinely stochastic. \emph{Pace}
Blatt, that the environment be genuinely stochastic (i.e. as
governed by indeterministic laws rather than classical mechanics) is
not an indispensable part of the interventionist programme. As
Ridderbos \& Redhead (1998, 1257) point out, all that is required is
that the system loses coherence, which can be achieved by
dissipating correlations into the environment. For observations
restricted to the actual system, this means that correlational
information is not available. But the information is not lost; it
has just been `dislocated' into the degrees of freedom pertaining to
the environment.

The question then becomes whether the universe as a whole is
expected to approach equilibrium, or whether thermodynamic behaviour
is only required to hold for a subsystem of the universe. Those who
hold that the `dissipation' of correlational information into
environmental degrees of freedom is enough to explain the approach
to equilibrium are committed to this view. Ridderbos \& Redhead are
explicit about this (1998, 1261-2). They hold that the fine-grained
Gibbs entropy of the universe is indeed constant since the universe
as a whole has no outside, and that there is no approach to
equilibrium at the level of the universe. Moreover, this does not
stand in conflict with the fact that cosmology informs us that the
entropy of the universe is increasing; cosmological entropies are
coarse-grained entropies and, as we have seen above, there is no
conflict between an increase in coarse-grained entropy and the
constancy of the fine-grained Gibbs entropy. Ridderbos \& Redhead
acknowledge that the question now is whether the claim that the
Gibbs entropy of the universe is constant is true, which is an issue
that has to be settled empirically.


\subsubsection{Changing the notion of Equilibrium}

One of the main problems facing Gibbsian non-equilibrium theory is
that under a Hamiltonian time evolution a non-stationary
distribution cannot evolve into a stationary one (see Unit
\ref{Gibbs Issue 4}). Hence strict stationarity is too stringent a
requirement for equilibrium. Nevertheless, it seems plausible to
assume that an equilibrium distribution has to approximate a
stationary distribution in some relevant sense. What is this
relevant sense?

Van Lith suggested turning the desired result into a definition and
replacing strict stationarity with the requirement that the
distribution be such that the phase average of every function in a
physically relevant set of functions only fluctuates mildly around
its average value (van Lith 1999, 114). More precisely, let $\Omega$
be a class of phase functions $f(x)$ corresponding to
macroscopically relevant quantities. Then the system is in
equilibrium from time $\tau$ onwards iff for every function $f(x)
\in \Omega$ there is a constant $c_{f}$ such that:

\begin{equation}\label{approx equ}
    \mid \int f(x) \rho_{t}(x) d\Gamma - c_{f} \mid \, \leq \, \varepsilon_{f},
\end{equation}
where $\varepsilon_{f}$ is a small number (which can be different
for every $f$). This definition of equilibrium seems to have the
advantage of preserving all the desirable features of equilibrium
while no longer running into the problem that equilibrium can never
be reached.

However, from the fact that an arbitrary non-equilibrium
distribution \emph{can} reach equilibrium thus defined it does not
follow that it actually \emph{does}. What conditions does the
dynamics of a system have to meet in order for the approach to
equilibrium to take place? Van Lith points out that being mixing is
a sufficient condition (van Lith 1999, 114) because the Convergence
Theorem (see Subsection \ref{Ergodic Theory}) states that in the
limit all time averages converge to the microcanonical averages, and
hence they satisfy the above definition.

But this proposal suffers from various problems. First, as van Lith
herself points out (1999, 115), the proposal does not contain a
recipe to get the (fine-grained) Gibbs entropy moving; hence the
approach to equilibrium need not be accompanied by a corresponding
increase in the Gibbs entropy.

Second, as we have seen above, mixing is too stringent a condition:
it is not met by many systems of interest. Remedy for this might be
found in the realisation that less than full-fledged mixing is
needed to make the above suggestion work. In fact, all we need is a
condition that guarantees that the Convergence Theorem holds (Earman
\& Redei 1996, 74; van Lith 1999, 115). One condition of that sort
is that the system has to be mixing for all $f \in \Omega$. The
question then is, what this involves. This question is difficult, if
not impossible to answer, before $\Omega$ is precisely specified.
And even then there is the question of whether the convergence is
sufficiently rapid to account for the fact that thermodynamic
systems reach equilibrium rather quickly.\footnote{Another
alternative definition of equilibrium, which also applies to open
systems, has been suggested by Pitowsky (2001, 2006), but for a lack
of space I cannot further discuss this suggestion here.}


\subsubsection{Alternative Approaches}

Before turning to the epistemic approach, I would like to briefly
mention three other approaches to non-equilibrium SM; lack of space
prevents me from discussing them in more detail.

\emph{Stochastic Dynamics}. \index{Stochastic dynamics} The leading
idea of this approach is to replace the Hamiltonian dynamics of the
system with an explicity probabilistic law of evolution.
Characteristically this is done by coarse-graining the phase space
and then postulating a probabilistic law describing the transition
from one cell of the partition to another one. Liouville's theorem
is in general not true for such a dynamics and hence the problem of
the constancy of the Gibbs entropy does not arise. Brief
introductions can be found in Kreuzer (1981, Ch. 10), Reif (1985,
Ch. 15) and Honerkamp (1998, Ch. 5); detailed expositions of the
approach include Penrose (1970; 1979), Mackey (1989; 1992), and
Streater (1995).

The main problem with this approach is that its probabilistic laws
are put in `by hand' and are not derived from the underlying
dynamics of the system; that is, it is usually not possible to
derive the probabilistic laws from the underlying deterministic
evolution and hence the probabilistic laws are introduced as
independent postulates. However, unless one can show how the
transition probabilities postulated in this approach can be derived
from the Hamiltonian equations of motion governing the system, this
approach does not shed light on how thermodynamical behaviour
emerges from the fundamental laws governing a system's constituents.
For critical discussions of the stochastic dynamics programme see
Sklar (1993, Chs. 6 and 7), Callender (1999, 358-364) and Uffink
(2007, 1038-1063).


\emph{The Brussels School} (sometimes also `Brussels-Austin
School').\index{Brussels-Austin School} An approach closely related
to the Stochastic Dynamics programme has been put forward by the
so-called Brussels School, led by Ilya Prigogine. The central
contention of this programme is that if the system exhibits
sensitive dependence on initial conditions (and most systems do) the
very idea of a precise micro-state given by a point in phase space
ceases to be meaningful and should be replaced by an explicitly
probabilistic description of the system in terms of open regions of
the phase space, i.e. by a Gibbs distribution function. This
programme, if successful, can be seen as providing the sought after
justification for the above-mentioned shift from a Hamiltonian micro
dynamics to an explicitly probabilistic scheme. These claims have
been challenged on different grounds; for presentations and critical
discussions of the ideas of the Brussels School see Batterman
(1991), Bricmont (1996), Karakostas (1996), Lombardi (1999, 2000),
Edens (2001) and Bishop (2004).

An approach that is similar to the programm of the Brussels School
in that it denies that the conceptual framework of classical
mechanics, in particular the classical notion of a state, is
adequate to understand SM, has been suggested by Krylov.
Unfortunately he died before he could bring his programme to
completion, and so it not clear what form his ideas would have taken
in the end. For philosophical discussions Krylov's programme see
Batterman (1990), R\'{e}dei (1992) and Sklar (1993, 262-69).


\emph{The BBGKY Hierarchy}.\index{BBGKY hierarchy} The main idea of
the BBGKY (after Bogolyubov, Born, Green, Kirkwood, and Yvon)
approach is to describe the evolution of an ensemble by dint of a
reduced probability density and then derive (something like) a
Boltzmann equation for this density, which yields the approach to
equilibrium. The problem with the approach is that, just as in the
case of the Boltzmann's (early) theory, the irreversibility is a
result of (something like) the \emph{Stosszahlansatz}, and hence all
its difficulties surface again at this point. For a discussion of
this approach see Uffink (2007, 1034-1038).


\subsection{The Epistemic Approach}\label{The Epistemic Approach}

The approaches discussed so far are based on the assumption that SM
probabilities are ontic (see Unit \ref{Probabilities}). It is this
assumption that those who argue for an epistemic interpretation
deny. They argue that SM probabilities are an expression of what we
know about a system, rather than a feature of a system itself. This
view can be traced back to Tolman (1938) and has been developed into
an all-encompassing approach to SM by Jaynes in a series of papers
published (roughly) between 1955 and 1980, some of which are
gathered in Jaynes \index{Jaynes, E. T.!epistemic approach to
statistical mechanics} (1983).\footnote{In this subsection I focus
on Jaynes' approach. Tolman's view is introduced in his (1938,
59-70); for a discussion of Tolman's interpretation of probability
see Uffink (1995b, 166-7).}

At the heart of Jaynes' approach to SM lies a radical
reconceptualisation of what SM is. On his view, SM is about our
knowledge of the world, not about the world itself. The probability
distribution represents our state of knowledge about the system at
hand and not some matter of fact about the system itself. More
specifically, the distribution represents our lack of knowledge
about a system's micro-state given its macro condition; and, in
particular, entropy becomes a measure of how much knowledge we lack.
As a consequence, Jaynes regards SM as a part of general statistics,
or `statistical inference', as he puts it:

\begin{quote}
`Indeed, I do not see Predictive Statistical Mechanics and
Statistical Inference as different subjects at all; the former is
only a particular realization of the latter [...] Today, not only do
Statistical Mechanics and Statistical Inference not appear to be two
different fields, even the term `statistical' is not entirely
appropriate. Both are special cases of a simple and general
procedure that ought to be called, simply, ``inference''.' (Jayenes
1983, 2 - 3)
\end{quote}

\noindent The questions then are: in what way a probability
distribution encodes a lack of knowledge; according to what
principles the correct distribution is determined; and how this way
of thinking about probabilities sheds any light on the foundation of
SM. The first and the second of these questions are addressed in
Unit \ref{The Shannon Entropy}; the third is discussed in \ref{MEP
and SM}.


\subsubsection{The Shannon Entropy}\label{The Shannon Entropy}

Consider a random variable $x$ which can take any of the $m$
discrete values in $X=\{x_{1}, ..., x_{m}\}$ with probabilities
$p(x_{i})$; for instance, $x$ can be the number of spots showing on
the next roll of a die, in which case $X=\{1, 2, 3, 4, 5, 6\}$ and
the probability for each even is $1/6$. The Shannon entropy
\index{Shannon entropy} of the probability distribution $p(x_{i})$
is defined (Shannon 1949) as:

\begin{equation}\label{Discrete Entropy}
      S_{_{S}}(p)\, := \, - \sum_{i=1}^{m} p(x_{i}) \log(p(x_{i})),
\end{equation}

\noindent which is a quantitative measure for the uncertainty of the
outcome. If the probability for one particular outcome is one while
the probabilities for all other outcomes are zero, then there is no
uncertainty and $S_{_{S}}$ equals zero; $S_{_{S}}$ reaches its
maximum for a uniform probability distribution, i.e. $p(x_{i}) =
1/m$ for all $i$, in which case we are maximally uncertain about the
outcome; an accessible discussion of the relation between the
Shannon entropy and uncertainty can be found Jaynes (1994, Ch. 11);
see Cover and Thomas (1991) for a detailed treatment.

Sometimes we are given $X$ but fail to know the $p(x_{i})$. In this
case Jaynes's maximum entropy principle (MEP) \index{Maximum entropy
principle} instructs us to choose that distribution $p(x_{i})$ for
which the Shannon entropy is maximal (under the constraint
$\sum_{i=1}^{m}p(x_{i})=1$). For instance, from this principle it
follows immediately that we should assign $p=1/6$ to each number of
spots when rolling the die. If there are constraints that need to be
taken into account then MEP instructs us to choose that distribution
for which $S_{_{S}}$ is maximal under the given constraints. The
most common type of constraint is that the expectation value for a
particular function $f$ has a given value $c$:

\begin{equation}\label{expectation value}
      \langle f\rangle\, := \, \sum_{i=1}^{m} f(x_{i}) p(x_{i})
      \, = \, c.
\end{equation}

This can be generalised to the case of a continuous variable, i.e.
$X=(a, \, b)$, where $(a, \, b)$ is an interval of real numbers (the
boundaries of this interval can be finite or infinite). The
continuous Shannon entropy is

\begin{equation}\label{Continuous Entropy}
      S_{_{S}}(p)\, := \, - \int_{a}^{b} p(x) \log[p(x)] dx,
\end{equation}

\noindent where $p(x)$ is a probability density over $(a, \,
b)$.\footnote{This generalisation is problematic in many respects;
and for the continuum limit to be taken properly, a background
measure and the relative entropy need to be introduced. In the
simplest case where the background measure is the Lebesgue measure
we retrieve Equation (\ref{Continuous Entropy}). For a discussion of
this issue see Uffink (1995a, 235-239).} So for a continuous
variable the most common type of constraint is

\begin{equation}\label{continuous expectation value}
      \langle f\rangle\, := \, \int_{a}^{b} f(x) p(x)dx
      \, = \, c,
\end{equation}

\noindent and MEP tells us to choose $p(x)$ such that it maximises
$S_{_{S}}(p)$ under the given constraints.

Why is MEP compelling? The intuitive idea is that we should always
choose the distribution that corresponds to a maximal amount of
uncertainty, i.e. is maximally non-committal with respect to the
missing information. But why is this a sound rule? In fact MEP is
fraught with controversy; and, to date, no consensus on its
significance, or even cogency, has been reached. However, debates
over the validity of MEP belong to the foundations of statistical
inference in general and as such they are beyond the scope of this
review; for discussions see, for instance, Lavis (1977), Denbigh and
Denbigh (1985), Lavis and Milligan (1985, Sec. 5),  Shimony (1985),
Seidenfeld (1986), Uffink (1995a; 1996a), Howson and Urbach (2006,
276-288).

In what follows let us, for the sake of argument, assume that MEP
can be justified satisfactorily and discuss what it has to offer for
the foundations of SM. But before moving on, a remark about the
epistemic probabilities here employed is in place. On the current
view, epistemic probabilities are not subjective, i.e. they do not
reduce to the personal opinion of individual observers, as would be
the case in a personalist Bayesian theory (such as de Finetti's). On
the contrary Jaynes advocates an `impersonalism' that bases
probability assignments solely on the available data and MEP;
anybody's personal opinions do not enter the scene at any point.
Hence, referring to Jaynes' position as `subjectivism' is
a---frequently used---misnomer.


\subsubsection{MEP and SM}\label{MEP and SM}

The appeal of MEP for equilibrium SM lies in the fact that the
continuous Shannon entropy is equivalent to the Gibbs entropy
(\ref{Gibbs Entropy}) up to the multiplicative constant $k_{_{B}}$
if in Equation (\ref{Continuous Entropy}) we take $X$ to be the
phase space and $\rho$ the probability distribution. Gibbsian
equilibrium distributions are required to maximise $S_{_{G}}$ under
certain constraints and hence, trivially, they also satisfy MEP. For
an isolated system, for instance, the maximum entropy distribution
is the microcanonical distribution. In fact, even more has been
achieved: MEP not only coincides with the Gibbsian maximum entropy
principle introduced in Subsection \ref{Gibbs Formalism}; on the
current view, this principle, which above has been postulated
without further explanation, is justified because it can be
understood as a version of MEP.

As we have seen at the beginning of this subsection, Jaynes sees the
aim of SM as making predictions, as drawing inferences. This opens a
new perspective on non-equilibrium SM, which, according to Jaynes,
should refrain from trying to explain the approach to equilibrium by
appeal to dynamical or other features of the system and only aim to
make predictions about the system's future behaviour (1983, 2). Once
this is understood, the puzzle of the approach to equilibrium has a
straightforward two-step answer. In Sklar's (1993, 255-257)
reconstruction, the argument runs as follows.

The first step consists in choosing the initial distribution.
Characteristic non-equilibrium situations usually arise from the
removing of a constraint (e.g. the opening of a shutter) in a
particular equilibrium situation. Hence the initial distribution is
chosen in the same way as an equilibrium distribution, namely by
maximising the Shannon entropy $S_{_{S}}$ relative to the known
macroscopic constraints. Let $\rho_{0}(q, p, t_{0})$ be that
distribution, where $t_{0}$ is the instant of time at which the
constraint in question is removed. Assume now that the experimental
set-up is such that a set of macroscopic parameters corresponding to
the phase functions $f_{i}$, $i=1, ..., k$, are measured. At time
$t_{0}$ these have as expected values

\begin{equation}\label{Jaynes mean value}
    \bar{f}_{i}(t_{0}) \, =  \, \int f_{i}(q, p)\, \rho_{0}(q, p, t_{0}) d\Gamma, \, i=1,
    ..., k.
\end{equation}

\noindent Furthermore we assume that the entropy which we determine
in an actual experiment, the experimental entropy $S_{_{e}}$, at
time $t_{0}$ is equal to the Gibbs entropy: $S_{_{e}}(t_{0}) =
S_{_{S}}(\rho_{0}(t_{0}))$.

The second step consists in determining the distribution and the
entropy of a system at some time $t_{1} > t_{0}$. To this end we
first use Liouville's equation to determine the image of the initial
distribution under the dynamics of the system, $\rho_{0}(t_{1})$,
and then calculate the expectation values of the observable
parameters at time $t_{1}$:

\begin{equation}
    \bar{f}_{i}(t_{1}) \, = \, \int f_{i}(q, p) \, \rho_{0}(q, p, t_{1}) d\Gamma  , \, i=1,
    ..., k.
\end{equation}

\noindent Now we calculate a new density $\rho_{1}(q, p, t_{1})$,
which maximises the Shannon entropy under the constraints that

\begin{equation}\label{Constraints t1}
     \int f_{i}(q, p) \, \rho_{1}(q, p, t_{1}) d\Gamma \, = \, \bar{f}_{i}(t_{1}), \, \, i=1,
    ..., k.
\end{equation}

\noindent The experimental entropy of the system at $t_{1}$ then is
$S_{_{e}}(t_{1})= S_{_{S}}(\rho_{1}(t_{1}))$. This entropy is
greater than or equal to $S_{_{e}}(t_{0})$ for the following reason.
By Liouville's theorem we have $S_{_{S}}(\rho_{0}(t_{0}))=
S_{_{S}}(\rho_{0}(t_{1}))$. Both $S_{_{S}}(\rho_{0}(t_{1}))$ and
$S_{_{S}}(\rho_{1}(t_{1}))$ satisfy the constraints in Equation
(\ref{Constraints t1}). By construction, $S_{_{S}}(\rho_{1}(t_{1}))$
is maximal relative to these constraints; this need not be the case
for $S_{_{S}}(\rho_{0}(t_{1}))$. Therefore $S_{_{e}}(t_{0}) \leq
S_{_{e}}(t_{1})$. This is Jaynes' justification of the Second Law.

Jaynes' epistemic approach to SM has several interesting features.
Unlike the approaches that we have discussed so far, it offers a
clear and cogent interpretation of SM probabilities, which it views
as rational degrees of belief. This interpretation enjoys the
further advantage over its ontic competitors that it can dispense
with ensembles. On Jaynes' approach there is only one system, the
one on which we are performing our experiments, and viewing
probabilities as reflecting our lack of knowledge about this system
rather than some sort of frequency renders ensembles superfluous.
And most importantly, the problems so far that beset non-equilibrium
theory no longer arise: the constancy of the Gibbs entropy becomes
irrelevant because of the `re-maximising' at time $t_{1}$, and the
stationarity of the equilibrium distribution is no longer an issue
because the dynamics of the probability distribution is now a
function of both our epistemic situation and the dynamics of the
system, rather than \emph{only} Liouville's equation. And last but
not least---and this is a point that Jaynes himself often
emphasised---all this is achieved without appealing to complex
mathematical properties like ergodicity or even mixing.


\subsubsection{Problems}

Let us discuss Jaynes' approach to non-equilibrium SM first.
Consider a sequence $t_{0} < t_{1} < t_{2} < ...$ of increasing
instants of time and consider the entropy $S_{_{e}}(t_{j})$, $j = 0,
1, ...$ at these instants; all the $S_{_{e}}(t_{j})$, $j \geq 2$ are
calculated with Equation (\ref{Constraints t1}) after substituting
$\rho_{j}$ for $\rho_{1}$. Conformity with the Second Law would
require that $S_{_{e}}(t_{0}) \leq S_{_{e}}(t_{1}) \leq
S_{_{e}}(t_{2}) \leq ...$. However, this is generally not the case
(Lavis and Milligan 1985, 204; Sklar 1993, 257-258) because the
experimental entropy $S_{_{e}}$ is not necessarily a monotonically
increasing function. Jaynes' algorithm to calculate the
$S_{_{e}}(t_{j})$ can only establish that $S_{_{e}}(t_{0}) \leq
S_{_{e}}(t_{j})$, for all $j > 0$ but it fails to show that
$S_{_{e}}(t_{i}) \leq S_{_{e}}(t_{j})$, for all $0 < i < j$; in
fact, it is indeed possible that $S_{_{e}}(t_{i}) > S_{_{e}}(t_{j})$
for some $i < j$.

A way around this difficulty would be to use $\rho_{j-1}(t_{j-1})$
to calculate $\rho_{j}(t_{j})$, rather than $\rho_{0}(t_{0})$. This
would result in the sequence becoming monotonic, but it would have
the disadvantage that the entropy curve would become dependent on
the sequence of instants of time chosen (Lavis and Milligan
\emph{ibid.}). This seems odd even from a radically subjectivist
point of view: why should the value of $S_{_{e}}$ at a particular
instant of time, depend on earlier instants of time at which we
chose to make predictions, or worse, why should it depend on us
having made any predictions at all?

In equilibrium theory, a problem similar to the one we discussed in
connection with the ergodic approach (Subsection \ref{Why Does Gibbs
Phase Averaging Work?}) arises. As we have seen in Equation
(\ref{Jaynes mean value}), Jaynes also assumes that experimental
outcomes correspond to phase averages as given in Equation
(\ref{Phase average}). But why should this be the case? It is
correct that we should rationally expect the mean value of a
sequence of measurements to coincide with the phase average, but
\emph{prima facie} this does not imply anything about individual
measurements. For instance, when throwing a die we expect the mean
of a sequence of events to be 3.5; but we surely don't expect the
die to show 3.5 spots after any throw! So why should we expect the
outcome of a measurement of a thermodynamic parameter to coincide
with the phase average? For this to be the case a further assumption
seems to be needed, for instance (something like) Khinchin's
assumption that the relevant phase functions assume almost the same
value for almost all points of phase space (see Subsection
\ref{Khinchin's Programme and the Thermodynamic Limit})

A further problem is that the dynamics of the system does not play
any r\^{o}le in Jaynes' derivation of the microcanonical
distribution (or any other equilibrium distribution that can be
derived using MEP). This seems odd because even if probability
distributions are eventually about our (lack of) knowledge, it seems
that what we can and cannot know must have something to do with how
the system behaves. This point becomes particularly clear from the
following considerations (Sklar 1993, 193-194). Jaynes repeatedly
emphasised that ergodicity---or the failure thereof---does not play
any r\^{o}le in his account. This cannot be quite true. If a system
is not ergodic then the phase space decomposes into two (or more)
invariant sets (see Unit \ref{Ergodic Theory}). Depending on what
the initial conditions are, the system's state may be confined to
some particular invariant set, where the relevant phase functions
have values that differ from the phase average; as a consequence MEP
leads to wrong predictions. This problem can be solved by searching
for the `overlooked' constants of motion and then controlling for
them, which yields the correct results.\footnote{Quay (1978, 53-54)
discusses this point in the context of ergodic theory.} However, the
fact remains that our \emph{original} probability assignment was
wrong, and this was because we have ignored certain important
dynamical features of the system. Hence the correct application of
MEP depends, after all, on dynamical features of the system. More
specifically, the micro canonical distribution seems to be correct
only if there are no such invariant subsets, i.e. if the system is
ergodic.

A final family of objections has to do with the epistemic
interpretation of probabilities itself (rather than with `technical'
problems in connection with the application of the MEP formalism).
First, the Gibbs entropy is defined in terms of the distribution
$\rho$, and if $\rho$ pertains to our epistemic situation rather
than to (aspects of) the system, it, strictly speaking, does not
make any sense to say that entropy is a property of the
\emph{system}; rather, entropy is a property of our knowledge of the
system. Second, in the Gibbs approach equilibrium is defined in
terms of specific properties that the distribution $\rho$ must
possess at equilibrium (see Subsection \ref{Gibbs Formalism}). Now
the same problem arises: if $\rho$ reflects our epistemic situation
rather than facts about the system, then it does not make sense to
say that the \emph{system} is in equilibrium; if anything, it is our
knowledge that is in equilibrium. This carries over the
non-equilibrium case. If $\rho$ is interpreted epistemically, then
the approach to equilibrium also pertains to our knowledge and not
to the system. This has struck many commentators as outright wrong,
if not nonsensical. Surely, the boiling of kettles or the spreading
of gases has something to do with how the molecules constituting
these systems behave and not with what we happen (or fail) to know
about them (Redhead 1995, 27-28; Albert 2000, 64; Goldstein 2001,
48; Loewer 2001, 611). Of course, nothing is sacred, but further
explanation is needed if such a radical conceptual shift is to
appear plausible.

Against the first point Jaynes argues that entropy is indeed
epistemic even in TD (1983, 85-86) because here is no such thing as
\emph{the} entropy of a physical system. In fact, the entropy is
relative to what variables one chooses to describe the system;
depending on how we describe the system, we obtain different entropy
values. From this Ben-Menahem (2001, Sec. 3) draws the conclusion
that, Jaynes' insistence on knowledge notwithstanding, one should
say that entropy is relative to descriptions rather than to
knowledge, which would mitigate considerably the force of the
objection. This ties in with the fact (mentioned in Sklar 1999, 195)
that entropy is only defined by its function in the theory (both in
TD and in SM); we neither have a phenomenal access to it nor are
there measurement instruments to directly measure entropy. These
points do, to some extent, render an epistemic (or descriptive)
understanding of entropy more plausible, but whether they in anyway
mitigate the implausibility that attaches to an epistemic
understanding of equilibrium and the approach to equilibrium remains
an open question.


\subsection{Reductionism}\label{Reductionism - Gibbs}

How does the Gibbsian approach fare with reducing TD to SM? The
aim\index{Reduction!of thermodynamics to statistical mechanics} of a
reduction is the same as in the Boltzmannian case: deduce a revised
version of the laws of TD from SM (see Subsection \ref{Reductionism
- Boltzmann}). The differences lie in the kind of revisions that are
made. I first discuss those approaches that proffer an ontic
understanding of probabilities and then briefly discuss how
reduction could be construed.

Boltzmann took over from TD the notion that entropy and equilibrium
are properties of an individual system and sacrificed the idea that
equilibrium (and the associated entropy values) are stationary.
Gibbs, on the contrary, retains the stationarity of equilibrium, but
at the price of making entropy and equilibrium properties of an
ensemble rather than an individual system. This is because both
equilibrium and entropy are defined in terms of the probability
distribution $\rho$, which is a distribution over an ensemble and
not over an individual system. Since a particular system can be a
member of many different ensembles one can no longer assert that an
individual system is in equilibrium. This `ensemble character'
carries over to other physical quantities, most notably temperature,
which are also properties of an ensemble and not of an individual
system.

This is problematic because the state of an individual system can
change considerably as time evolves while the ensemble average does
not change at all; so we cannot infer from the behaviour of an
ensemble to the behaviour of an individual system. However, what we
are dealing with in experimental contexts are individual systems;
and so the shift to ensembles has been deemed inadequate by some.
Maudlin (1995, 147) calls it a `Pyrrhic victory' and Callender
(1999) argues that this and related problems disqualify the Gibbs
approach as a serious contender for a reduction of TD.

It is worth observing that Gibbs himself never claimed to have
reduced TD to SM and only spoke about `thermodynamic analogies' when
discussing the relation between TD and SM; see Uffink (2007,
994-996) for a discussion. The notion of analogy is weaker than that
of reduction, but it is at least an open question whether this is an
advantage. If the analogy is based on purely algebraic properties of
certain variables then it is not clear what, if anything, SM
contributes to our understanding of thermal phenomena; if the
analogy is more than a merely formal one, then at least some of the
problems that we have been discussing in connection with reduction
are bound to surface again.


\section{Conclusion}\label{Conclusion}

Before drawing some general conclusions from the discussion in
Sections \ref{The Boltzmann Approach} and \ref{The Gibbs Approach},
I would like to briefly mention some of the issues, which, for lack
of space, I could not discuss.


\subsection{Sins of Omission}\label{Omissions}

\emph{SM and the Direction of Time}.\index{Direction of time} The
discussion of irreversibility so far has focussed on the problem of
the directionality of change \emph{in} time. One can take this one
step further and claim that this directionality in fact
\emph{constitutes} the direction of time itself (the `arrow of
time'). Attempts to underwrite the arrow of time by an appeal to the
asymmetries of thermodynamics and SM can be traced back to
Boltzmann, and have been taken up by many since. The literature on
the problem of the direction of time is immense and it is impossible
to give a comprehensive bibliography here; instead I mention just
some approaches that are closely related to SM. The modern locus
classicus for a view that seeks to ground the arrow of time on the
flow of entropy is Reichenbach (1956). Earman (1974) offers a
sceptical take on this approach and provides a categorisation of the
different issues at stake. These are further discussed in Sklar
(1981; 1993, Ch. 10), Price (1996; 2002a; 2002b), Horwich (1987),
Callender (1998), Albert (2000, Ch. 6), Brown (2000), North (2002),
Castagnino \& Lombardi (2005), Hagar (2005) and Frisch (2006).


\emph{The Gibbs paradox}.\index{Gibbs' paradox} Consider a container
that is split in two halves by a barrier in the middle. The left
half is filled with gas $G_{1}$, the right half with a different gas
$G_{2}$; both gases have the same temperature. Now remove the
shutter. As a result both gases start to spread and get mixed. We
then calculate the entropy of the initial and the final state and
find that the entropy of the mixture is greater than the entropy of
the gases in their initial compartments. This is the result that we
would expect. The paradox arises from the fact that the calculations
do \emph{not} depend on the fact that the gases are different; that
is, if we assume that we have air of the same temperature on both
sides of the barrier the calculations still yield an increase in
entropy when the barrier is removed. This seems wrong since it would
imply that the entropy of a gas depends on its history and cannot be
a function of its thermodynamic state alone (as thermodynamics
requires).

What has gone wrong? The standard `textbook solution' of this
problem is that classical SM gets the entropy wrong because it makes
a mistake when counting states (see for instance Huang 1963, 153-4;
Greiner \emph{et al.} 1993, 206-8). The alleged mistake is that we
count states that differ only by a permutation of two
indistinguishable particles as distinct, while we should not do
this. Hence the culprit is a flawed notion of individuality, which
is seen as inherent to classical mechanics. The solution, so the
argument goes, is provided by quantum mechanics, which treats
indistinguishable particles in the right way.

This argument raises a plethora of questions concerning the nature
of individuality in classical and quantum mechanics, the way of
counting states in both the Boltzmann and the Gibbs approach, and
the relation of SM to thermodynamics. These issues are discussed in
Rosen (1964), Lande (1965), van Kampen (1984), Denbigh and Denbigh
(1985, Ch. 4), Denbigh and Redhead (1989), Jaynes (1992), Redhead
and Teller (1992), Mosini (1995), Costantini \& Garibaldi (1997,
1998), Huggett (1999), Gordon (2002) and Saunders (2006).


\emph{Maxwell's Demon}. \index{Maxwell's demon} Imagine the
following scenario, originating in a letter of Maxwell's written in
1867. Take two gases of different temperature that are separated
from one another only by a wall. This wall contains a shutter, which
is operated by a demon who carefully observes all molecules.
Whenever a particle moves towards the shutter from the \emph{colder}
side and the particle's velocity is greater than the mean velocity
of the particles in the \emph{hotter} gas, then the demon opens the
shutter, and so lets the particle pass through. Similarly, when a
particle heads for the shutter from within the \emph{hotter} gas and
the particle's velocity is lower than the mean velocity of the
particles of the \emph{colder} gas, then the demon lets the particle
pass through the shutter. The net effect of this is that the hotter
gas becomes even hotter and the colder one even colder. So we have a
heat transfer from the cooler to the hotter gas, and this without
doing any work; it is only the skill and intelligence of the demon,
who is able to sort molecules, that brings about the heat transfer.
But this sort of heat transfer is not allowed according to the
Second Law of thermodynamics. So the conclusion is that the demon
has produced a violation of the second law of thermodynamics.

In Maxwell's own interpretation, this thought experiment shows that
the second law is not an exceptionless law; it rather describes a
general tendency for systems to behave in a certain way, or, as he
also puts it, it shows that the second law has only `statistical
certainty'. Since Maxwell. the demon had a colourful history. In
particular, in the wake of Szilard's work, much attention has been
paid to the entropy costs of processing and storing information.
These issues are discussed in Daub (1970), Klein (1970), Leff \& Rex
(1990; 2003), Shenker (1997; 1999), Earman \& Norton (1998, 1999),
Albert (2000, Ch. 5), Bub (2001), Bennett (2003), Norton (2005),
Maroney (2005) and Ladyman \emph{et al.} (2007).


\emph{Entropy}. \index{Entropy} There are a number of related but
not equivalent concepts denoted by the umbrella term `entropy':
thermodynamic entropy, Shannon entropy, Boltzmann entropy
(fine-grained and coarse-grained), Gibbs entropy (fine-grained and
coarse-grained), Kolmogorov-Sinai entropy, von Neumann entropy and
fractal entropy, to mention just the most important ones. It is not
always clear how these relate to one another as well as to other
important concepts such as algorithmic complexity and informational
content. Depending on how these relations are construed and on how
the probabilities occurring in most definitions of entropy are
interpreted, different pictures emerge. Discussions of these issues
can be found in Grad (1961), Jaynes (1983), Wehrl (1978), Denbigh \&
Denbigh (1985), Denbigh (1989b), Barrett \& Sober (1992; 1994;
1995), Smith \emph{et al.} (1992), Denbigh (1994), Frigg (2004),
Balian (2005), and with a particular focus on entropy in quantum
mechanics in Shenker (1999), Henderson (2003), Timpson (2003),
Campisi (2005, 2008), Sorkin (2005) and Hemmo \& Shenker (2006). The
relation between entropy and counterfactuals is discussed in Elga
(2001) and Kutach (2002).


\emph{Quantum Mechanics and Irreversibility}. \index{Irreversible
processes} This review was concerned with the problem of somehow
`extracting' time-asymmetric macro laws from time-symmetric
classical micro laws. How does this project change if we focus on
quantum rather than classical mechanics? \emph{Prima facie} we are
faced with the same problems because the Schr\"{o}dinger equation is
time reversal invariant (if we allow replacing the wave function by
its complex conjugate when evolving it backwards in time). However,
in response to the many conceptual problems of quantum mechanics new
interpretations of quantum mechanics or even alternative quantum
theories have been suggested, some of which are not time reversal
invariant. Dynamical reduction theories (such as GRW theory) build
state collapses into the fundamental equation, which thereby becomes
non time-reversal invariant. Albert (1994a; 1994b; 2000, Ch. 7) has
suggested that this time asymmetry can be exploited to underwrite
thermodynamic irreversibility; this approach is discussed in Uffink
(2002). Another approach has been suggested by Hemmo and Shenker
who, in a series of papers, develop the idea that we can explain the
approach to equilibrium by environmental decoherence (2001; 2003;
2005).


\emph{Phase Transitions}. \index{Phase transitions} Most substances,
for instance water, can exist in different phases (liquid, solid,
gas). Under suitable conditions, so-called phase transitions can
occur, meaning that the substance changes from, say, the liquid to
the solid phase. How can the phenomenon of phase transitions be
understood from a microscopic point of view? This question is
discussed in Sewell (1986, Chs. 5-7), Lebowitz (1999), Liu (2001)
and Emch \& Liu (2002, Chs. 11-14).


\emph{SM methods outside physics}. Can the methods of SM be used to
deal with problems outside physics? In some cases it seems that this
is the case. Constantini \& Garibaldi (2004) present a generalised
version of the Ehrenfest flea model and show that it can be used to
describe a wide class of stochastic processes, including problems in
population genetics and macroeconomics. The methods of SM have also
been applied to markets, a discipline now known as `econophysics';
see Voit (2005) and Rickles (2008).


\subsection{Summing Up}

The foremost problem of the foundation of SM is the lack of a
generally accepted and universally used formalism, which leads to a
kind of schizophrenia in the field. The Gibbs formalism has a wider
range of application and is mathematically superior to the
Boltzmannian approach and is therefore the practitioner's workhorse.
In fact, virtually all practical applications of SM are based on the
Gibbsian machinery. The weight of successful applications
notwithstanding, a consensus has emerged over the last decade and a
half that the Gibbs formalism cannot explain why SM works and that
when it comes to foundational issues the Boltzmannian approach is
the only viable option (see Lavis (2005) and references therein).
Hence, whenever the question arises of why SM is so successful, an
explanation is given in Boltzmannian terms .

This is problematic for at least two reasons. First, at least in its
current form, the Boltzmann formalism has a very limited range of
applicability. The Boltzmann formalism only applies to non (or very
weakly) interacting particles and at the same time it is generally
accepted that the Past Hypothesis, an assumption about the
\emph{universe as a whole}, is needed to make it work. But the
universe as a whole is not an a collection of weakly interacting
system, not even approximately.

Second, even if the internal problems of the Boltzmann approach can
be solved, we are left with the fact that what delivers the goodies
in `normal science' is the Gibbs rather than the Boltzmann approach.
This would not be particularly worrisome if the two formalisms were
intertranslatable or equivalent in some other sense (like, for
instance, the Schr\"{o}dinger and the Heisenberg picture in quantum
mechanics). However, as we have seen above, this is not the case.
The two frameworks disagree fundamentally over what the object of
study is, the definition of equilibrium, and the nature of entropy.
So even if all the internal difficulties of either of these
approaches were to find a satisfactory solution, we would still be
left with the question of how the two relate.

A suggestion of how these two frameworks could be reconciled has
recently been presented by Lavis (2005). His approach involves the
radical suggestion to give up the notion of equilibrium, which is
binary in that systems either are or not in equilibrium, and to
replace it by the continuous property of `commonmess'. Whether this
move is justified and whether it solves the problem is a question
that needs to be discussed in the future.


\section*{Acknowledgements}
\addcontentsline{toc}{section}{Acknowledgements}

I would like to thank Jeremy Butterfield, Craig Callender, Adam
Caulton, Jos\'{e} D\'{i}ez, Foad Dizadji-Bahmani, Isabel Guerra
Bobo, Stephan Hartmann, Carl Hoefer, David Lavis, Olimpia Lombardi,
Wolfgang Pietsch, Jos Uffink, and Charlotte Werndl, for invaluable
discussions and comments on earlier drafts. Thanks to Dean Rickles
for all the hard editorial work, and for his angelic patience with
my ever changing drafts. I would also like to acknowledge financial
support from two project grants of the Spanish Ministry of Science
and Education (SB2005-0167 and HUM2005-04369).


\section*{Appendix}

\addcontentsline{toc}{section}{Appendix}

\subsection*{A. Classical Mechanics}

\addcontentsline{toc}{subsection}{A. Classical Mechanics}

CM can be presented in various more or less but not entirely
equivalent formulations: Newtonian mechanics, Lagrangean mechanics,
Hamiltonian mechanics and Hamiltion-Jacobi theory; for comprehensive
presentations of these see Arnold (1978), Goldstein (1980),  Abraham
\& Marsden (1980) and Jos\'{e} \& Saletan (1998). Hamiltonian
Mechanics (HM) is best suited to the purposes of SM; hence this
appendix focuses entirely on HM.

CM describes the world as consisting of point-particles, which are
located at a particular point in space and have a particular
momentum. A system's state is fully determined by a specification of
each particle's position and momentum. Conjoining the space and
momentum dimension of all particles of a system in one vector space
yields the so-called phase space $\Gamma$ of the system. The phase
space of a system with $m$ degrees of freedom is $2m$ dimensional;
for instance, the phase space of a system consisting of $n$
particles in three-dimensional space has $6n$ dimensions. Hence, the
state of a mechanical system is given by the $2m$-tuple
$x:=(q,p):=(q_{_{1}}, \ldots, q_{_{m}}, p_{_{1}}, \ldots, p_{_{m}})
\in \Gamma$. The phase space $\Gamma$ is endowed with a Lebesgue
measure $\mu_{_{L}}$, which, in the context of SM, is also referred
to as the `standard measure' or the `natural measure'.

The time evolution of the system is governed by Hamilton's equation
of motion:\footnote{The dot stands for the total time derivative:
$\dot{f}:=df/dt$.}

\begin{equation}
    \dot{q}_{_{i}} = \frac{\partial H}{\partial p_{_{i}}}  \hspace{1cm}
    \textnormal{and} \hspace{1cm}
    \dot{p}_{_{i}} = - \frac{\partial H}{\partial q_{_{i}}}, \hspace{1cm}
    i=1, \ldots, m,
\end{equation}
where $H(q,p,t)$ is the so-called `Hamiltonian' of the system. Under
most circumstances the Hamiltonian is the energy of the system (this
is not true in systems with time dependent boundary conditions, but
these do not play a r\^{o}le in the present discussion).

If the Hamiltonian satisfies certain conditions (see Arnold (2006)
for a discussion of these) CM is deterministic in the sense that the
state $x_{0}$ of the system \textit{at some particular} instant of
time $t_{0}$ (the so-called `initial condition') uniquely determines
the state of the system at any other time $t$. Hence, each point in
$\Gamma$ lies on exactly one trajectory (i.e. no two trajectories in
phase space can ever cross) and $H(q,p,t)$ defines a one parameter
group of transformations $\phi_{t}$, usually referred to as `phase
flow', mapping the phase space onto itself: $x \rightarrow
\phi_{t}(x)$ for all $x \in \Gamma$ and all $t$.

A quantity $f$ of the system is a function of the coordinates and
(possibly) time: $f(q,p,t)$. The time evolution of $f$ is given by

\begin{equation}\label{f-dot}
    \dot{f} = \{f, H\} + \frac{\partial f}{\partial t},
\end{equation}
where $\{\,\, , \,\, \}$ is the so-called Poisson bracket:
\begin{equation}
    \{g, h\} := \sum_{i} \bigg{[} \frac{\partial g}{\partial q_{i}}
    \frac{\partial h}{\partial p_{i}} - \frac{\partial h}{\partial q_{i}}
    \frac{\partial g}{\partial p_{i}}\bigg{]},
\end{equation}
for any two differentiable functions $g$ and $h$ on $\Gamma$.

From this it follows that $H$ is a conserved quantity iff it does
not explicitly depend on time. In this case we say that the motion
is {\it stationary}, meaning that the phase flow depends only on the
\textit{time interval} between the beginning of the motion and `now'
but not on the choice of the initial time. If $H$ is a conserved
quantity, the motion is confined to a $2m-1$ dimensional
hypersurface $\Gamma_{E}$, the so-called `energy hypersurface',
defined by the condition $H(q,p)=E$, where $E$ is the value of the
total energy of the system.

Hamiltonian dynamics has three distinctive features, which we will
now discuss.

\textit{Liouville's theorem} \index{Liouville's theorem} asserts that the Lebesgue measure (in
this context also referred to as `phase volume') is invariant under
the Hamiltonian flow:

\begin{quote}
    For any Lebesgue measurable region $R \subseteq \Gamma $ and for any time
    $t$: $R$ and the image of $R$ under the Hamiltonian flow,
    $\phi_{t}(R)$, have the same Lebesgue measure; i.e.
    $\mu_{_{L}}(R)=\mu_{_{L}}(\phi_{t}(R))$.
\end{quote}
\noindent In geometrical terms, a region $R$ can (and usually will)
change its shape but not its volume under the Hamiltonian time
evolution.

This also holds true if we restrict the motion of the system to the
energy hypersurface $\Gamma_{E}$, provided we choose the `right'
measure on $\Gamma_{E}$. We obtain this measure, $\mu_{_{L, \, E}}$,
by restricting $\mu_{_{L}}$ to $\Gamma_{E}$ so that the $6n-1$
dimensional hypervolume of regions in $\Gamma_{E}$ is conserved
under the dynamics. This can be achieved by dividing the surface
element $d\sigma_{_{E}}$ on $\Gamma_{E}$ by the gradient of $H$ (Kac
1959, 63):

\begin{equation}\label{Restricted Lebesgue Measure}
    \mu_{_{L, \, E}}(R_{E}) \, := \, \int_{R_{E}}
    \frac{d\sigma_{_{E}}}{\| {\textnormal{grad}} H \|}
\end{equation}

\noindent for any $R_{E} \subseteq \Gamma_{E}$, where

\begin{equation}\label{Grad}
    \| {\textnormal{grad}} H \| := {\Bigg [} \sum_{k=1}^{n} {\Big (}
    \frac{\partial H}{\partial p_{k}}{\Big )}^{2} +
    {\Big (}
    \frac{\partial H}{\partial q_{k}}{\Big )}^{2}
    {\Bigg ]}^{1/2}.
\end{equation}

\noindent We then have $\mu_{_{L, \,E}}(R_{E})= \mu_{_{L,
\,E}}(\phi_{t}(R_{E}))$ for all $R_{E} \subseteq \Gamma_{E}$ and for
all $t$.

\textit{Poincar\'{e}'s recurrence theorem}: Roughly speaking,
Poincar\'{e}'s recurrence theorem says that any system that has
finite energy and is confined to a finite region of space must, at
some point, return arbitrarily close to its initial state, and does
so infinitely many times. The time that it takes the system to
return close to its initial condition is called `Poincar\'{e}
recurrence time'. Using the abstract definition of a dynamical
system introduced in Unit \ref{Ergodic Theory}, the theorem can be
stated as follows:

\begin{quote}
Consider an area-preserving mapping of the phase space $X$ of a
system onto itself, $\phi_{t}(X)=X$, and suppose that its measure is
finite, $\mu(X)< \infty$. Then, for any measurable subset $A$ with
$\mu_{_{L}}(A) > 0$ of $X$, almost every point $x$ in $A$ returns to
$A$ infinitely often; that is, for all finite times $\tau$ the set
$B:= \{x | x \in A {\textnormal{ and for all times}} \ t \geq \tau :
\phi_{t}x \notin A\}$ has measure zero.
\end{quote}

\noindent The Hamiltonian systems that are of interest in SM satisfy
the requirements of the theorem if we associate $X$ with the
accessible region of the energy hypersurface.

\textit{Time reversal invariance}. \index{Time reversal invariance}
Consider, say, a ball moving from left to right and record this
process on videotape. Intuitively, time reversal amounts to playing
the tape backwards, which makes us see a ball moving from right to
left. Two ingredients are needed to render this idea precise, a
transformation reversing the direction of time, $t \rightarrow -t$,
and the reversal of a system's instantaneous state. In some contexts
it is not obvious what the instantaneous state of a system is and
what should be regarded as its reverse.\footnote{A recent
controversy revolves around this issue. Albert (2000, Ch. 1) claims
that, common physics textbook wisdom notwithstanding, neither
electrodynamics, nor quantum mechanics nor general relativity nor
any other fundamental theory turns out to be time reversal invariant
once the instantaneous states and their reversals are defined
correctly. This point of view has been challenged by Earman (2002),
Uffink (2002) and Malament (2004), who defend common wisdom; for a
further discussion see Leeds (2006).} In the case of HM, however,
the ball example provides a lead. The instantaneous state of a
system is given by $(q, p)$, and in the instant in which the time is
reversed the ball suddenly `turns around' and moves from right to
left. This suggests the sought-after reversal of the instantaneous
state amounts to changing the sign of the momentum: $R(q, p):=(q,
-p)$, where $R$ is the reversal operator acting on instantaneous
states.

Now consider a system in the initial state $(q_{_{i}}, p_{_{i}})$ at
time $t_{_{i}}$ that evolves, under the system's dynamics, into the
final state $(q_{_{f}}, p_{_{f}})$ at some later time $t_{_{f}}$.
The entire process (`history') is a parametrised curve containing
all intermediates states: $h := \{(q(t), p(t)) | t \in [t_{_{i}},
t_{_{f}}] \}$, where $(q(t_{_{i}}), p(t_{_{i}}))= (q_{_{i}},
p_{_{i}})$ and $(q(t_{_{f}}), p(t_{_{f}}))= (q_{_{f}}, p_{_{f}})$.
We can now define the time-reversed process of $h$ as follows: $Th
:= \{R(q(-t), p(-t)) | t \in [-t_{_{f}}, -t_{_{i}}] \}$, where $T$
is the time-reversal operator acting on histories. Introducing the
variable $\tau := -t$ and applying $R$ we have $Th = \{(q(\tau), -
p(\tau)) | \, \tau  \in  [t_{_{i}}, t_{_{f}}] \}$. Hence, $Th$ is a
process in which the system evolves from state $R(q_{_{f}},
p_{_{f}})$ to state $R(q_{_{i}}, p_{_{i}})$ when $\tau$ ranges over
$[t_{_{i}}, t_{_{f}}]$.

Call the class of processes $h$ that are allowed by a theory $A$; in
the case of HM $A$ contains all trajectories that are solutions of
Hamilton's equation of motion. A theory is time reversal invariant
(TRI) iff for every $h$: if $h \in A$ then $Th \in A$ (that is, if
$A$ is closed under time reversal). Coming back to the analogy with
videotapes, a theory is TRI iff a censor who has to ban films
containing scenes which violate the law of the theory issues a
verdict which is the same for either direction of playing the film
(Uffink 2001, 314). This, however, does not imply that the processes
allowed by a TRI theory are all palindromic in the sense that the
processes \emph{themselves} look the same when played backwards;
this can but need not be the case.

HM is TRI in this sense. This can be seen by time-reversing the
Hamiltonian equations: carry out the transformations  $t \rightarrow
\tau$ and $(q, p) \rightarrow R (q, p)$ and after some elementary
algebraic manipulations you find $dq_{i}/d\tau = \partial H
/\partial p_{i}$ and $dp_{i}/d\tau = - \partial H /\partial p_{i}$,
$i=1, ..., m$. Hence the equations have the same form in either
direction of time, and therefore what is allowed in one direction of
time is also allowed in the other.\footnote{There was some
controversy over the question of whether classical mechanics really
is TRI; see Hutchison (1993, 1995a, 1995b), Savitt (1994) and
Callender (1995). However, the moot point in this debate was the
status of frictional forces, which, unlike in Newtonian Mechanics,
are not allowed in HM. So this debate has no implications for the
question of whether HM is TRI.}

The upshot of this is that if a theory is TRI then the following
holds: if a transition from state  $(q_{_{i}}, p_{_{i}})$ to state
$(q_{_{f}}, p_{_{f}})$ in time span $\Delta := t_{_{f}} - t_{_{i}}$
is allowed by the lights of the theory, then the transition from
state $R(q_{_{f}}, p_{_{f}})$ to state $R(q_{_{i}}, p_{_{i}})$ in
time span $\Delta$ is allowed as well, and vice versa. This is the
crucial ingredient of Loschmitd's reversibility objection (see
Subsection \ref{Loschmidt's Reversibility Objection}).


\subsection*{B. Thermodynamics}

\addcontentsline{toc}{subsection}{B. Thermodynamics}

Thermodynamics is a theory about macroscopic quantities such as
pressure, volume and temperature and it is formulated solely in
terms of these; no reference to unobservable microscopic entities is
made. At its heart lie two laws, the First Law and Second Law of TD.
Classical presentations of TD include Fermi (1936), Callen (1960),
Giles (1964) and Pippard (1966).

\index{Thermodynamics!laws of}
\emph{The first law of thermodynamics}. The first law says that
there are two ways of exchanging energy with a system, putting heat
into it and doing work on it, and that energy is a conserved
quantity:

\begin{equation}
    \Delta U = \Delta Q + \Delta W,
    \label{eq:¥}
\end{equation}¥

\noindent where $\Delta U$ is the energy put into the system, and
$\Delta Q$ and $\Delta W$ are, respectively, the heat and work that
went into the system. Hence, put simply, the first law, says that
one cannot create energy and thereby rules out the possibility of a
perpetual motion machine.

\emph{The second law of thermodynamics}. The First Law does not
constrain the ways in which one form of energy can be transformed
into another one and how energy can be exchanged between systems or
parts of a system. For instance, according to the first law it is in
principle possible to transform heat into work or work into heat
according to one's will, provided the total amount of heat is
equivalent to the total amount of work. However, it turns out that
although one can always transform work in to heat, there are severe
limitations on the ways in which heat can be transformed into work.
These limitations are specified by the Second Law.

Following the presentation in Fermi (1936, 48-55), the main tenets
of the Second Law can be summarised as follows. Let $A$ and $B$ be
two equilibrium states of the system. Then consider a quasi-static
transformation (i.e. one that is infinitely gentle in the sense that
it proceeds only through equilibrium states), which takes the system
from $A$ to $B$ . Now consider the integral

\begin{equation}
    \int_{A}^{B}\frac{dQ}{T},
\end{equation}
where $T$ is the temperature of the system and $dQ$ is the amount of
heat quasi-statically absorbed by the system. One can then prove
that the value of this integral does not depend on the sequence by
which one gets from $A$ to $B$; it only depends on $A$ and $B$
themselves.

Now choose an arbitrary equilibrium state $E$ of the system and call
it the \textit{standard state}. Then we can define the entropy of
the state $A$ as

\begin{equation}
    S(A) = \int_{E}^{A}\frac{dQ}{T},
\end{equation}
where the integral is taken over a quasi-static transformation.

With this at hand we can formulate the Second Law of thermodynamics:

\begin{equation}
    \int_{A}^{B}\frac{dQ}{T} \leq S(B) - S(A).
\end{equation}

For a totally isolated system we have $dQ=0$. In this case the
second law takes the particularly intuitive form:

\begin{equation}
    S(A) \leq S(B).
\end{equation}

\noindent That is, for any transformation in an isolated system, the
entropy of the final state can never be less than that of the
initial state. The equality sign holds if, and only if, the
transformation is quasi-static.

Thermodynamics is not free of foundational problems. The status of
the Second Law is discussed in Popper (1957), Lieb \& Yngvason
(1999) and Uffink (2001); Cooper (1967), Boyling (1972), Moulines
(1975, 2000), Day (1977) and Garrido (1986) examine the formalism of
TD and possible axiomatisations. The nature of time in TD is
considered in Denbigh (1953) and Brown and Uffink (2001); Rosen
(1959), Roberts and Luce (1968) and Liu (1994) discuss the
compatibility of TD and relativity theory. Wickens (1981) addresses
the issue of causation in TD.

\end{document}